\documentclass[12pt]{iopart}

\usepackage{graphicx}
\usepackage{dcolumn}
\usepackage{bm}

\usepackage[utf8]{inputenc}
\usepackage[T1]{fontenc}
\usepackage{newtxtext}
\usepackage[varvw]{newtxmath}
\usepackage{longtable}
\usepackage{float}
\usepackage[dvipsnames]{xcolor}

\begin{document}

\title[A multi-component model of plasma turbulence and neutral dynamics]{A self-consistent multi-component model of plasma turbulence and kinetic neutral dynamics for the simulation of the tokamak boundary}

\author{A Coroado $^1$, P Ricci $^1$}

\address{$^1$ \'Ecole Polytechnique F\'ed\'erale de Lausanne (EPFL), Swiss Plasma Center (SPC), CH-1015 \\
Lausanne, Switzerland.}
\ead{andre.caladocoroado@epfl.ch}
\vspace{10pt}
\begin{indented}
\item[]June 2021
\end{indented}

\begin{abstract}
A self-consistent model is presented for the simulation of a multi-component plasma in the tokamak boundary. A deuterium plasma is considered, with the plasma species that include electrons, deuterium atomic ions and deuterium molecular ions, while the deuterium atoms and molecules constitute the neutral species. The plasma and neutral models are coupled via a number of collisional interactions, which include dissociation, ionization, charge-exchange and recombination processes. The derivation of the three-fluid drift-reduced Braginskii equations used to describe the turbulent plasma dynamics is presented, including its boundary conditions. The kinetic advection equations for the neutral species are also derived, and their numerical implementation discussed. The first results of multi-component plasma simulations carried out by using the GBS code are then presented and analyzed, being compared with results obtained with the single-component plasma model.
\end{abstract}

\section{\label{sec:level1}Introduction}

The boundary of a tokamak plays a crucial role in determining the overall performance of the device, as it sets the confinement of particles and heat, determines the heat exhaust to the vessel walls and controls the impurity level in the core \cite{Stangeby2000}. The boundary is also the region where the plasma is fueled and helium ashes generated by fusion reactions are removed.

The tokamak boundary is characterized by the presence of several ion and neutral species that interact through a complex set of collisional processes \cite{CSP1988,Stangeby2000}. In particular, neutral atoms and molecules are relevant in the boundary as they result from processes such as the plasma recycling at the vessel walls and gas puffs. Recycling occurs because ions and electrons, transported along the magnetic-field lines by the parallel flow or across them by turbulent motion, eventually end at the vessel, where they recombine and re-enter the plasma as neutral particles, either being reflected, in which case they keep the energy of the original ion or, following an absorption process, being reemitted at the wall temperature. In case of absorption, a significant fraction of the atoms may associate to form molecules before being reemitted back to the plasma \cite{Zhang2019}. The exact probability of reflection or reemission, as well as the probability that atoms associate into molecules, depends on the physical properties of the limiter or divertor plate material \cite{Stotler1994}. At the same time, external injection of neutral molecules can be used to fuel the plasma, reduce the heat load on the vessel wall (e.g., by reducing the temperature of the plasma and hence inducing volumetric recombination processes), or diagnose the plasma. 

The neutral atoms and molecules interact with the plasma in the tokamak boundary. Indeed, neutral atoms and molecules can be ionized, thus generating atomic and molecular ions, leading to a multi-component plasma. Molecular species also undergo dissociative processes that break them into mono-atomic species. Recombination, charge-exchange, elastic and inelastic collisions are also at play. These collisional interactions convert neutral particles into ions and electrons and vice versa, affect the temperature of the plasma species because of the energy required to trigger ionization and dissociation processes and modify the plasma velocity. As a result, since the dynamics of the plasma in the boundary are strongly influenced by its interaction with neutral species, it is important that simulations of the plasma dynamics in the tokamak boundary take into account its multi-component nature and the interactions between the different species to provide reliable quantitative predictions. 

The description of a multi-component plasma is usually addressed by means of a fluid-diffusive model, which typically considers a version of the Braginskii fluid equations for the plasma species simplified by modelling cross-field transport through empirical anomalous transport coefficients. This approach is used by codes such as B2 \cite{Braams1996,Baelmans1996}, EDGE2D \cite{Simonini1994}, EMC3 \cite{Feng1999}, SOLEDGE-2D \cite{Bufferand2011} and TECXY \cite{Zagorski1998}. Sometimes, neutral particle species are also modelled using a diffusive fluid approach \cite{Sawada1995}, for example in the UEDGE code \cite{Pigarov2003}. However, diffusive models are no longer valid when the neutral mean free path is large, i.e. of the order of the plasma gradient scale length, which is often the case in the tokamak boundary. For this reason, neutrals are more commonly modelled by using a kinetic description valid for all ranges of mean free path. These models, typically based on Monte Carlo methods for the numerical solution, are implemented in the DEGAS2 \cite{Stotler1994}, EIRENE \cite{Reiter2005}, GTNEUT \cite{Mandrekas2004} and NEUT2D \cite{Shimizu2003} codes. As a matter of fact, heat exhaust studies strongly rely on integrated neutral-plasma simulations of the tokamak boundary, which are most often based upon the coupling of the aforementioned fluid-diffusive models for the multi-component plasma and Monte Carlo-based models for the several neutral species, such as B2-EIRENE \cite{Reiter2005}, EDGE2D-EIRENE \cite{Lawson2017}, EMC3-EIRENE \cite{Lore2012} and SOLPS \cite{Schneider2006}).

In order to shed light on perpendicular transport processes, simulations of plasma turbulence in the tokamak boundary have been carried out since a decade by using fluid and gyrofluid models, implemented in codes such as BOUT++ \cite{Dudson2009} (and its module Hermes \cite{Dudson2017}), FELTOR \cite{Wiesenberger2019}, GBS \cite{Ricci2012,Halpern2016}, GDB \cite{Francisquez2017}, GRILLIX \cite{Stegmeir2018}, HESEL \cite{Thrysoe2018} and TOKAM3X \cite{Baudoin2018}. Kinetic models, implemented in other codes such as Gkeyll \cite{Shi2017} and XGC1 \cite{Chang2008,Ku2009}, have also been used. These simulations have allowed remarkable progress in the understanding of the mechanisms underlying turbulence and cross-field transport in a single-ion species boundary plasma. On the other hand, multi-component plasma simulations that include turbulent transport processes are still in their early days. Recent progress was made thanks to the synergy between the SOLEDGE2D \cite{Bufferand2017} and TOKAM3X \cite{Baudoin2018} codes. Based on the EIRENE Monte Carlo code for the kinetic simulation of the neutral species, multi-component plasma simulations are now enabled by the SOLEDGE3X code. The investigations carried out with SOLEDGE3X focused on the study of the dynamics of carbon impurities in the tokamak boundary \cite{Bufferand2019}. Progress was also made by coupling the two-dimensional fluid code HESEL \cite{Thrysoe2018} with a one-dimensional fluid-diffusive model for the neutral particles, which accounts for both atomic and molecular species. The resulting nHESEL \cite{Thrysoe2016,Thrysoe2018,Thrysoe2018a} code thus allows for the simulation of a single-ion plasma including the interactions with three neutral species: cold hydrogen molecules puffed into the system, warm atoms resulting from the dissociation of the hydrogen molecules and hot hydrogen atoms generated by charge-exchange processes. Such a model was used to study the plasma fueling in the presence of gas puffs and the formation of a density shoulder in the tokamak boundary at a high gas puffing rate.

In the present work, we describe the development and numerical implementation in the GBS code of a multi-component model that addresses the turbulent multi-ion species plasma dynamics through a set of fluid drift-reduced Braginskii equations, while each multiple neutral species are simulated by solving a kinetic equation. This work generalizes the implementation of the neutral-plasma interaction in GBS described in \cite{Wersal2015} for single-component plasmas. Single-component GBS simulations were used to study the electron temperature drop along the magnetic field \cite{Wersal2017} and to determine the influence of neutrals on gas puff imaging diagnostics \cite{Wersal2017a}. The model has been improved recently through the implementation of mass-conservation by taking toroidal geometry consistently into account and by making use of particle-conserving boundary conditions to properly describe the recycling processes \cite{Coroado2021}.

While the methodology presented in this work has the potential to include an arbitrary number of particle species and the corresponding complex scenarios, we consider a deuterium plasma, composed of five different particle species: three charged particle species, namely electrons ($e^-$), monoatomic deuterium ions ($D^+$) and diatomic deuterium ions ($D_2^+$), and two neutral species, namely deuterium atoms ($D$) and molecules ($D_2$).  

The model constitutes the first implementation of a kinetic multi-species model that avoids the statistical noise from the Monte Carlo method. In fact, the neutral kinetic equation, valid for any neutral mean free path, is solved by discretizing the kinetic equation integrated along the neutral path. The model has the potential to provide the fundamental elements necessary for the description and understanding of the key mechanisms taking place in the boundary, such as the fueling or gas puff imaging, where molecular species play an important role. 

The results of the first simulation carried out with the multi-component model are also described in the present work, shedding some light on the processes underlying plasma fueling. In particular, for the limited configuration and sheath limited regime considered, we show that molecular dissociation processes have an impact on the location of the ionization source and plasma profiles, with respect to single-component simulations.

The outline for the present paper is as follows. After the Introduction, the collisional processes at play within the multi-component deuterium plasma model now implemented in GBS is presented in Sec. 2. In Sec. 3, we guide the reader through the derivation of the set of drift-reduced Braginskii equations used to describe a multi-component plasma, extending the approach previously followed by the single-component version of GBS. In Sec. 4, we present the boundary conditions we apply at the tokamak wall. The kinetic model for the neutral species is discussed in Sec. 5, where the numerical approach is also described, based on discretizing the kinetic equation integrated along the neutral path, in a generalization of the approach developed in \cite{Wersal2015} for a single neutral species model. Finally, in Sec. 6 we present and discuss the preliminary results of the first multi-component plasma GBS simulations, analyzing the impact of the molecules on the plasma dynamics, also presenting comparison to results from previous single-ion plasma simulations of GBS. The summary follows. In App. A, we derive the average energy of the reaction products and the average electron energy loss for the dissociative processes considered in the model. App. B presents the derivation of the friction and thermal force terms in the velocity and temperature equations used for the multi-component plasma description, following the Zhdanov closure \cite{Zhdanov2002} and considering the approach described in \cite{Bufferand2019}. App. C features the list of kernel functions used to express the system of equations solved for the neutral species, while App. D presents the neutral system of equations in the matrix form implemented in GBS.

\section{\label{sec:level2} Collisional processes in multi-component deuterium plasmas}

In the present paper we aim at describing an experimentally relevant multi-component deuterium plasma. Similarly to the works in \cite{Sawada1995}, \cite{Stotler1996} and \cite{Wensing2019}, the plasma we consider is composed of the $e$, $\text{D}^+$ and $\text{D}_2^+$ species and we consider the $\text{D}$ and $\text{D}_2$ neutral species. The $\text{D}_2$ molecules are present as the result of the association of atoms at the vessel walls and external injection. The $\text{D}_2$ molecules can be ionized, thus giving rise to $\text{D}_2^+$ ions, while dissociative processes are responsible for generating mono-atomic ions, $\text{D}^+$, and neutrals, $\text{D}$, the later being possibly further ionized into $\text{D}^+$. In general, the presence of $\text{D}_2^{2+}$ ions is negligible in deuterium plasmas. Additionally, in the typical conditions of the tokamak boundary considered here, also the concentration of species that might be present in a deuterium plasma, such as the $\text{D}^-$ and $\text{D}_3^+$ ions, is negligible  \cite{Sawada1995,Reiter2011,Janev1987}. Considering five different species contrasts with the three species model used in the previous GBS simulations of a single-ion species plasma \cite{Wersal2015}, where only mono-atomic deuterium ions and neutrals are evolved. We highlight that, by introducing the tools necessary to deal with the fundamental processes at play in multi-component plasmas, the present model can be further extended to describe more complex scenarios that include a larger number of plasma and neutral species. 

The charged particle and neutral species are coupled by means of collisional processes, which include ionization, recombination, charge-exchange and dissociation processes, as well as electron-neutral collisions. These processes appear both in the neutral and plasma species model as particle and heat sources or sinks, as well as friction terms.  

We henceforth list the collisional processes considered in our multi-component model, as well as their respective reaction rates, in Table \ref{collisions}. We remark that we neglect the distinction between fundamental and excited states for atoms, molecules and ions. In particular, we use the total cross section for each process considering the sum over the accessible electronic states of the reactants and products, following \cite{Reiter2011} and \cite{Janev1987}. Based on momentum and energy considerations, we also compute the values of the velocity and energy of the collision products. Since these also depend on the electronic states of the reactants and products, we perform an average over the states relevant to a given reaction, taking into account the cross section of each state. 

We denote with $v_{\text{e}}$, $v_{\text{D}^+}$ and $v_{\text{D}_2^+}$ the modulus of the electron, $\text{D}^+$ and $\text{D}_2^+$ velocities, while $n_{\text{e}}$, $n_{\text{D}^+}$ and $n_{\text{D}_2^+}$ represent their densities. The cross sections $\sigma_{\text{iz,D}}$ and $\sigma_{\text{iz,D}_2}$ refer to the collisions leading to the ionization of $\text{D}$ and $\text{D}_2$ respectively, $\sigma_{\text{rec,D}^+}$ and $\sigma_{\text{rec,D}_2^+}$ are the cross sections for recombination of $\text{D}^+$ and $\text{D}_2^+$ with electrons, $\sigma_{\text{e-D}}$ and $\sigma_{\text{e-D}_2}$ stand for the cross sections of elastic collisions between electrons and $\text{D}$ and $\text{D}_2$ respectively, $\sigma_{\text{diss,D}_2}$ and $\sigma_{\text{diss,D}_2^+}$ represent the dissociation cross sections of $\text{D}_2$ and $\text{D}_2^+$, $\sigma_{\text{diss-iz,D}_2}$ and $\sigma_{\text{diss-iz,D}_2^+}$ are the cross sections for dissociative ionization of $\text{D}_2$ and $\text{D}_2^+$, $\sigma_{\text{diss-rec,D}_2^+}$ is the cross section of dissociative recombination of $\text{D}_2^+$ ions and, finally, $\sigma_{\text{cx,D}^+}$, $\sigma_{\text{cx,D}_2^+}$, $\sigma_{\text{cx,D-D}_2^+}$ and $\sigma_{\text{cx,D}_2-\text{D}^+}$ represent the cross sections for $\text{D}-\text{D}^+$, $\text{D}_2-\text{D}_2^+$, $\text{D}-\text{D}_2^+$ and $\text{D}_2-\text{D}^+$ charge-exchange interactions. 

By considering Krook collision operators, the collision frequencies for ionization, recombination, elastic collisions and dissociative processes are computed as the average over the electron velocity distribution function, neglecting therefore the velocity of the colliding massive particle ($\text{D}$, $\text{D}_2$, $\text{D}^+$ or $\text{D}_2^+$) when computing the relative velocity between the electron and the other particle. In fact, electrons have significantly larger thermal velocity than ions or neutrals. As for the charge-exchange interactions between $\text{D}^+$ ions and the neutral species $\text{D}$ and $\text{D}_2$, since the dependence of the cross section upon the ion-neutral relative velocity is weak \cite{Stangeby2000}, we neglect the velocity of the neutral particles ($\text{D}$ or $\text{D}_2$) when evaluating the relative velocity of the colliding particles (we note that the velocity of a neutral particle is typically smaller than the velocity of the ions). Thus, we compute the reaction rates $\sigma_{\text{cx,D}^+}$ and $\sigma_{\text{cx,D}_2-\text{D}^+}$ by averaging over the distribution function of the $\text{D}^+$ species, which we assume to be a Maxwellian with temperature $T_{\text{D}^+}$. Following the same approach when computing the cross section of charge-exchange interactions between $\text{D}_2^+$ ions and the $\text{D}_2$ and $\text{D}$ neutrals, we average the cross sections $\sigma_{\text{cx,D}_2^+}$ and $\sigma_{\text{cx,D}-\text{D}_2^+}$ over the $\text{D}_2^+$ distribution function, assumed to be a Maxwellian of temperature $T_{\text{D}_2^+}$.

\begin{table}[!ht]
\caption{\label{collisions}Collisional processes considered and their respective reaction rates.}
\begin{indented}
\item[]\begin{tabular}{@{}lll} 
\br
\textbf{Collisional process} & \textbf{Equation} & \textbf{Reaction Frequency} \\
\mr
Ionization of $\text{D}$ & $\text{e}^- +\text{D} \rightarrow 2\text{e}^- + \text{D}^+$ & $\nu_{\text{iz,D}} = n_{\text{e}}\left\langle v_{\text{e}}\sigma_{\text{iz,D}}(v_{\text{e}})\right\rangle$ \\ 
Recombination of $\text{D}^+$ and $\text{e}^-$ & $\text{e}^- + \text{D}^+ \rightarrow \text{D}$ & $\nu_{\text{rec,D}^+} = n_{\text{e}}\left\langle v_{\text{e}}\sigma_{\text{rec,D}^+}(v_{\text{e}})\right\rangle$ \\ 
$\text{e}^- -\text{D}$ elastic collisions & $\text{e}^- +\text{D} \rightarrow \text{e}^- +\text{D}$ & $\nu_{\text{e-D}} = n_{\text{e}}\left\langle v_{\text{e}}\sigma_{\text{\text{e-D}}}(v_{\text{e}})\right\rangle$ \\ 
Ionization of $\text{D}_2$ & $\text{e}^- + \text{D}_2 \rightarrow 2\text{e}^- + \text{D}_2^+$ & $\nu_{\text{iz,D}_2} = n_{\text{e}}\left\langle v_{\text{e}}\sigma_{\text{iz,D}_2}(v_{\text{e}})\right\rangle$ \\ 
Recombination of $\text{D}_2^+$ and $\text{e}^-$ & $\text{e}^- + \text{D}_2^+ \rightarrow \text{D}_2$ & $\nu_{\text{rec,D}_2^+} = n_{\text{e}}\left\langle v_{\text{e}}\sigma_{\text{rec,D}_2^+}(v_{\text{e}})\right\rangle$ \\ 
$\text{e}^- -\text{D}_2$ elastic collisions & $\text{e}^- + \text{D}_2 \rightarrow \text{e}^- + \text{D}_2$ & $\nu_{\text{e-D}_2} = n_{\text{e}}\left\langle v_{\text{e}}\sigma_{\text{e-D}_2}(v_{\text{e}})\right\rangle$ \\ 
Dissociation of $\text{D}_2$ & $\text{e}^- + \text{D}_2 \rightarrow \text{e}^- + \text{D} + \text{D}$ & $\nu_{\text{diss,D}_2} = n_{\text{e}}\left\langle v_{\text{e}}\sigma_{\text{diss,D}_2}(v_{\text{e}})\right\rangle$ \\ 
Dissociative ionization of $\text{D}_2$ & $\text{e}^- + \text{D}_2 \rightarrow 2\text{e}^- + \text{D} +  \text{D}^+$ & $\nu_{\text{diss-iz,D}_2} = n_{\text{e}}\left\langle v_{\text{e}}\sigma_{\text{diss-iz,D}_2}(v_{\text{e}})\right\rangle$ \\ 
Dissociation of $\text{D}_2^+$ & $\text{e}^- + \text{D}_2^+ \rightarrow \text{e}^- + \text{D} +  \text{D}^+$ & $\nu_{\text{diss,D}_2^+} = n_{\text{e}}\left\langle v_{\text{e}}\sigma_{\text{diss,D}_2^+}(v_{\text{e}})\right\rangle$ \\ 
Dissociative ionization of $\text{D}_2^+$ & $\text{e}^- + \text{D}_2^+ \rightarrow 2\text{e}^- + 2\text{D}^+$ & $\nu_{\text{diss-iz,D}_2^+} = n_{\text{e}}\left\langle v_{\text{e}}\sigma_{\text{diss-iz,D}_2^+}(v_{\text{e}})\right\rangle$ \\ 
Dissociative recombination of $\text{D}_2^+$ & $\text{e}^- + \text{D}_2^+ \rightarrow 2\text{D}$ & $\nu_{\text{diss-rec,D}_2^+} = n_{\text{e}}\left\langle v_{\text{e}}\sigma_{\text{diss-rec,D}_2^+}(v_{\text{e}})\right\rangle$ \\ 
Charge-exchange of $\text{D}^+,\text{D}$ & $\text{D}^+ +\text{D} \rightarrow \text{D} +  \text{D}^+$ & $\nu_{\text{\text{cx,D}}} = n_{\text{D}^+}\left\langle v_{\text{D}^+}\sigma_{\text{cx,D}^+}(v_{\text{D}^+})\right\rangle$ \\ 
Charge-exchange of $\text{D}_2^+,\text{D}_2$ & $\text{D}_2^+ + \text{D}_2 \rightarrow \text{D}_2 + \text{D}_2^+$ & $\nu_{\text{cx,D}_2} = n_{\text{D}_2^+}\left\langle v_{\text{D}_2^+}\sigma_{\text{cx,D}_2^+}(v_{\text{D}_2^+})\right\rangle$ \\ 
Charge-exchange of $\text{D}_2^+,\text{D}$ & $\text{D}_2^+ +\text{D} \rightarrow \text{D}_2 + \text{D}^+$ & $\nu_{\text{cx,D-D}_2^+} = n_{\text{D}_2^+}\left\langle v_{\text{D}_2^+}\sigma_{\text{cx,D-D}_2^+}(v_{\text{D}_2^+})\right\rangle$ \\ 
Charge-exchange of $\text{D}_2,\text{D}^+$ & $\text{D}_2 + \text{D}^+ \rightarrow \text{D}_2^+ +\text{D}$ & $\nu_{\text{cx,D}_2-\text{D}^+} = n_{\text{D}^+}\left\langle v_{\text{D}^+}\sigma_{\text{cx,D}_2-\text{D}^+}(v_{\text{D}^+})\right\rangle$ \\ 
\br
\end{tabular}
\end{indented}
\end{table}

\begin{figure}[H]
\centering
\includegraphics[keepaspectratio=t,width=0.65\textwidth]{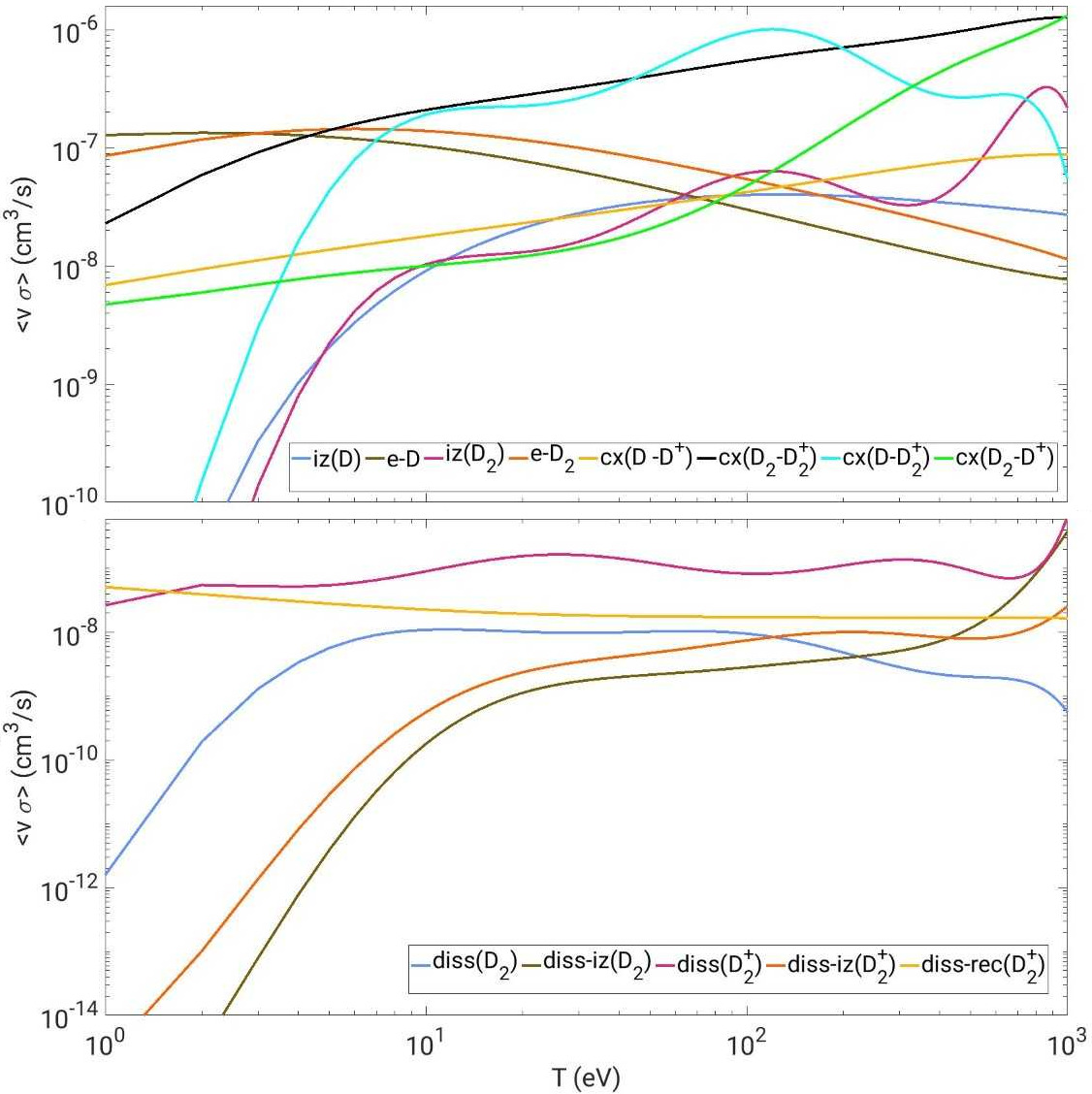}
\caption{$\left\langle v \sigma \right\rangle$ product for the collisional processes considered in this work. Ionization processes, elastic collisions and charge-exchange processes are displayed on the top panel, dissociative reactions on the bottom panel. The $\left\langle v \sigma \right\rangle$ product is plotted as a function of the temperature of the colliding particle.}
\label{reactions}
\end{figure}

We highlight that the values of the $\left\langle v \sigma \right\rangle$ product for most of the reactions considered in Table \ref{collisions} are obtained from the AMJUEL \cite{Reiter2011} and HYDEL \cite{Janev1987} databases (precise references for each cross section are listed in Table 1 of \cite{Wensing2019}). While these databases list the cross sections for ordinary hydrogen plasmas, we assume here that they apply also to deuterium. More precisely, the cross section for the $\text{e}^- -\text{D}$ elastic collisions is obtained from \cite{Janev1995} (page 40, Table 2), while for the $\text{e}^- -\text{D}_2$ elastic collision we use \cite{Yoon2008} (page 917, Table 13). The cross section for the $\text{D}_2-\text{D}_2^+$ charge-exchange reaction is taken from the HYDEL database (H.4, reaction 4.3.1), while for the $\text{D}-\text{D}_2^+$ charge-exchange we use the cross section values in the ALADDIN database \cite{Chung2017}, which are obtained from \cite{Krstic2002,Krstic2005}. For all the other reactions, we use the cross sections from the AMJUEL database \cite{Reiter2011}. The $\left\langle v \sigma \right\rangle$ product for the collisional processes considered in this work is plotted as a function of the temperature in Fig. \ref{reactions}.

Having listed the collisional processes, we now focus on the velocity and energy of their products. For charge-exchange interactions of the kind $A + B^+ \rightarrow A^+ + B$, we assume that, while $A$ and $B^+$ exchange an electron, their velocities are not affected and energy is conserved. As a consequence, the ion $A^+$ is released from the charge-exchange collision with the velocity of $A$, and $B$ is released with the velocity of $B^+$. For the $\text{e}^- +\text{D} \rightarrow \text{e}^- +\text{D}$ elastic collisions, given the large electron to deuterium mass ratio, we consider that the $\text{D}$ velocity is not affected by the collision, while the electron is emitted isotropically in the reference frame of the massive particle according to a Maxwellian distribution function, $\Phi_{e\left[\mathbf{v_{\text{D}}},T_{\text{e},\text{e-D}}\right]} = \left[m_{\text{e}}/(2 \pi T_{\text{e},\text{e-D}})\right]^{3/2} \exp\left[- m_{\text{e}} (\mathbf{v}-\mathbf{v}_{\text{D}})^2/(2 T_{\text{e},\text{e-D}})\right]$, centered at the velocity of the incoming $\text{D}$ particle, $\mathbf{v}_{\text{D}} = \int \mathbf{v} f_{\text{D}} d\mathbf{v} / \int f_{\text{D}} d\mathbf{v}$. The temperature $T_{\text{e},\text{e-D}}$ is established by energy conservation considerations. Precisely, we observe that the average energy of the incoming electrons consists of the sum of the kinetic energy associated with the electron fluid velocity, $\mathbf{v_{\text{e}}}$, and the thermal contribution, given by $(3/2) T_{\text{e}}$. On the other hand, the energy of the outcoming electrons has a contribution given by the collective re-emission velocity, $\mathbf{v_{\text{D}}}$, and a thermal contribution, $T_{\text{e,e-D}}$. It follows that $T_{\text{e},\text{e-D}}$ satisfies the following balance, $3 T_{\text{e}}/2+m_{\text{e}} v_{\text{e}}^2/2 = T_{\text{e,e-D}}+m_{\text{e}} v_{\text{D}}^2/2$. The elastic collisions between electronds and $\text{D}_2$ can be described similarly. The re-emitted electrons have a distribution $\Phi_{e\left[\mathbf{v_{\text{D}_2}},T_{\text{e},\text{e-D}_2}\right]}$, with $T_{\text{e},\text{e-D}_2}$ obtained from an analogous conservation law, $3 T_{\text{e}}/2+m_{\text{e}} v_{\text{e}}^2/2 = T_{\text{e,e-D}_2}+m_{\text{e}} v_{\text{D}_2}^2/2$. 

We now consider the electrons generated by ionization of $\text{D}$. We assume that they are described by the Maxwellian distribution function $\Phi_{e\left[\mathbf{v_{\text{D}}},T_{\text{e},\text{iz(D)}}\right]}$ centered at the fluid velocity of the $\text{D}$ atom $\mathbf{v_{\text{D}}}$ with $T_{\text{e},\text{iz(D)}}$ that takes into account the ionization energy loss, $\left\langle E_{\text{iz}}\right\rangle$, whose value is presented in Table \ref{energies}. More precisely, $T_{\text{e,iz}(\text{D})}$ satisfies the energy conservation law, $3 T_{\text{e}}/2+m_{\text{e}} v_{\text{e}}^2/2 = 2\left[T_{\text{e,iz(D)}}+m_{\text{e}} v_{\text{D}}^2/2\right]+\left\langle E_{\text{iz,D}}\right\rangle$, as the reaction gives rise to two electrons with the same properties. The same approach is followed for the ionization of $\text{D}_2$, with the two released electrons being described by a Maxwellian $\Phi_{e\left[\mathbf{v_{\text{D}_2}},T_{\text{e},\text{iz(D)}_2}\right]}$ centered at the velocity of the $\text{D}_2$ molecules, $\mathbf{v_{\text{D}_2}}$, and with temperature $T_{\text{e},\text{iz(D)}_2}$ obtained from $3 T_{\text{e}}/2+m_{\text{e}} v_{\text{e}}^2/2 = 2\left[T_{\text{e,iz}(\text{D}_2)}+m_{\text{e}} v_{\text{D}_2}^2/2\right]+\left\langle E_{\text{iz,D}_2}\right\rangle$, with $\left\langle E_{\text{iz,D}_2}\right\rangle$ the average energy loss due to ionization of $\text{D}_2$ (see Table \ref{energies}). We highlight that we neglect multi-step ionization processes when computing the cross section for ionization of $\text{D}$ and $\text{D}_2$, and we do the same for all other electron impact-induced reactions, such as the dissociative processes considered here.

We apply the procedure used for ionization processes to describe the properties of the electrons resulting from dissociative processes, with the electron generated by dissociation of $\text{D}_2$ being described by the Maxwellian $\Phi_{e\left[\mathbf{v_{\text{D}_2}},T_{\text{e},\text{diss(D)}_2}\right]}$ centered around $\mathbf{v_{\text{D}_2}}$ and with temperature $T_{\text{e},\text{diss(D)}_2}$ obtained from $3 T_{\text{e}}/2+m_{\text{e}} v_{\text{e}}^2/2 = T_{\text{e,diss}(\text{D}_2)}+m_{\text{e}} v_{\text{D}_2}^2/2+\left\langle E_{\text{diss,D}_2}\right\rangle$. Regarding dissociation of $\text{D}_2^+$, the resulting electron is similarly modelled by a Maxwellian $\Phi_{e\left[\mathbf{v_{\text{D}_2}^+},T_{\text{e},\text{diss(D)}_2^+}\right]}$ centered at the velocity of the $\text{D}_2^+$ ion and with temperature $T_{\text{e},\text{diss(D)}_2^+}$ given by energy conservation, $3 T_{\text{e}}/2+m_{\text{e}} v_{\text{e}}^2/2 = T_{\text{e,diss}(\text{D}_2^+)}+m_{\text{e}} v_{\text{D}_2^+}^2/2+\left\langle E_{\text{diss,D}_2^+}\right\rangle$. On the other hand, dissociative ionization of $\text{D}_2$ generates two electrons, whose Maxwellian distribution function, $\Phi_{e\left[\mathbf{v_{\text{D}_2}},T_{\text{e},\text{diss-iz(D)}_2}\right]}$, is centered around the $\text{D}_2$ velocity, $\mathbf{v_{\text{D}_2}}$, and characterized by a temperature $T_{\text{e,diss-iz}(\text{D}_2)}$, obtained from $3 T_{\text{e}}/2+m_{\text{e}} v_{\text{e}}^2/2 = 2\left[T_{\text{e,diss-iz}(\text{D}_2)}+m_{\text{e}} v_{\text{D}_2}^2/2\right]+\left\langle E_{\text{diss-iz,D}_2}\right\rangle$. Similarly, the electrons generated by dissociative ionization of $\text{D}_2^+$ are assumed to follow a Maxwellian $\Phi_{e\left[\mathbf{v_{\text{D}_2^+}},T_{\text{e},\text{diss-iz(D)}_2^+}\right]}$ centered at $\mathbf{v_{\text{D}_2^+}}$ and with temperature $T_{\text{e,diss-iz}(\text{D}_2^+)}$ obtained from the corresponding energy conservation law, $3 T_{\text{e}}/2+m_{\text{e}} v_{\text{e}}^2/2 = 2\left[T_{\text{e,diss-iz}(\text{D}_2^+)}+m_{\text{e}} v_{\text{D}_2^+}^2/2\right]+\left\langle E_{\text{diss-iz,D}_2^+}\right\rangle$. 

The evaluation of the temperature of the $\text{D}$ atoms and $\text{D}^+$ ions released from dissociative reactions are based on modelling these reactions as Franck-Condon dissociation processes. These temperatures are summarized in Table \ref{energies} and rely on data from \cite{Janev1987}. The detailed calculations are presented in the App. A. We also remark that these particles are emitted isotropically in the frame of the centre of mass of the incoming $\text{D}_2$ or $\text{D}_2^+$ particle. Thus, the $\text{D}$ atoms generated in dissociation of $\text{D}_2$ molecules, for instance, are assumed to have a Maxwellian distribution $\Phi_{\text{D}\left[\mathbf{v_{\text{D}_2}},T_{\text{D},\text{diss(D}_2\text{)}}\right]}$. Analogously, we describe the neutral $\text{D}$ atoms and $\text{D}^+$ ions generated by dissociative-ionization of $\text{D}_2$ molecules by the Maxwellian distributions $\Phi_{\text{D}\left[\mathbf{v}_{\text{D}_2},T_{\text{D,diss-iz}\left(\text{D}_2\right)}\right]}$ and $\Phi_{\text{D}^+\left[\mathbf{v}_{\text{D}_2},T_{\text{D,diss-iz}\left(\text{D}_2\right)}\right]}$ respectively, with the temperature $T_{\text{D,diss-iz(D}_2\text{)}}$ listed in Table \ref{energies} and evaluated in App. A. Similarly, $\Phi_{\text{D}\left[\mathbf{v}_{\text{D}^+_2},T_{\text{D,diss}\left(D_2^+\right)}\right]}$ and $\Phi_{\text{D}^+\left[\mathbf{v}_{\text{D}^+_2},T_{\text{D,diss}\left(D_2^+\right)}\right]}$ are the Maxwellian distributions of $\text{D}$ atoms and $\text{D}^+$ ions generated by dissociation of $\text{D}_2^+$ ions, where $\mathbf{v}_{\text{D}^+_2}$ is the fluid velocity of the $\text{D}^+_2$ ion population that includes the leading order components (see Sec. 3). Finally, dissociative-ionization of $\text{D}_2^+$ generates $\text{D}^+$ ions that are described by a Maxwellian distribution $\Phi_{\text{D}^+\left[\mathbf{v}_{\text{D}^+_2},T_{\text{D,diss-iz}\left(D_2^+\right)}\right]}$. To conclude, we note that the $\text{D}$ atoms and $\text{D}^+$ generated by dissociative-recombination of $\text{D}_2^+$ are described by the Maxwellian distributions $\Phi_{\text{D}\left[\mathbf{v}_{\text{D}^+_2},T_{\text{D,diss-rec}\left(\text{D}_2^+\right)}\right]}$ and $\Phi_{\text{D}^+\left[\mathbf{v}_{\text{D}^+_2},T_{\text{D,diss-rec}\left(\text{D}_2^+\right)}\right]}$ respectively, with $T_{\text{D,diss-rec(D}_2^+\text{)}}$ the average thermal energy of the reaction products.

\begin{table}[!ht]
\caption{\label{energies}Average electron energy loss and average energy of reaction products for the ionization and dissociative processes included in the model.}
\begin{indented}
\item[]\begin{tabular}{@{}lll} 
\br
\textbf{Collisional process} & \textbf{Electron energy loss} & \textbf{Reaction product temperature} \\ 
\mr
Ionization of $\text{D}$ & $\left\langle E_{\text{iz,D}}\right\rangle = 13.60 \text{eV}$ & ------------------------------------\\ 
Ionization of $\text{D}_2$ & $\left\langle E_{\text{iz,D}_2}\right\rangle = 15.43 \text{eV}$ & ------------------------------------\\ 
Dissociation of $\text{D}_2$ & $\left\langle E_{\text{diss,D}_2}\right\rangle \simeq 14.3 \text{eV}$ & $T_{\text{D}\text{,diss}\left(\text{D}_2\right)} \simeq 1.95 \text{eV}$\\ 
Dissociative-ionization of $\text{D}_2$ ($E_{\text{e}} < 26 \text{eV}$) & $E_{\text{diss-iz,D}_2} = 18.25 \text{eV}$ & $T_{\text{D}\text{,diss-iz}\left(\text{D}_2\right)} \simeq 0.25 \text{eV}$\\
Dissociative-ionization of $\text{D}_2$ ($E_{\text{e}} > 26 \text{eV}$) & $E_{\text{diss-iz,D}_2} = 33.6 \text{eV}$ & $T_{\text{D}\text{,diss-iz}\left(\text{D}_2\right)} \simeq 7.8 \text{eV}$\\
Dissociation of $\text{D}_2^+$ & $\left\langle E_{\text{diss,D}_2^+}\right\rangle \simeq 13.7 \text{eV}$ & $T_{\text{D}\text{,diss}\left(\text{D}_2^+\right)} \simeq 3.0 \text{eV}$\\
Dissociative-ionization of $\text{D}_2^+$ & $\left\langle E_{\text{diss-iz,D}_2^+}\right\rangle \simeq 15.5 \text{eV}$ & $T_{\text{D}\text{,diss-iz}\left(\text{D}_2^+\right)} \simeq 0.4 \text{eV}$\\
Dissociative-recombination of $\text{D}_2^+$ & ------------------------------------ & $T_{\text{D}\text{,diss-rec}\left(\text{D}_2^+\right)} \simeq 11.7 \text{eV}$\\
\end{tabular}
\end{indented}
\end{table}

\section{\label{sec:level3} The three-fluid drift-reduced Braginskii equations}

The kinetic equations for $\text{e}^-$, $\text{D}^+$ and $\text{D}_2^+$, which include the terms associated with the neutral-plasma interactions, are the starting point for the derivation of the Braginskii set of equations, used here to model the plasma dynamics. These equations generalise the ones considered in the single-ion species model described in \cite{Halpern2016,Wersal2015}, by adding the new collisional neutral-plasma interaction terms listed in Table \ref{collisions}, as well as an equation for the description of molecular ions, $\text{D}_2^+$. The kinetic equations are

\begin{flalign}\label{f_e}
\begin{aligned}
& \frac{\partial f_{\text{e}}}{\partial t} + \mathbf{v}\cdot\frac{\partial f_{\text{e}}}{\partial \mathbf{x}} + \mathbf{a}\cdot\frac{f_{\text{e}}}{\partial \mathbf{v}} = \nu_{\text{iz,D}} n_{\text{D}} \left[2 \Phi_{e\left[\mathbf{v}_{\text{D}},T_{\text{e,iz(D)}}\right]}-\frac{f_{\text{e}}}{n_{\text{e}}}\right]\\
& +\nu_{\text{e-D}}n_{\text{D}}\left[\Phi_{e\left[\mathbf{v}_{\text{D}},T_{e,en(D)}\right]}-\frac{f_{\text{e}}}{n_{\text{e}}}\right]-\nu_{\text{rec,D}^+}\frac{n_{\text{D}^+}}{n_{\text{e}}}f_{\text{e}}+\nu_{\text{iz,D}_2}n_{\text{D}_2}\left[2 \Phi_{e\left[\mathbf{v}_{\text{D}_2},T_{\text{e,iz}(\text{D}_2)}\right]}-\frac{f_{\text{e}}}{n_{\text{e}}}\right]\\
& +\nu_{\text{e-D}_2}n_{\text{D}_2}\left[\Phi_{e\left[\mathbf{v}_{\text{D}_2},T_{e,en(\text{D}_2)}\right]}-\frac{f_{\text{e}}}{n_{\text{e}}}\right]-\nu_{\text{rec,D}_2^+}\frac{n_{\text{D}_2^+}}{n_{\text{e}}}f_{\text{e}}\\
& +\nu_{\text{diss,D}_2} n_{\text{D}_2} \left[\Phi_{e\left[\mathbf{v}_{\text{D}_2},T_{\text{e,diss}(\text{D}_2)}\right]}-\frac{f_{\text{e}}}{n_{\text{e}}}\right]+\nu_{\text{diss-iz,D}_2} n_{\text{D}_2} \left[2\Phi_{e\left[\mathbf{v}_{\text{D}_2},T_{\text{e,diss-iz}(\text{D}_2)}\right]}-\frac{f_{\text{e}}}{n_{\text{e}}}\right]\\
& +\nu_{\text{diss-iz,D}_2^+} n_{\text{D}_2^+} \left[2\Phi_{e\left[\mathbf{v}_{\text{D}_2^+},T_{\text{e,diss-iz}(\text{D}_2^+)}\right]}-\frac{f_{\text{e}}}{n_{\text{e}}}\right]\\
& +\nu_{\text{diss,D}_2^+} n_{\text{D}_2^+} \left[\Phi_{e\left[\mathbf{v}_{\text{D}_2^+},T_{\text{e,diss}(\text{D}_2^+)}\right]}-\frac{f_{\text{e}}}{n_{\text{e}}}\right]-\nu_{\text{diss-rec,D}_2^+}n_{\text{D}_2^+}\frac{f_{\text{e}}}{n_{\text{e}}}+C(f_{\text{e}}),\\
\end{aligned}
\end{flalign}  

\vspace{0.2cm}

\begin{flalign}\label{f_Dp}
\begin{aligned}
& \frac{\partial f_{\text{D}^+}}{\partial t} + \mathbf{v}\cdot\frac{\partial f_{\text{D}^+}}{\partial \mathbf{x}} + \mathbf{a}\cdot\frac{f_{\text{D}^+}}{\partial \mathbf{v}} = \nu_{\text{iz,D}} f_{\text{D}} -\nu_{\text{rec,D}^+}f_{\text{D}^+}\\
& - \nu_{\text{cx,D}}\left(\frac{n_{\text{D}}}{n_{\text{D}^+}}f_{\text{D}^+}-f_{\text{D}}\right)+\nu_{\text{cx,D-D}_2^+} f_{\text{D}} - \nu_{\text{cx,D}_2-\text{D}^+}\frac{n_{\text{D}_2}}{n_{\text{D}^+}}f_{\text{D}^+}\\
& +\nu_{\text{diss-iz,D}_2} f_{\text{D}_2}+2\nu_{\text{diss-iz,D}_2^+} f_{\text{D}_2^+} + \nu_{\text{diss,D}_2^+}f_{\text{D}_2^+} + C(f_{\text{D}^+}),\\
\end{aligned}
\end{flalign}  

\noindent and

\begin{flalign}\label{f_D2p}
\begin{aligned}
& \frac{\partial f_{\text{D}_2^+}}{\partial t} + \mathbf{v}\cdot\frac{\partial f_{\text{D}_2^+}}{\partial \mathbf{x}} + \mathbf{a}\cdot\frac{f_{\text{D}_2^+}}{\partial \mathbf{v}} = \nu_{\text{iz,D}_2} f_{\text{D}_2} -\nu_{\text{rec,D}_2^+}f_{\text{D}_2^+}\\
& - \nu_{\text{cx,D}_2}\left(\frac{n_{\text{D}_2}}{n_{\text{D}_2^+}}f_{\text{D}_2^+}-f_{\text{D}_2}\right)-\nu_{\text{cx,D}_2-\text{D}^+} f_{\text{D}_2} - \nu_{\text{cx,D-D}_2^+} \frac{n_\text{D}}{n_{\text{D}_2^+}} f_{\text{D}_2^+}\\
& -\left(\nu_{\text{diss-iz,D}_2^+} + \nu_{\text{diss,D}_2^+} + \nu_{\text{diss-rec,D}_2^+}\right)f_{\text{D}_2^+} + C(f_{\text{D}_2^+}).\\
\end{aligned}
\end{flalign} 

\vspace{0.2cm}

\noindent In Eqs. (\ref{f_e}-\ref{f_D2p}), $\mathbf{v}$ is the particle velocity, $\mathbf{a}$ is the particle acceleration due to the Lorentz Force, $\partial / \partial \mathbf{x}$ is the gradient in real space and $\partial / \partial \mathbf{v}$ in the velocity space. The $C(f_{\text{e}})$, $C(f_{\text{D}^+})$ and $C(f_{\text{D}_2^+})$ terms represent Coulomb collisions between charged particles affecting the $e$, $\text{D}^+$ and $\text{D}_2^+$ distribution functions, respectively.

The Braginskii equations for the 3-species plasma ($\text{e}^-$, $\text{D}^+$ and $\text{D}_2^+$) are then obtained by taking the first three moments of the kinetic equations for each species in the limit $\Omega_{\text{cD}^+} \tau_{\text{D}^+} \gg 1$, with $\Omega_{\text{cD}^+} = eB/m_{\text{D}^+}$ the cyclotron frequency ($m_{\text{D}^+}$ denotes the $\text{D}^+$ ion mass and $e$ is the elementary charge) and $\tau_{\text{D}^+}$ the characteristic Coulomb collision time for $\text{D}^+$ ions. The Braginskii equations, including the neutral-plasma interaction terms, can be derived by following the steps presented in \cite{Zeiler1997}, and take the following form

      \begin{equation}\label{Brag_n_e}
      \begin{aligned}
      & \frac{\partial n_{\text{e}}}{\partial t} + \nabla\cdot\left(n_{\text{e}}\mathbf{v}_{\text{e}}\right) = n_{\text{D}} \nu_{\text{iz,D}} - n_{\text{D}^+} \nu_{\text{rec,D}^+} + n_{\text{D}_2} \nu_{\text{iz,D}_2} - n_{\text{D}_2^+} \nu_{\text{rec,D}_2^+}\\
      & + n_{\text{D}_2} \nu_{\text{diss-iz,D}_2} + n_{\text{D}_2^+} \nu_{\text{diss-iz,D}_2^+} - n_{\text{D}_2^+} \nu_{\text{diss-rec,D}_2^+},\\
      \end{aligned}
      \end{equation}  
      
      \begin{equation}\label{Brag_n_D}
      \begin{aligned}
      & \frac{\partial n_{\text{D}^+}}{\partial t} + \nabla\cdot\left(n_{\text{D}^+}\mathbf{v}_{\text{D}^+}\right) = n_{\text{D}}\nu_{\text{iz,D}}-n_{\text{D}^+}\nu_{\text{rec,D}^+}+n_{\text{D}}\nu_{\text{cx,D-D}_2^+}-n_{\text{D}_2}\nu_{\text{cx,D}_2-\text{D}^+}\\
      & +n_{\text{D}_2}\nu_{\text{diss-iz,D}_2}+n_{\text{D}_2^+}\left(2\nu_{\text{diss-iz,D}_2^+}+\nu_{\text{diss,D}_2^+}\right),\\
      \end{aligned}
      \end{equation}  

      \begin{equation}\label{Brag_n_D2}
      \begin{aligned}
      & \frac{\partial n_{\text{D}_2^+}}{\partial t} + \nabla\cdot\left(n_{\text{D}_2^+}\mathbf{v}_{\text{D}_2^+}\right) = n_{\text{D}_2}\nu_{\text{iz,D}_2}-n_{\text{D}_2^+}\nu_{\text{rec,D}_2^+}+n_{\text{D}_2}\nu_{\text{cx,D}_2-\text{D}^+}-n_{\text{D}}\nu_{\text{cx,D-D}_2^+}\\
      & -n_{\text{D}_2^+}\left(\nu_{\text{diss-iz,D}_2^+}+\nu_{\text{diss,D}_2^+}+\nu_{\text{diss-rec,D}_2^+}\right),\\
      \end{aligned}
      \end{equation}  

      \begin{equation}\label{Brag_vpare}
      \begin{aligned}
      & m_{\text{e}} n_{\text{e}} \frac{d_{\text{e}} v_{\text{e}\alpha}}{dt} = -\frac{\partial p_{\text{e}}}{\partial x_{\alpha}} -\frac{\partial \Pi_{\text{e}\alpha\beta}}{\partial x_{\beta}}-e n_{\text{e}}\left[E_{\alpha}+\left(\mathbf{v}_{\text{e}} \times \mathbf{B}\right)_{\alpha}\right]+R_{e\alpha}\\
      & +m_{\text{e}}\left[n_{\text{D}}\left(2\nu_{\text{iz,D}}+\nu_{\text{e-D}}\right)\left(v_{\text{D}\alpha}-v_{\text{e}\alpha}\right)+n_{\text{D}_2}\left(2\nu_{\text{iz,D}_2}+\nu_{\text{e-D}_2}\right)\left(v_{\text{D}_2\alpha}-v_{\text{e}\alpha}\right)\right.\\
      & +2n_{\text{D}_2}\nu_{\text{diss-iz,D}_2}\left(v_{\text{D}_2\alpha}-v_{\text{e}\alpha}\right)+2\nu_{\text{diss-iz,D}_2^+}n_{\text{D}_2^+}\left(v_{\text{D}_2^+\alpha}-v_{\text{e}\alpha}\right)\\
      &\left.+n_{\text{D}_2^+}\nu_{\text{diss,D}_2^+}\left(v_{\text{D}_2^+\alpha}-v_{\text{e}\alpha}\right)+n_{\text{D}_2}\nu_{\text{diss,D}_2}\left(v_{\text{D}_2\alpha}-v_{\text{e}\alpha}\right)\right],\\
      \end{aligned}
      \end{equation}  

      \begin{equation}\label{Brag_vparD}
      \begin{aligned}
      & m_{\text{D}} n_{\text{D}^+} \frac{d_{\text{D}^+} v_{\text{D}^+\alpha}}{dt} = -\frac{\partial p_{\text{D}^+}}{\partial x_{\alpha}} -\frac{\partial \Pi_{\text{D}^+\alpha\beta}}{\partial x_{\beta}}+e n_{\text{D}^+}\left[E_{\alpha}+\left(\mathbf{v}_{\text{D}^+} \times \mathbf{B}\right)_{\alpha}\right]+R_{\text{D}^+\alpha}\\
      & +m_{\text{D}}\left[n_{\text{D}}(\nu_{\text{iz,D}}+\nu_{\text{cx,D}}+\nu_{\text{cx,D-D}_2^+})(v_{\text{D}\alpha}-v_{\text{D}^+\alpha})+n_{\text{D}_2}\nu_{\text{diss-iz,D}_2}(v_{\text{D}_2\alpha}-v_{\text{D}^+\alpha}) \right.\\
      &\left.+n_{\text{D}_2^+}\left(2\nu_{\text{diss-iz,D}_2^+}+\nu_{\text{diss,D}_2^+}\right)(v_{\text{D}_2^+\alpha}-v_{\text{D}^+\alpha})\right],\\
      \end{aligned}
      \end{equation}  
      
      \begin{equation}\label{Brag_vparD2}
      \begin{aligned}
      & m_{\text{D}_2} n_{\text{D}_2^+} \frac{d_{\text{D}_2^+} v_{\text{D}_2^+\alpha}}{dt} = -\frac{\partial p_{\text{D}_2^+}}{\partial x_{\alpha}} -\frac{\partial \Pi_{\text{D}^+\alpha\beta}}{\partial x_{\beta}}+e n_{\text{D}_2^+}\left[E_{\alpha}+\left(\mathbf{v}_{\text{D}_2^+} \times \mathbf{B}\right)_{\alpha}\right]+R_{\text{D}_2^+\alpha}\\
      & +m_{\text{D}_2}n_{\text{D}_2}(\nu_{\text{iz,D}_2}+\nu_{\text{cx,D}_2}+\nu_{\text{cx,D}_2-\text{D}^+})(v_{\text{D}_2\alpha}-v_{\text{D}_2^+\alpha}),\\
      \end{aligned}
      \end{equation}  

      \begin{equation}\label{Brag_Te}
      \begin{aligned}
      & \frac{3}{2} n_{\text{e}} \frac{d_{\text{e}} T_{\text{e}}}{dt} + p_{\text{e}} \nabla\cdot\mathbf{v}_{\text{e}} = -\nabla\cdot\mathbf{q}_{\text{e}}-\Pi_{\text{e}\alpha\beta}\frac{\partial v_{\text{e} \beta}}{\partial x_{\alpha}}+Q_{\text{e}}\\
      &+n_{\text{D}} \nu_{\text{iz,D}}\left[-E_{\text{iz,D}}-\frac{3}{2}T_{\text{e}}+\frac{3}{2}m_{\text{e}} \mathbf{v}_{\text{e}}\cdot\left(\mathbf{v}_{\text{e}}-\frac{4}{3}\mathbf{v}_{\text{D}}\right)\right]-n_{\text{D}}\nu_{\text{e-D}}m_{\text{e}} \mathbf{v}_{\text{e}}\cdot(\mathbf{v}_{\text{D}}-\mathbf{v}_{\text{e}})\\
      &+n_{\text{D}_2}\nu_{\text{iz,D}_2}\left[-E_{\text{iz,D}_2}-\frac{3}{2}T_{\text{e}}+\frac{3}{2}m_{\text{e}} \mathbf{v}_{\text{e}}\cdot\left(\mathbf{v}_{\text{e}}-\frac{4}{3}\mathbf{v}_{\text{D}_2}\right)\right]-n_{\text{D}_2}\nu_{\text{e-D}_2}m_{\text{e}}\mathbf{v}_{\text{e}}\cdot(\mathbf{v}_{\text{D}_2}-\mathbf{v}_{\text{e}})\\
      &+n_{\text{D}_2}\nu_{\text{diss,D}_2}\left[-E_{\text{diss,D}_2}+m_{\text{e}}\mathbf{v}_{\text{e}}\cdot\left(\mathbf{v}_{\text{e}}-\mathbf{v}_{\text{D}_2}\right)\right]\\
      &+n_{\text{D}_2}\nu_{\text{diss-iz,D}_2}\left[-E_{\text{diss-iz,D}_2}-\frac{3}{2}T_{\text{e}}+\frac{3}{2}m_{\text{e}} \mathbf{v}_{\text{e}}\cdot\left(\mathbf{v}_{\text{e}}-\frac{4}{3}\mathbf{v}_{\text{D}_2}\right)\right]\\
      &+n_{\text{D}_2^+}\nu_{\text{diss,D}_2^+}\left[-E_{\text{diss,D}_2^+}+m_{\text{e}} \mathbf{v}_{\text{e}}\cdot\left(\mathbf{v}_{\text{e}}-\mathbf{v}_{\text{D}_2^+}\right)\right]\\
      &+n_{\text{D}_2^+}\nu_{\text{diss-iz,D}_2^+}\left[-E_{\text{diss-iz,D}_2^+}-\frac{3}{2}T_{\text{e}}+\frac{3}{2}m_{\text{e}} \mathbf{v}_{\text{e}}\cdot\left(\mathbf{v}_{\text{e}}-\frac{4}{3}\mathbf{v}_{\text{D}_2^+}\right)\right],\\
      \end{aligned}
      \end{equation}  
      
      \begin{equation}\label{Brag_TD}
      \begin{aligned}
      & \frac{3}{2} n_{\text{D}^+} \frac{d_{\text{D}^+} T_{\text{D}^+}}{dt} + p_{\text{D}^+} \nabla\cdot\mathbf{v}_{\text{D}^+} = -\nabla\cdot\mathbf{q}_{\text{D}^+}-\Pi_{\text{D}^+\alpha\beta}\frac{\partial v_{\text{D}^+ \beta}}{\partial x_{\alpha}}+Q_{\text{D}^+}\\
      &+ n_{\text{D}}(\nu_{\text{iz,D}}+\nu_{\text{cx,D}}+\nu_{\text{cx,D-D}_2^+})\left[\frac{3}{2}\left(T_{\text{D}}-T_{\text{D}^+}\right)+\frac{m_{\text{D}^+}}{2}\left(\mathbf{v}_{D}-\mathbf{v}_{\text{D}^+}\right)^2\right]\\
      &+n_{\text{D}_2}\nu_{\text{diss-iz,D}_2}\left[\frac{3}{2}\left(T_{\text{D}^+\text{,diss-iz}\left(\text{D}_2\right)}-T_{\text{D}^+}\right)+\frac{m_{\text{D}^+}}{2}\left(\mathbf{v}_{\text{D}_2}-\mathbf{v}_{\text{D}^+}\right)^2\right]\\
      &+2n_{\text{D}_2^+}\nu_{\text{diss-iz,D}_2^+}\left[\frac{3}{2}\left(T_{\text{D}^+\text{,diss-iz}\left(\text{D}_2^+\right)}-T_{\text{D}^+}\right)+\frac{m_{\text{D}^+}}{2}\left(\mathbf{v}_{\text{D}_2^+}-\mathbf{v}_{\text{D}^+}\right)^2\right] \\
      &+n_{\text{D}_2^+}\nu_{\text{diss,D}_2^+}\left[\frac{3}{2}\left(T_{\text{D}^+\text{,diss}\left(\text{D}_2^+\right)}-T_{\text{D}^+}\right)+\frac{m_{\text{D}^+}}{2}\left(\mathbf{v}_{\text{D}_2^+}-\mathbf{v}_{\text{D}^+}\right)^2\right],\\
      \end{aligned}
      \end{equation}  
      
      \begin{equation}\label{Brag_TD2}
      \begin{aligned}
      & \frac{3}{2} n_{\text{D}_2^+} \frac{d_{\text{D}_2^+} T_{\text{D}_2^+}}{dt} + p_{\text{D}_2^+} \nabla\cdot\mathbf{v}_{\text{D}_2^+} = -\nabla\cdot\mathbf{q}_{\text{D}_2^+}-\Pi_{\text{D}_2^+\alpha\beta} \frac{\partial v_{\text{D}_2^+ \beta}}{\partial x_{\alpha}}+Q_{\text{D}_2^+}\\
      &+n_{\text{D}_2}(\nu_{\text{cx,D}_2}+\nu_{\text{iz,D}_2}+\nu_{\text{cx,D}_2-\text{D}^+})\left[\frac{3}{2}\left(T_{\text{D}_2^+}-T_{\text{D}_2^+}\right)+\frac{m_{\text{D}_2^+}}{2}(\mathbf{v}_{\text{D}_2}-\mathbf{v}_{\text{D}_2^+})^2\right],\\
      \end{aligned}
      \end{equation}  

\noindent where $\Pi_{\text{e}\alpha\beta}$ is the component of the stress tensor along the $\alpha$ and $\beta$ directions, $\mathbf{R}_{\text{e}}$ is the friction force acting on the electrons, $\mathbf{q}_{\text{e}}$ is the electron heat flux density, $Q_{\text{e}}$ is the electron heat generated by Coulomb collisions and $d_{\text{e}}/dt = \partial/\partial t +(\mathbf{v}_{\text{e}}\cdot\nabla)$ is the electron advective derivative. The equivalent notation is used for the $\text{D}^+$ and $\text{D}_2^+$ species.

The drift-limit of the Braginskii equations is finally derived by applying the $d/dt\ll\Omega_{\text{cD}^+}$ ordering, valid in typical conditions of the tokamak boundary. Only leading order components in $(1/\Omega_{\text{cD}^+}) d/dt$ are retained in the electron perpendicular velocity, i.e. $\mathbf{v}_{\perp \text{e}} = \mathbf{v}_{\perp \text{e0}} = \mathbf{v}_{\text{E}\times\text{B}} + \mathbf{v}_{\text{d}\text{e}}$, with $\mathbf{v}_{\text{E}\times\text{B}} = (\mathbf{E} \times \mathbf{B})/B^2$ the $E \times B$ drift and $\mathbf{v}_{\text{d}\text{e}} = (\mathbf{B} \times \nabla p_{\text{e}})/(en_{\text{e}} B^2)$ the electron diamagnetic drift, thus neglecting electron inertia. Similarly, the $\text{D}^+$ perpendicular velocity is decomposed as $\mathbf{v}_{\perp \text{D}^+} = \mathbf{v}_{\perp \text{D}^+ 0} + \mathbf{v}_{\text{pol},\text{D}^+} + \mathbf{v}_{\text{fric,D}^+}$, where the leading order perpendicular velocity, 

      \begin{equation}
      \begin{aligned}
      \mathbf{v}_{\perp \text{D}^+ 0} = \mathbf{v}_{\text{E}\times\text{B}} + \mathbf{v}_{\text{d}\text{D}^+},\\
      \end{aligned}\label{v_Dp_perp_0}
      \end{equation}  

\noindent is the sum of the $E\times B$ drift and the diamagnetic drift, $\mathbf{v}_{\text{d}\text{D}^+} = (\mathbf{B} \times \nabla p_{\text{D}^+})/(en_{\text{D}^+} B^2)$. The polarization drift,

      \begin{equation}
      \begin{aligned}
      & \mathbf{v}_{\text{pol},\text{D}^+} = - \frac{1}{n_{\text{D}^+} \Omega_{\text{cD}^+}}\frac{d_{\text{D}^+}}{dt}\left(\frac{n_{\text{D}^+}}{B}\nabla_{\perp} \phi + \frac{1}{B} \nabla_{\perp} p_{\text{D}^+}\right) + \frac{1}{m_{\text{D}^+} n_{\text{D}^+} \Omega_{\text{cD}^+}} \mathbf{b} \times \left[G_{\text{D}^+} \mathbf{k} - \frac{\nabla G_{\text{D}^+}}{3}\right],\\
      \end{aligned}\label{v_pol_Dp}
      \end{equation} 

\noindent is of higher order than $\mathbf{v}_{\perp \text{D}^+ 0}$ in the $d/dt \ll \Omega_{\text{cD}^+}$ expansion, as shown in \cite{Zeiler1997}. Similarly, the drift arising from friction between $\text{D}^+$ ions and other species,
      
      \begin{equation}
      \begin{aligned}
      & \mathbf{v}_{\text{fric},\text{D}^+} = \frac{n_{\text{D}}}{n_{\text{D}^+}} \frac{\nu_{\text{cx,D}} + \nu_{\text{iz,D}} + \nu_{\text{cx,D}-\text{D}_2^+}}{\Omega_{\text{cD}^+}} \left(\mathbf{v}_{\perp \text{D}}-\mathbf{v}_{\perp \text{D}^+ 0}\right) \times \mathbf{b} + \frac{n_{\text{D}_2}}{n_{\text{D}^+}} \frac{\nu_{\text{iz}-\text{diss,D}_2}}{\Omega_{\text{cD}^+}} \left(\mathbf{v}_{\perp \text{D}_2}-\mathbf{v}_{\perp \text{D}^+ 0}\right) \times \mathbf{b} \\
      & + \frac{n_{\text{D}_2^+}}{n_{\text{D}^+}} \frac{2 \nu_{\text{diss-iz,D}_2^+} + \nu_{\text{diss,D}_2^+}}{\Omega_{\text{cD}^+}} \left(\mathbf{v}_{\perp \text{D}_2^+ 0}-\mathbf{v}_{\perp \text{D}^+ 0}\right) \times \mathbf{b},\\
      \end{aligned}\label{v_fric_Dp}
      \end{equation} 
      
\noindent is also of higher order in $(1/\Omega_{\text{cD}^+}) d/dt$. This term includes contributions from collisions of $\text{D}^+$ with $\text{D}$, $\text{D}_2$ and $\text{D}_2^+$ particles. Assuming $v_{\text{D}} \lesssim v_{\text{D}^+}$, $v_{\text{D}_2} \lesssim v_{\text{D}^+}$ and $v_{\text{D}_2^+} \lesssim v_{\text{D}^+}$, and noticing that $\nu / \Omega_{\text{cD}^+} \ll 1$, one obtains $v_{\text{fric},\text{D}^+} \sim (\nu/\Omega_{\text{cD}^+}) v_{\text{D}^+} \ll v_{\text{D}^+}$. For this reason, $\mathbf{v}_{\text{D}^+}$ and $\mathbf{v}_{\text{D}^+}$ are approximated with their leading order components, i.e. $\mathbf{v}_{\perp \text{D}^+} \simeq \mathbf{v}_{\perp \text{D}^+ 0}$ and $\mathbf{v}_{\perp \text{D}_2^+} \simeq \mathbf{v}_{\perp \text{D}_2^+ 0}$, in Eq. (\ref{v_fric_Dp}). In Eqs. (\ref{v_pol_Dp}) and (\ref{v_fric_Dp}) we introduce the giroviscous term for $\text{D}^+$ ions $G_{\text{D}^+} = -\eta_{0\text{D}^+} \left[2\nabla_{\|}v_{\| \text{D}^+} +C(\phi)/B+C(p_{\text{D}^+})/(Z_{\text{D}^+} n_{\text{D}^+} B)\right]$, the $\text{D}^+$ viscosity $\eta_{0\text{D}^+}$, the magnetic field curvature vector $\mathbf{k} = \left(\mathbf{b} \cdot \nabla \right) \mathbf{b}$, the gradient along the magnetic field $\nabla_{\|} = \mathbf{b}\cdot \nabla$, the gradient perpendicular to the magnetic field $\nabla_{\perp} = \nabla - \mathbf{b} \nabla_{\|}$, and the magnetic field unit vector $\mathbf{b} = \mathbf{B}/B$.

For the derivation of the drift-limit of the $\text{D}_2^+$ velocity, we follow a similar approach, as the $d/dt\ll\Omega_{\text{cD}_2^+}$ ordering is also valid in typical tokamak boundary conditions. The $\text{D}_2^+$ ions perpendicular velocity is thus given by $\mathbf{v}_{\perp \text{D}_2^+} = \mathbf{v}_{\perp \text{D}_2^+ 0} + \mathbf{v}_{\text{pol},\text{D}_2^+} + \mathbf{v}_{\text{fric,D}_2^+ }$, with 

      \begin{equation}
      \begin{aligned}
      \mathbf{v}_{\perp \text{D}_2^+ 0} = \mathbf{v}_{\text{E}\times\text{B}} + \mathbf{v}_{\text{d}\text{D}_2^+}\\
      \end{aligned}\label{v_D2p_perp_0}
      \end{equation} 

\noindent the leading order component, with $\mathbf{v}_{\text{d}\text{D}_2^+} = (\mathbf{B} \times \nabla p_{\text{D}_2^+})/(en_{\text{D}_2^+} B^2)$. The velocity $\mathbf{v}_{\text{pol},\text{D}_2^+}$ denotes the polarization drift and $\mathbf{v}_{\text{fric},\text{D}_2^+}$ stands for the drift velocity arising from friction between $\text{D}_2^+$ ions and other species. Their expressions are given by

      \begin{equation}
      \begin{aligned}
      & \mathbf{v}_{\text{pol},\text{D}_2^+} = - \frac{1}{n_{\text{D}_2^+} \Omega_{\text{cD}_2^+}}\frac{d_{\text{D}_2^+}}{dt}\left(\frac{n_{\text{D}_2^+}}{B}\nabla_{\perp} \phi + \frac{1}{B} \nabla_{\perp} p_{\text{D}_2^+}\right) + \frac{1}{m_{\text{D}_2^+} n_{\text{D}_2^+} \Omega_{\text{cD}_2^+}} \mathbf{b} \times \left[G_{\text{D}_2^+} \mathbf{k} - \frac{\nabla G_{\text{D}_2^+}}{3}\right],\\
      \end{aligned}
      \end{equation} 
      
\noindent and
      
      \begin{equation}
      \begin{aligned}
      & \mathbf{v}_{\text{fric},\text{D}_2^+} = \frac{n_{\text{D}_2}}{n_{\text{D}_2^+}} \frac{\nu_{\text{iz,D}_2} + \nu_{\text{cx,D}_2} + \nu_{\text{cx,D}_2-\text{D}^+}}{\Omega_{\text{cD}_2^+}} \left(\mathbf{v}_{\perp \text{D}_2}-\mathbf{v}_{\perp \text{D}_2^+ 0}\right) \times \mathbf{b},\\
      \end{aligned}\label{v_fric_D2p}
      \end{equation} 

\noindent with $G_{\text{D}_2^+} = -\eta_{0\text{D}_2^+} \left[2\nabla_{\|}v_{\| \text{D}_2^+} +C(\phi)/B+C(p_{\text{D}_2^+})/(n_{\text{D}_2^+} B)\right]$ the $\text{D}_2^+$ giroviscous term and $\eta_{0\text{D}_2^+}$ the related viscosity. The approximation $\mathbf{v}_{\perp \text{D}_2^+} \simeq \mathbf{v}_{\perp \text{D}_2^+ 0}$ is used in Eq. (\ref{v_fric_D2p}). 

To obtain an expression for the parallel friction forces and parallel heat fluxes and close the Braginskii equations, we use the collisional closure proposed by Zhdanov in \cite{Zhdanov2002}, leveraging the formulation presented in \cite{Bufferand2019,Raghunathan2021}, more suitable for numerical implementation. The application of this procedure to the particular case of the multispecies plasma considered here is described in App. B, where we take advantage of the fact that the $\text{D}_2^+$ density is considerably smaller than the $\text{D}^+$ density, i.e. $n_{\text{D}_2^+}/n_{\text{D}^+} \ll 1$, for typical tokamak boundary conditions, which leads to $n_{\text{e}} \simeq n_{\text{D}^+}$ because of quasi-neutrality. On the other hand, the contributions from the perpendicular heat flux arising from $\nabla \cdot \mathbf{q}_{\text{e}}$ and $\nabla \cdot \mathbf{q}_{\text{D}^+}$ in the $T_{\text{e}}$ and $T_{\text{D}^+}$ equations, respectively, can be evaluated as in the single-ion species model \cite{Ricci2012,Halpern2016}, in particular following the derivation presented in \cite{Zeiler1997}. This approach can be generalised to evaluate the term arising from the perpendicular component of $\nabla \cdot \mathbf{q}_{\text{D}_2^+}$ in the $T_{\text{D}_2^+}$ equation. Thus, the drift-reduced Braginskii system of equations is composed of the continuity equation for the electron species, the continuity equation for the $\text{D}_2^+$ species, the vorticity equations that ensures quasi-neutrality, $n_{\text{e}} = n_{\text{D}^+} + n_{\text{D}_2^+}$, and the equations for the parallel velocities and temperature of all species. They take the form
      
      \begin{equation}\label{n_e}
      \begin{aligned}
      & \frac{\partial n_{\text{e}}}{\partial t} = -\frac{\rho_*^{-1}}{B}[\phi,n_{\text{e}}] + \frac{2}{B} \left[C(p_{\text{e}})
      - n_{\text{e}} C(\phi)\right] - \nabla\cdot(n_{\text{e}} v_{\|\text{e}}\mathbf{b}) + \mathcal{D}_{n_{\text{e}}}\nabla_{\perp}^2 n_{\text{e}} + S_{n_{\text{e}}} \\
      & + n_{\text{D}} \nu_{\text{iz,D}} - n_{\text{D}^+} \nu_{\text{rec,D}^+} + n_{\text{D}_2} \nu_{\text{iz,D}_2} - n_{\text{D}_2^+} \nu_{\text{rec,D}_2^+}\\
      & + n_{\text{D}_2} \nu_{\text{diss-iz,D}_2} + n_{\text{D}_2^+} \nu_{\text{diss-iz,D}_2^+} - n_{\text{D}_2^+} \nu_{\text{diss-rec,D}_2^+},\\
      \end{aligned}
      \end{equation}
      
      \begin{equation}\label{n_D2}
      \begin{aligned}
      & \frac{\partial n_{\text{D}_2^+}}{\partial t} = -\frac{\rho_*^{-1}}{B}[\phi,n_{\text{D}_2^+}] - \nabla\cdot(n_{\text{D}_2^+} v_{\|\text{D}_2^+} \mathbf{b}) - \frac{2}{B} \left[n_{\text{D}_2^+}C(T_{\text{D}_2^+})+T_{\text{D}_2^+}C(n_{\text{D}_2^+})+n_{\text{D}_2^+}C(\phi)\right]\\ 
      & + \mathcal{D}_{n_{\text{D}_2^+}}\nabla_{\perp}^2 n_{\text{D}_2^+} + S_{n_{\text{D}_2^+}} + n_{\text{D}_2}\nu_{\text{iz,D}_2}-n_{\text{D}_2^+}\nu_{\text{rec,D}_2^+}+n_{\text{D}_2}\nu_{\text{cx,D}_2-\text{D}^+}-n_{\text{D}}\nu_{\text{cx,D-D}_2^+}\\
      & -n_{\text{D}_2^+}\left(\nu_{\text{diss-iz,D}_2^+}+\nu_{\text{diss,D}_2^+}+\nu_{\text{diss-rec,D}_2^+}\right),\\
      \end{aligned}
      \end{equation} 

      \begin{equation}\label{vort}
      \begin{aligned}
      & \frac{\partial \Omega}{\partial t} = -\nabla \cdot \left[\frac{\rho_*^{-1}}{B} \left( \left[\phi, B \bm{\Omega_{\text{D}^+}}\right] + 2\left[\phi, B \bm{\omega_{\text{D}_2^+}}\right] \right) \right] 
      -\nabla \cdot \left[ \frac{v_{\|\text{D}^+}}{B} \nabla_{\|} \left(B \bm{\Omega_{\text{D}^+}} \right) + \frac{v_{\|\text{D}_2^+}}{B} \nabla_{\|} \left(B \bm{\omega_{\text{D}_2^+}} \right) \right] \\
      & + \frac{2}{B} \left[n_{\text{e}} C(T_{\text{e}})+T_{\text{e}} C(n_{\text{e}})+n_{\text{D}^+} C(T_{\text{D}^+})+T_{\text{D}^+} C(n_{\text{D}^+})+n_{\text{D}_2^+} C(T_{\text{D}_2^+})+T_{\text{D}_2^+} C(n_{\text{D}_2^+}) \right] \\
      & +\nabla \cdot \left(j_{\|} \mathbf{b}\right) + \left[\nabla G_{\text{D}^+} \cdot \left(\frac{\bm{b}\times \bm{k}}{B}\right)+G_{\text{D}^+} \nabla \cdot \left(\frac{\bm{b}\times \bm{k}}{B}\right)-\frac{2}{3 B} C\left(G_{\text{D}^+}\right)\right] \\
      & + \left[\nabla G_{\text{D}_2^+} \cdot \left(\frac{\bm{b}\times \bm{k}}{B}\right)+G_{\text{D}_2^+} \nabla \cdot \left(\frac{\bm{b}\times \bm{k}}{B}\right)-\frac{2}{3 B} C\left(G_{\text{D}_2^+}\right)\right] + \eta_{0 \Omega}\nabla_{\|}^2 \Omega + \mathcal{D}_{\perp \Omega}\nabla_{\perp}^2 \Omega \\
      & - \nabla \cdot \left[\frac{2 n_{\text{D}_2}}{n_{\text{D}_2^+}}\left(\nu_{\text{cx,D}_2} + \nu_{\text{iz,D}_2} + \nu_{\text{cx,D}_2-\text{D}^+}\right)\bm{\omega_{\text{D}_2^+}}\right] - \nabla \cdot \left[\frac{n_{\text{D}}}{n_{\text{D}^+}}\left(\nu_{\text{cx,D}} + \nu_{\text{iz,D}} + \nu_{\text{cx,D-D}_2^+}\right)\bm{\omega_{\text{D}^+}}\right] \\
      & - \nabla \cdot \left[\frac{n_{\text{D}_2}}{n_{\text{D}^+}}\nu_{\text{di-iz,D}_2}\bm{\omega_{\text{D}^+}}\right] + \nabla \cdot \left[\frac{n_{\text{D}_2^+}}{n_{\text{D}^+}}\left(2\nu_{\text{di-iz},\text{D}_2^+}+\nu_{\text{di,D}_2^+}\right)\left(\bm{\omega_{\text{D}_2^+}}-\bm{\omega_{\text{D}^+}}\right)\right],\\
      \end{aligned}
      \end{equation} 
      
      \begin{equation}\label{vpare}
      \begin{aligned}
      & \frac{\partial v_{\|\text{e}}}{\partial t} = -\frac{\rho_*^{-1}}{B}[\phi,v_{\|\text{e}}]-v_{\| \text{e}}\nabla_\| v_{\|\text{e}}
      +\frac{m_{\text{D}}}{m_{\text{e}}}\left[\nabla_{\|}\phi-\frac{\nabla_\| p_{\text{e}}}{n_{\text{e}}} -\frac{2}{3n_{\text{e}}}\nabla_\|G_{\text{e}}-0.71\nabla_{\|}T_{\text{e}}\right] \\
      & -\frac{m_{\text{D}}}{m_{\text{e}}}\nu \left(v_{\| \text{e}}-v_{\|\text{D}^+}\right)
      +\mathcal{D}_{v_{\|\text{e}}}\nabla_{\perp}^2 v_{\| \text{e}} + \frac{1}{n_{\text{e}}}\left[n_{\text{D}}\left(2\nu_{\text{iz,D}}+\nu_{\text{e-D}}\right)\left(v_{\|\text{D}}-v_{\| \text{e}}\right) \right.\\ 
      & + n_{\text{D}_2}\left(2\nu_{\text{iz,D}_2}+\nu_{\text{e-D}_2}\right)\left(v_{\|\text{D}_2}-v_{\| \text{e}}\right) + n_{\text{D}_2}\left(2\nu_{\text{diss-iz,D}_2}+\nu_{\text{diss,D}_2}\right)\left(v_{\|\text{D}_2}-v_{\| \text{e}}\right) \\ 
      & \left. + n_{\text{D}_2^+} \left(2\nu_{\text{diss-iz,D}_2^+}+\nu_{\text{diss,D}_2^+}\right)\left(v_{\|\text{D}_2^+}-v_{\| \text{e}}\right)\right],\\
      \end{aligned}
      \end{equation}
      
      \vspace{0.3cm}
      
      \begin{equation}\label{vparD}
      \begin{aligned}
      & \frac{\partial v_{\|\text{D}^+}}{\partial t} = -\frac{\rho_*^{-1}}{B}[\phi,v_{\|\text{D}^+}]-v_{\|\text{D}^+}\nabla_\| v_{\|\text{D}^+}-\nabla_{\|}\phi-\frac{\nabla_\| p_\text{D}^+}{n_{\text{D}^+}}-\frac{2}{3n_{\text{D}^+}}\nabla_\|G_{\text{D}^+}+0.71\frac{n_{\text{e}}}{n_{\text{D}^+}}\nabla_{\|}T_{\text{e}} \\
      & -\nu\frac{n_{\text{e}}}{n_{\text{D}^+}}\left(v_{\|\text{D}^+}-v_{\| \text{e}}\right)+ \mathcal{D}_{v_{\|\text{D}^+}}\nabla_{\perp}^2 v_{\|\text{D}^+} + \frac{1}{n_{\text{D}^+}}\left[n_{\text{D}}(\nu_{\text{iz,D}}+\nu_{\text{cx,D}}+\nu_{\text{cx,D-D}_2^+})(v_{\|\text{D}}-v_{\|\text{D}^+}) \right. \\
      & \left. + n_{\text{D}_2^+}\nu_{\text{diss-iz},\text{D}_2}(v_{\|\text{D}_2}-v_{\|\text{D}^+})
      + n_{\text{D}_2^+}\left(2\nu_{\text{diss-iz,D}_2^+}+\nu_{\text{diss,D}_2^+}\right)(v_{\|\text{D}_2^+}-v_{\|\text{D}^+})\right],\\
      \end{aligned}
      \end{equation}
      \vspace{0.3cm}
      
      \begin{equation}\label{vparD2}
      \begin{aligned}
      & \frac{\partial v_{\|\text{D}_2^+}}{\partial t} = -\frac{\rho_*^{-1}}{B}[\phi,v_{\|\text{D}_2^+}]-v_{\|\text{D}_2^+}\nabla_\| v_{\|\text{D}_2^+} +\frac{1}{2}\left[-\nabla_{\|}\phi -\frac{\nabla p_{\text{D}_2^+}}{n_{\text{D}_2^+}} -\frac{2}{3 n_{\text{D}_2^+}}\nabla_{\|}G_{\text{D}_2^+}\right] \\
      & + \mathcal{D}_{v_{\|\text{D}_2^+}}\nabla_{\perp}^2 v_{\|\text{D}_2^+} +\frac{n_{\text{D}_2}}{n_{\text{D}_2^+}}(\nu_{\text{iz,D}_2}+\nu_{\text{cx,D}_2}+\nu_{\text{cx,D}_2-\text{D}^+})(v_{\|\text{D}_2}-v_{\|\text{D}_2^+}),\\ 
      \end{aligned}
      \end{equation}
      
      \vspace{0.3cm}
      
      \begin{equation}\label{Te}
      \begin{aligned}
      &\frac{\partial T_{\text{e}}}{\partial t} = -\frac{\rho_*^{-1}}{B}[\phi,T_{\text{e}}] - v_{\|\text{e}}\nabla_\| T_{\text{e}} +
      \frac{4 T_{\text{e}}}{3B} \left[ \frac{C(p_{\text{e}})}{n_{\text{e}}} + \frac{5}{2}C(T_{\text{e}}) - C(\phi) \right] - \frac{2 T_{\text{e}}}{3} \nabla\cdot\left(v_{\| \text{e}} \mathbf{b}\right)\\
      & +\frac{2}{3 n_{\text{e}}} \frac{1.62}{\nu}\left[n_{\text{e}} T_{\text{e}}\left(\nabla_{\|}T_{\text{e}}\right) \nabla\cdot\mathbf{b}
      + \nabla_{\|}\left(n_{\text{e}} T_{\text{e}} \nabla_{\|} T_{\text{e}}\right)\right] -\frac{2}{3}0.71 T_{\text{e}}\nabla\cdot(v_{\| \text{e}}-v_{\|\text{D}^+})\mathbf{b}\\
      & -\frac{2}{3}0.71\left(\frac{T_{\text{e}}}{n_{\text{e}}} \nabla_\| n_{\text{e}} + \nabla_\|T_{\text{e}}\right) \left(v_{\| \text{e}}-v_{\|\text{D}^+}\right)+\mathcal{\chi}_{\perp \text{e}}\nabla_{\perp}^2 T_{\text{e}} + \nabla_{\|}\left(\mathcal{\chi}_{\|\text{e}}\nabla_{\|} T_{\text{e}}\right) + S_{T_{\text{e}}}\\
      & + \frac{n_{\text{D}}}{n_{\text{e}}} \nu_{\text{iz,D}}\left[-\frac{2}{3}E_{\text{iz,D}}-T_{\text{e}}+\frac{m_{\text{e}}}{m_{\text{D}}} v_{\| \text{e}}\left(v_{\| \text{e}}-\frac{4}{3}v_{\|\text{D}}\right)\right]-\frac{n_{\text{D}}}{n_{\text{e}}}\nu_{\text{e-D}}\frac{m_{\text{e}}}{m_{\text{D}}}\frac{2}{3}v_{\| \text{e}}(v_{\|\text{D}}-v_{\| \text{e}})\\
      & +\frac{n_{\text{D}_2}}{n_{\text{e}}}\nu_{\text{iz,D}_2}\left[-\frac{2}{3} E_{\text{iz,D}_2} - T_{\text{e}} + \frac{m_{\text{e}}}{m_{\text{D}}} v_{\| \text{e}}\left(v_{\| \text{e}}-\frac{4}{3}v_{\|\text{D}_2}\right)\right]-\frac{n_{\text{D}_2}}{n_{\text{e}}}\nu_{\text{e-D}_2}\frac{m_{\text{e}}}{m_{\text{D}}}\frac{2}{3}v_{\| \text{e}}(v_{\|\text{D}_2}-v_{\| \text{e}})\\
      & + \frac{n_{\text{D}_2}}{n_{\text{e}}}\nu_{\text{diss,D}_2}\left[-\frac{2}{3}E_{\text{diss,D}_2}+\frac{2}{3}\frac{m_{\text{e}}}{m_{\text{D}}} v_{\| \text{e}}\left(v_{\| \text{e}}-v_{\|\text{D}_2}\right)\right]\\
      & +\frac{n_{\text{D}_2}}{n_{\text{e}}}\nu_{\text{diss-iz,D}_2}\left[-\frac{2}{3} E_{\text{diss-iz,D}_2} - T_{\text{e}} + \frac{m_{\text{e}}}{m_{\text{D}}} v_{\| \text{e}}\left(v_{\| \text{e}}-\frac{4}{3}v_{\|\text{D}_2}\right)\right]\\
      & + \frac{n_{\text{D}_2^+}}{n_{\text{e}}}\nu_{\text{diss,D}_2^+}\left[-\frac{2}{3}E_{\text{diss,D}_2^+}+\frac{2}{3}\frac{m_{\text{e}}}{m_{\text{D}}} v_{\| \text{e}}\left(v_{\| \text{e}}-v_{\|\text{D}_2^+}\right)\right]\\
      & +\frac{n_{\text{D}_2^+}}{n_{\text{e}}}\nu_{\text{diss-iz,D}_2^+}\left[-\frac{2}{3} E_{\text{diss-iz,D}_2^+} - T_{\text{e}} + \frac{m_{\text{e}}}{m_{\text{D}}} v_{\| \text{e}}\left(v_{\| \text{e}}-\frac{4}{3}v_{\|\text{D}_2^+}\right)\right],\\
      \end{aligned}
      \end{equation}             
      
      \vspace{0.5cm}
      
      \begin{equation}\label{TD}
      \begin{aligned}
      &\frac{\partial T_{\text{D}^+}}{\partial t} = -\frac{\rho_*^{-1}}{B}[\phi,T_{\text{D}^+}] - v_{\|\text{D}^+}\nabla_\| T_{\text{D}^+} +
      \frac{4}{3} \frac{T_{\text{D}^+}}{B} \left[-C(\phi)+\frac{C(p_{\text{e}} + p_{\text{D}_2^+})}{n_{\text{D}^+}}\right] \\
      & -\frac{2 T_{\text{D}^+}}{3 n_{\text{D}^+}} \left[n_{\text{e}} \nabla \cdot \left(v_{\|\text{e}} \mathbf{b}\right) - n_{\text{D}_2^+} \nabla \cdot \left(v_{\|\text{D}_2^+} \mathbf{b}\right) + v_{\|\text{e}} \nabla_{\|} n_{\text{e}} - v_{\|\text{D}_2^+} \nabla_{\|} n_{\text{D}_2^+} - v_{\|\text{D}^+} \nabla_{\|} n_{\text{D}^+}\right] \\
      & -\frac{10}{3} \frac{T_{\text{D}^+}}{B} C(T_{\text{D}^+})
      +\frac{2}{3 n_{\text{D}^+}}\frac{2.32}{\sqrt{2} \nu} \sqrt{\frac{m_{\text{e}}}{m_{\text{D}}}}\nabla\cdot\left(n_{\text{e}} T_{\text{D}^+}\nabla_\|T_{\text{D}^+}\right)\mathbf{b} \\
      & + \mathcal{\chi}_{\perp \text{D}^+}\nabla_{\perp}^2 T_{\text{D}^+} + \nabla_{\|}\left(\mathcal{\chi}_{\|\text{D}^+}\nabla_{\|} T_{\text{D}^+}\right) + S_{T_{\text{D}^+}} \\
      & + \frac{1}{n_{\text{D}^+}}\left\{n_{\text{D}}\left(\nu_{\text{iz,D}}+\nu_{\text{cx,D}}+\nu_{\text{cx,D-D}_2^+}\right)\left[T_{\text{D}}-T_{\text{D}^+}+\frac{1}{3}\left(v_{\|\text{D}}-v_{\|\text{D}^+}\right)^2\right]\right.\\
      &+n_{\text{D}_2}\nu_{\text{diss-iz,D}_2}\left[T_{\text{D}^+\text{,diss-iz}\left(\text{D}_2\right)}-T_{\text{D}^+}+\frac{1}{3}\left(v_{\|\text{D}_2}-v_{\|\text{D}^+}\right)^2\right]\\
      &+ 2 n_{\text{D}_2^+}\nu_{\text{diss-iz,D}_2^+}\left[T_{\text{D}^+\text{,diss-iz}\left(\text{D}_2^+\right)}-T_{\text{D}^+}+\frac{1}{3}\left(v_{\|\text{D}_2^+}-v_{\|\text{D}^+}\right)^2\right]\\
      & \left. + n_{\text{D}_2^+}\nu_{\text{diss,D}_2^+}\left[T_{\text{D}^+\text{,diss}\left(\text{D}_2^+\right)}-T_{\text{D}^+}+\frac{1}{3}\left(v_{\|\text{D}_2^+}-v_{\|\text{D}^+}\right)^2\right]\right\}\\
      \end{aligned}
      \end{equation}
      
\noindent and
      
      \begin{equation}\label{TD2}
      \begin{aligned}
       & \frac{\partial T_{\text{D}_2^+}}{\partial t} = -\frac{\rho_*^{-1}}{B}[\phi,T_{\text{D}_2^+}] - v_{\|\text{D}_2^+}\nabla_\| T_{\text{D}_2^+} -
      \frac{4}{3} \frac{T_{\text{D}_2^+}}{B} \left[C(\phi)+\frac{C(p_{\text{D}_2^+})}{n_{\text{D}_2^+}}\right] - \frac{10}{3} \frac{T_{\text{D}_2^+}}{B} C(T_{\text{D}_2^+})\\
      & - \frac{2 T_{\text{D}_2^+}}{3} \nabla\cdot\left(v_{\|\text{D}_2^+} \mathbf{b}\right) +\frac{2}{3 n_{\text{D}_2^+}}\frac{0.92}{\sqrt{2} \nu}\sqrt{\frac{m_{\text{e}}}{m_{\text{D}}}}\nabla\cdot \left(n_{\text{e}} T_{\text{D}^+}\nabla_\|T_{\text{D}^+} \right)\mathbf{b} \\
      & + \mathcal{\chi}_{\perp \text{D}_2^+}\nabla_{\perp}^2 T_{\text{D}_2^+} + \nabla_{\|}\left(\mathcal{\chi}_{\|\text{D}_2^+}\nabla_{\|} T_{\text{D}_2^+}\right) + S_{T_{\text{D}_2^+}} \\
      & + \frac{n_{\text{D}_2}}{n_{\text{D}_2^+}}(\nu_{\text{cx,D}_2}+\nu_{\text{iz,D}_2}+\nu_{\text{cx,D}_2-\text{D}^+})\left[T_{\text{D}_2^+}-T_{\text{D}_2^+}+\frac{2}{3}(v_{\|\text{D}_2}-v_{\|\text{D}_2^+})^2\right].\\
      \end{aligned}
      \end{equation}
      \vspace{0.3cm}
  
\noindent In Eqs. (\ref{n_e}-\ref{TD2}) we introduce $[A,B] = \mathbf{b}\cdot(\nabla A \times \nabla B)$, $C(A) = (B/2)\left[\nabla \times (\mathbf{b}/B)\right]\cdot\nabla A$ and the plasma vorticity $\Omega = \Omega_{\text{D}^+} + 2 \Omega_{\text{D}_2^+}$, with the $\text{D}^+$ contribution given by $\Omega_{\text{D}^+} = \nabla\cdot\bm{\omega_{\text{D}^+}} = \nabla\cdot\left[\left(n_{\text{D}^+}/B^2\right) \nabla_\perp\phi+\left(1/B^2\right)\nabla_\perp p_{\text{D}^+}\right]$ and an analogous $\text{D}_2^+$ contribution, $\Omega_{\text{D}_2^+}$. The system is thus closed by the generalized Poisson equation, which is obtained by inverting the definition of the plasma vorticity, $\Omega$, yielding

\begin{flalign}\label{poisson}
&\begin{aligned}
\nabla_\perp \cdot \left[\frac{n_{\text{D}^+} + 2 n_{\text{D}_2^+}}{B^2}\nabla_\bot\phi\right]=\Omega-\nabla_\perp \cdot \left[\frac{1}{B^2}\nabla_\perp \left(p_{\text{D}^+}+2p_{\text{D}_2^+}\right)\right]. 
\end{aligned}&
\end{flalign}

\noindent We remark that the electron giroviscous term in Eq. (\ref{vpare}) is defined similarly to the ion giroviscous terms, $G_{\text{e}} = -\eta_{0e}\left[2\nabla_{\|}v_{\|\text{e}}+C(\phi)/B-C(p_{\text{e}})/\left(n_{\text{e}} B\right)\right]$. Eq. (\ref{vort}) is written avoiding the Boussinesq approximation and taking into account all components of the velocity of the ion species $\text{D}^+$ and $\text{D}_2^+$, including the higher order polarization and friction contributions. On the other hand, in order to express the advective derivative for the ion species, $d_{\text{D}^+}/dt$ and $d_{\text{D}_2^+}/dt$, we only consider the leading order components of the perpendicular velocity, $v_{\perp \text{D}^+ 0}$ and $v_{\perp \text{D}_2^+ 0}$, therefore neglecting $\mathbf{v}_{\text{pol}}$ and $\mathbf{v}_{\text{fric}}$. Similarly, we neglect the friction and polarization drifts in the continuity equation for $\text{D}_2^+$. We also remark that the terms of higher order in $1/\Omega_{\text{cD}_2^+} d/dt$ in the perpendicular velocity of $\text{D}_2^+$ ions are neglected when writing $\nabla \cdot \mathbf{v}_{\text{D}_2^+}$ in the temperature equations, Eqs. (\ref{TD}) and (\ref{TD2}), which is a necessary assumption in order to avoid explicit time derivatives arising from the polarization drift velocity, $\mathbf{v}_{\text{pol,D}_2^+}$. Nevertheless, all terms are considered in the divergence of the perpendicular velocity of $\text{D}^+$ ions in Eq. (\ref{TD}), as we make use of $\nabla \cdot \mathbf{j} = 0$ to write $\nabla \cdot \mathbf{v}_{\text{D}^+}$ in terms of $\nabla \cdot \mathbf{v}_{\text{e}}$ and $\nabla \cdot \mathbf{v}_{\text{D}_2^+}$. Finally, when taking the divergence of these terms, we make use of $\nabla \cdot \mathbf{v}_{\text{D}} \ll \nabla \cdot \mathbf{v}_{\text{D}^+}$ to neglect the contribution of the velocity of $\text{D}$ atoms, which is valid since $\rho_{\text{s},\text{D}^+} \ll \lambda_{\text{mfp},\text{D}}$ (with $\rho_{\text{s,D}^+} = c_{\text{s,D}^+}/\Omega_{\text{c,D}^+}$ the sound Larmor radius of $\text{D}^+$ ions, $c_{\text{s,D}^+} = \sqrt{T_e/m_{\text{D}^+}}$ the $\text{D}^+$ ions sound speed and $\lambda_{\text{mfp,\text{D}}}$ the mean free path of a $\text{D}$ atoms). This relation can be generalized to the other neutral and ion species, namely $\text{D}_2$ molecules and $\text{D}_2^+$ ions. Thus we neglect the contribution of the divergence of neutral particle velocities when compared to the divergence of ion velocities. 

We note that dimensionless units are used in Eqs. (\ref{n_e}-\ref{TD2}) and in the rest of the paper. The densities, $n_{\text{e}}$, $n_{\text{D}^+}$ and $n_{\text{D}_2^+}$, are normalized to the reference value $n_0$, while temperatures, $T_{\text{e}}$, $T_{\text{D}^+}$ and $T_{\text{D}_2^+}$, are normalized to the respective reference values, $T_{\text{e}0}$, $T_{\text{D}^+ 0}$ and $T_{\text{D}_2^+ 0} = T_{{\text{D}^+}0}$, which are related by the dimensionless quantity $\tau = T_{\text{D}^+0}/T_{e0}$. Conversely, lengths parallel to the magnetic field are normalized to the tokamak major radius, $R_0$, lengths perpendicular to the magnetic field are normalized to the ion sound Larmor radius, $\rho_{\text{s}0}=c_{\text{s}0}/\Omega_{\text{cD}^+ 0}$, where $c_{\text{s}0}=T_{\text{e}0}/m_{\text{D}^+}$ is the normalized $\text{D}^+$ ion sound speed and $\Omega_{\text{cD}^+ 0} = e B_0/m_{\text{D}^+}$ is the $\text{D}^+$ ion cyclotron frequency at the magnetic axis, and time is normalized to $R_0/c_{\text{s}0}$. All other normalizations follow, namely the parallel velocities, $v_{\| \text{e}}$, $v_{\| \text{D}^+}$ and $v_{\| \text{D}_2^+}$, are normalized to $c_{\text{s}0}$, the plasma vorticity $\Omega$ is normalized to $n_0 T_{\text{e}0}/(\rho_{\text{s}0}^2 B_0^2)$, perpendicular diffusion coefficients $D_{\perp}$ and conductivities $\chi_{\perp}$ are normalized to $c_{\text{s}0} \rho_{\text{s}0}^2 / R_0$, while the parallel diffusion coefficients $D_{\|}$ and conductivities $\chi_{\|}$ are normalized to $c_{\text{s}0} R_0$. Normalized quantities are used in the rest of the paper, except when explicitly mentioned. The parameter $\rho_\star = \rho_{\text{s}0}/R_0$ is the ratio between the $\text{D}^+$ ion sound Larmor radius and the tokamak major radius $R_0$. We also note that $\nu$ is the dimensionless resistivity given by $\nu=(e^2 n_{\text{e}0} R_0)/(m_{\text{D}} c_{\text{s}0} \sigma_{\|})$, with the parallel conductivity defined in terms of the electron characteristic time $\tau_{\text{e}}$ as $\sigma_{\|} = e^2 n_{\text{e}} \tau_{\text{e}}/(0.51 m_{\text{e}})$.

We conclude with a few final remarks on Eqs. (\ref{n_e}-\ref{TD2}). We first note that the parallel conductivity appearing in the temperature equations for electrons is expressed in the form $\chi_{\|,\text{e}} = \chi_{\|0,\text{e}} T_{\text{e}}^{5/2}$, where we retain the Spitzer temperature dependence while we neglect the weaker space and time variation of the $2/(3n_{\text{e}})$ factor, similarly to the approach followed in the single-component plasma model previously implemented in GBS (\cite{Ricci2012,Halpern2016}). A similar approach is followed for $\chi_{\|,\text{D}^+}$ and $\chi_{\|,\text{D}_2^+}$. This is not expected to impact the simulation results in the sheath-limited regime, where conductivity-related contributions are small. Finally, we point that, since the $\text{D}_2^+$ density may drop to a very low value, numerical issues may arise in the equations for $v_{\| \text{D}_2^+}$ and $T_{\text{D}_2^+}$ due to terms featuring a $1/n_{\text{D}_2^+}$ dependence. For a robust numerical approach, we evolve the parallel flux and pressure of the $\text{D}_2^+$ ion species, $\Gamma_{\| \text{D}_2^+} = n_{\text{D}_2^+} v_{\| \text{D}_2^+}$ and $p_{\text{D}_2^+} = n_{\text{D}_2^+} T_{\text{D}_2^+}$, instead of $v_{\| \text{D}_2^+}$ and $T_{\text{D}_2^+}$. The equations for the time evolution of $\Gamma_{\| \text{D}_2^+}$ and $p_{\text{D}_2^+}$ are

      \begin{equation}\label{fparD2}
      \begin{aligned}
      & \frac{\partial \Gamma_{\|\text{D}_2^+}}{\partial t} = \frac{\partial n_{\text{D}_2^+}}{\partial t} 
      v_{\|\text{D}_2^+} + n_{\text{D}_2^+} \frac{\partial v_{\|\text{D}_2^+}}{\partial t},\\ 
      \end{aligned}
      \end{equation}
      
\noindent and
      
      \begin{equation}\label{pD2}
      \begin{aligned}
      & \frac{\partial p_{\text{D}_2^+}}{\partial t} = \frac{\partial n_{\text{D}_2^+}}{\partial t} T_{\text{D}_2^+} + n_{\text{D}_2^+} \frac{\partial T_{\text{D}_2^+}}{\partial t},\\ 
      \end{aligned}
      \end{equation}
      
\noindent with $\partial_t n_{\text{D}_2^+}$, $\partial_t v_{\|\text{D}_2^+}$ and $\partial_t T_{\text{D}_2^+}$ given, respectively, by Eqs. (\ref{n_D2}), (\ref{vparD2}) and (\ref{TD2}). We focus on the parallel flux, $\Gamma_{\|\text{D}_2^+}$, and pressure, $p_{\text{D}_2^+}$, when presenting the simulation results.

\section{Boundary conditions}

The boundary conditions implemented in the previous GBS models for single-ion species plasma are extended in the present work to include the molecular ion species $\text{D}_2^+$. In the case considered here, of a plasma with a toroidal limiter, the domain boundary includes the limiter plates, the outer wall and the interface with the core, where the low plasma collisionality questions the application of a fluid model.

We first consider the boundary conditions at the limiter plates, where most of the plasma ends by flowing along the magnetic field lines. Those are the most important boundary conditions to impact the simulation dynamics. The boundary conditions are imposed at the interface between the collisional pre-sheath (CP) and the magnetic pre-sheath (MP), derived from the Bohm-Chodura boundary conditions, following the approach described in Ref. \cite{Loizu2012} in the cold ion limit and generalized in Ref. \cite{Mosetto2015} to account for finite ion temperature. Here, we further extend this procedure to the case of a multi-ion species plasma. For this purpose, we use the $(y,x,z)$ coordinates, with $z$ the direction of the magnetic field, $x$ the direction perpendicular to the magnetic field and parallel to the limiter surface, and $y$ the direction perpendicular to both $x$ and $z$ (here all spatial coordinates are normalized to $\rho_{\text{s}0}$, while the other quantities are normalized as in Eqs. (\ref{n_e}-\ref{TD2})). We also introduce $s = y \text{cos}\alpha + z \text{sin}\alpha$, the coordinate perpendicular to the limiter plate, with $\alpha$ the angle between the magnetic field line and the plane of the limiter. 

As a first step in the derivation of the boundary conditions, we note that the steady-state dynamics of the multispecies plasma in the CP is described by means of the continuity equation for the $\text{D}^+$ and $\text{D}_2^+$ species (quasi-neutrality provides the electron density) and the parallel momentum equations for $\text{e}^-$, $\text{D}^+$ and $\text{D}_2^+$. In steady state, these can be written as 

      \begin{equation}\label{bc_cont_Dp}
      \begin{aligned}
      & \nabla \cdot (n_{\text{D}^+} \mathbf{v}_{\text{D}^+}) = S_{\text{p,D}^+},\\ 
      \end{aligned}
      \end{equation}
      
      \begin{equation}\label{bc_cont_D2p}
      \begin{aligned}
      & \nabla \cdot (n_{\text{D}_2^+} \mathbf{v}_{\text{D}_2^+}) = S_{\text{p,D}_2^+},\\ 
      \end{aligned}
      \end{equation}
      
      \begin{equation}\label{bc_mom_e}
      \begin{aligned}
      & n_{\text{e}} (\mathbf{v}_{\text{e}} \cdot \nabla) \mathbf{v}_{\text{e}} = - \mu (n_{\text{e}} \mathbf{E} + n_{\text{e}} \mathbf{v}_{\text{D}^+}) \times \mathbf{B} + \nabla p_{\text{e}} + \mathbf{S}_{\text{m,e}},\\ 
      \end{aligned}
      \end{equation}
      
      \begin{equation}\label{bc_mom_Dp}
      \begin{aligned}
      & n_{\text{D}^+} \left(\mathbf{v}_{\text{D}^+} \cdot \nabla \mathbf{v}_{\text{D}^+}\right) = \left(n_{\text{D}^+} \mathbf{E} + n_{\text{D}^+} \mathbf{v}_{\text{D}^+}\right) \times \mathbf{B} - \nabla p_{\text{D}^+} + \mathbf{S}_{\text{m,D}^+}\\ 
      \end{aligned}
      \end{equation}
      
\noindent and
      
      \begin{equation}\label{bc_mom_D2p}
      \begin{aligned}
      & n_{\text{D}_2^+} \left(\mathbf{v}_{\text{D}_2^+} \cdot \nabla \mathbf{v}_{\text{D}_2^+}\right) = \left(n_{\text{D}_2^+} \mathbf{E} + n_{\text{D}_2^+} \mathbf{v}_{\text{D}_2^+}\right) \times \mathbf{B} - \nabla p_{\text{D}_2^+} + \mathbf{S}_{\text{m,D}_2^+},\\ 
      \end{aligned}
      \end{equation}

\noindent with $\mu = m_{\text{e}} / m_{\text{D}^+}$, $S_{\text{p,D}^+}$ and $S_{\text{p,D}_2^+}$ the particle sources for $\text{D}^+$ and $\text{D}_2^+$, and $\mathbf{S}_{\text{m,e}}$, $\mathbf{S}_{\text{m,D}^+}$ and $\mathbf{S}_{\text{m,D}_2^+}$ the momentum sources for $\text{e}^-$, $\text{D}^+$ and $\text{D}_2^+$. 

From Eqs. (\ref{bc_cont_Dp}-\ref{bc_mom_D2p}) and following the approach described in \cite{Mosetto2015}, a system of five equations for $\partial_{\text{s}} n_{\text{D}^+}$, $\partial_{\text{s}} n_{\text{D}_2^+}$, $\partial_{\text{s}} v_{\| \text{D}^+}$ and $\partial_{\text{s}} v_{\| \text{D}_2^+}$, $\partial_{\text{s}} \phi$ is hence obtained for the interface between the CP and the MP border, considering the $\mu \ll 1$ limit and isothermal ions and electrons. For this purpose, at the MP entrance, gradients along the $x$ direction are assumed weaker than gradients along $s$ by a factor $\epsilon = \rho_{\text{s}0}/L_{\text{n}} \simeq \rho_{\text{s}0}/L_{\text{T}_{\text{e}}} \simeq \rho_{\text{s}0}/L_{\phi} \ll 1$, with $L_{\text{n}}$, $L_{\text{T}_{\text{e}}}$ and $L_{\phi}$ respectively the scale lengths of $n_{\text{e}}$, $T_{\text{e}}$ and $\phi$ along the $x$ direction. In addition, finite Larmor radius (FLR) effects are neglected and, to express the $y$ and $x$ components of the velocity of each ion species, $\text{D}^+$ and $\text{D}_2^+$, we consider only the leading order terms in $(1/\Omega_{\text{cD}^+})d/dt$ (see Eqs. (\ref{v_Dp_perp_0}) and (\ref{v_D2p_perp_0})). This yields

      \begin{equation}\label{vy_Dp}
      \begin{aligned}
      v_{\text{y,D}^+} = v_{\text{y,E}\times\text{B}} + v_{\text{y,dD}^+},\\
      \end{aligned}
      \end{equation}

      \begin{equation}\label{vx_Dp}
      \begin{aligned}
      v_{\text{x,D}^+} = v_{\text{x,E}\times\text{B}} + v_{\text{x,dD}^+},\\
      \end{aligned}
      \end{equation}
      
      \begin{equation}\label{vy_D2p}
      \begin{aligned}
      v_{\text{y,D}_2^+} = v_{\text{y,E}\times\text{B}} + v_{\text{y,dD}_2^+}\\
      \end{aligned}
      \end{equation}
      
\noindent and

      \begin{equation}\label{vx_D2p}
      \begin{aligned}
      v_{\text{x,D}_2^+} = v_{\text{x,E}\times\text{B}} + v_{\text{x,dD}_2^+},\\
      \end{aligned}
      \end{equation}

\noindent where $v_{\text{y,E}\times\text{B}}$ and $v_{\text{y,E}\times\text{B}}$ are respectively the $y$ and $x$ components of the $E \times B$ drift velocity, $v_{\text{y,dD}^+}$ and $v_{\text{x,dD}^+}$ are the $y$ and $x$ components of the $\text{D}^+$ diamagnetic velocity and $v_{\text{y,dD}_2^+}$ and $v_{\text{x,dD}_2^+}$ are the $y$ and $x$ components of the $\text{D}_2^+$ diamagnetic velocity. Finally, we define the velocity of the $\text{D}^+$ ions along the $s$ direction as $v_{\text{s,D}^+} = v_{\| \text{D}^+} \text{sin}\alpha + v_{\text{y,D}^+} \text{cos}\alpha$. We also introduce the velocity of the $\text{D}^+$ ions along the $s$ direction that excludes the diamagnetic contribution, that is $v'_{\text{s},\text{D}^+} = v_{\text{s},\text{D}^+} - v_{\text{y,dD}^+} \text{cos}\alpha$ and $v'_{\text{s},\text{D}_2^+} = v_{\text{s},\text{D}_2^+} - v_{\text{y,dD}_2^+} \text{cos}\alpha$ for the $\text{D}_2^+$ ions. The system in Eqs. (\ref{bc_cont_Dp}-\ref{bc_mom_D2p}) yields

      \begin{equation}\label{D_cont}
      \begin{aligned}
      v_{\text{s,D}^+} \partial_{\text{s}} n_{\text{D}^+} + n_{\text{D}^+} \text{sin}\alpha \partial_{\text{s}}v_{\|\text{D}^+} - \partial_{\text{x}}n_{\text{D}^+} \text{cos}\alpha \partial_{\text{s}} \phi = S_{\text{p,D}^+},\\
      \end{aligned}
      \end{equation}
      
      \begin{equation}\label{D2_cont}
      \begin{aligned}
      v_{\text{s,D}_2^+} \partial_{\text{s}} n_{\text{D}_2^+} + n_{\text{D}_2^+} \text{sin}\alpha \partial_{\text{s}}v_{\|\text{D}_2^+} - \partial_{\text{x}}n_{\text{D}_2^+} \text{cos}\alpha \partial_{\text{s}} \phi = S_{\text{p,D}_2^+},\\
      \end{aligned}
      \end{equation}
      
      \begin{equation}\label{D_mom}
      \begin{aligned}
      n_{\text{D}^+} v_{\text{s,D}^+} \partial_{\text{s}} v_{\| \text{D}^+} + n_{\text{D}^+} (\text{sin}\alpha - \partial_{\text{x}} v_{\|\text{D}^+} \text{cos}\alpha) \partial_{\text{s}} \phi + T_{\text{D}^+} \text{sin}\alpha \partial_{\text{s}} n_{\text{D}^+} = S_{\| \text{m,D}^+},\\
      \end{aligned}
      \end{equation}
      
      \begin{equation}\label{D2_mom}
      \begin{aligned}
      n_{\text{D}_2^+} v_{\text{s,D}_2^+} \partial_{\text{s}} v_{\| \text{D}_2^+} + n_{\text{D}_2^+} (\text{sin}\alpha - \partial_{\text{x}} v_{\|\text{D}_2^+} \text{cos}\alpha) \partial_{\text{s}} \phi + T_{\text{D}_2^+} \text{sin}\alpha \partial_{\text{s}} n_{\text{D}_2^+} = S_{\| \text{m,D}_2^+}\\
      \end{aligned}
      \end{equation}
      
\noindent and
      
      \begin{equation}\label{e_mom}
      \begin{aligned}
      \mu \text{sin}\alpha T_{\text{e}} \partial_{\text{s}} n_{\text{e}} - \mu \text{sin}\alpha n_{\text{e}} \partial_{\text{s}} \phi = S_{\| \text{m,e}},\\
      \end{aligned}
      \end{equation}
      
\noindent where $S_{\| \text{m,D}^+} = \mathbf{S_{\text{m,D}^+}} \cdot \mathbf{b}$, $S_{\| \text{m,D}_2^+} = \mathbf{S_{\text{m,D}_2^+}} \cdot \mathbf{b}$ and $S_{\| \text{m,e}} = \mathbf{S_{\text{m,e}}} \cdot \mathbf{b}$. We then make use of the quasi-neutrality condition, $n_{\text{e}} = n_{\text{D}^+} + n_{\text{D}_2^+}$, to obtain a system of five linear equations that we express in matrix form as $\text{M} \mathbf{x} = \mathbf{S}$, with
      
      \begin{equation}\label{M}   
      \mathbf{M} = 
      \begin{pmatrix}
      v'_{\text{s,D}^+} & n_{\text{D}^+} \text{sin} \alpha & 0 & 0 & - \text{cos} \alpha \partial_{\text{x}} n_{\text{D}^+} \\
      T_{\text{D}^+} \text{sin} \alpha & n_{\text{D}^+} v'_{\text{s,D}^+} & 0 & 0 & n_{\text{D}^+} (\text{sin} \alpha - \partial_{\text{x}} v_{\| \text{D}^+} \text{cos} \alpha) \\
      0 & 0 & v'_{\text{s,D}_2^+} & n_{\text{D}_2^+} \text{sin} \alpha & - \text{cos} \alpha \partial_{\text{x}} n_{\text{D}_2^+} \\
      0 & 0 & T_{\text{D}_2^+} \text{sin} \alpha & n_{\text{D}_2^+} v'_{\text{s,D}_2^+} & n_{\text{D}_2^+} (\text{sin} \alpha - \partial_{\text{x}} v_{\| \text{D}_2^+} \text{cos} \alpha) \\
      \mu \text{sin} \alpha T_{\text{e}} & 0 & \mu \text{sin} \alpha T_{\text{e}} & 0 & - \mu (n_{\text{D}^+} + n_{\text{D}_2^+}) \text{sin} \alpha \\
      \end{pmatrix},
      \end{equation}
      
      \begin{equation}\label{x}   
      \mathbf{x} = 
      \begin{pmatrix}
      \partial_{\text{s}} n_{\text{D}^+} \\
      \partial_{\text{s}} n_{\text{D}_2^+} \\
      \partial_{\text{s}} v_{\| \text{D}^+} \\
      \partial_{\text{s}} v_{\| \text{D}_2^+} \\
      \partial_{\text{s}} \phi \\
      \end{pmatrix}
      \end{equation}
      
\noindent and
      
      \begin{equation}\label{S}   
      \mathbf{S} = 
      \begin{pmatrix}
      S_{\text{p,D}^+} \\
      S_{\text{p,D}_2^+} \\
      S_{\| \text{m,D}^+} \\
      S_{\| \text{m,D}_2^+} \\
      S_{\| \text{m,e}} \\
      \end{pmatrix}.
      \end{equation}

\noindent Following \cite{Loizu2012,Mosetto2015}, we observe that, while the source terms are important in the CP, they are small at the MP entrance with respect to the gradient terms. This allows one to assume $|\Sigma_{\text{j}} \text{M}_{\text{ij}} \text{X}_{\text{j}}| \gg |\text{S}_{\text{i}}|$ at the MP entrance. Thus, the linear system $\text{M} \mathbf{x} = \mathbf{S}$ reduces to $\text{M} \mathbf{x} = 0$ at the MP entrance. We solve $\text{det}(\text{M}) = 0$ with respect to $v'_{\text{sD}^+}$ to obtain the non-trivial solution at the MP entrance. For this purpose, following \cite{Tskhakaya2006}, the parallel velocity of the $\text{D}_2^+$ ion species, $v_{\|\text{D}_2^+}$, is related to $v_{\|\text{D}^+}$, 

\begin{equation}
    v_{\|\text{D}_2^+} = \sqrt{\frac{m_{\text{D}^+}}{m_{\text{D}_2^+}}} v_{\|\text{D}^+} = \frac{v_{\|\text{D}^+}}{\sqrt{2}}.
    \label{vparD2_bc}
\end{equation}  

In addition, we assume $n_{\text{D}_2^+}/n_{\text{e}} \ll 1$ (and therefore $n_{\text{D}_2^+} \simeq n_{\text{e}}$) and keep only zero order terms in $\epsilon$, neglecting therefore all derivatives along the $x$ direction. The condition $\text{det}(\text{M}) = 0$ then yields

\begin{equation}
    v'_{s\text{D}^+} = \pm \sqrt{T_{\text{e}} F_T} \text{sin}\alpha
    \label{vparsD_bc_hot_ion}
\end{equation}

\noindent where the $\pm$ signs refer to the magnetic field lines entering/leaving the vessel and we have defined $F_T = 1 + \tau T_{\text{D}^+}/T_{\text{e}}$. We now note that $v'_{\text{s,D}^+} = v_{\| \text{D}^+} \text{sin} \alpha$, since we neglect $v_{\text{y},\text{E}\times\text{B}} \text{cos} \alpha = \partial_{\text{x}} \phi \text{cos} \alpha$, to obtain the boundary condition for $v_{\|\text{D}^+}$ at the limiter,

\begin{equation}
    v_{\|\text{D}^+} = \pm \sqrt{T_{\text{e}} F_T}.
    \label{vparD_bc}
\end{equation}

\noindent The expressions of the boundary conditions for the other plasma quantities then follow. In fact, Eq. (\ref{D_mom}) can be inverted to express $\partial_{\text{s}} \phi$ in terms of $\partial_{\text{s}} v_{\| D^+}$, which yields

\begin{equation}
    \partial_{\text{s}} \phi = - \frac{v'_{\text{s}\text{D}^+}\partial_{\text{s}} v_{\| D^+}}{F_T \text{sin}\alpha} = \mp \frac{\sqrt{T_{\text{e}}}}{\sqrt{F_T}} \partial_{\text{s}} v_{\| D^+}.
    \label{phi_bc}
\end{equation}

We then use Eq. (\ref{e_mom}) to express $\partial_{\text{s}} n_{\text{e}}$ in terms of $\partial_{\text{s}} \phi$, that is

\begin{equation}
    \partial_{\text{s}} n_{\text{e}} = \frac{n_{\text{e}}}{T_{\text{e}}} \partial_{\text{s}} \phi = \mp \frac{n_{\text{e}}}{\sqrt{T_{\text{e}}F_T}} \partial_{\text{s}} v_{\| D^+}
    \label{ne_bc}
\end{equation}

\noindent and, making use of $n_{\text{D}^+} = n_{\text{e}}$, we also obtain

\begin{equation}
    \partial_{\text{s}} n_{\text{D}^+} = n_{\text{e}}/T_{\text{e}} \partial_{\text{s}} \phi = \mp \frac{n_{\text{e}}}{\sqrt{T_{\text{e}}F_T}} \partial_{\text{s}} v_{\| D^+}
    \label{nD_bc}
\end{equation}

Regarding the density of the $\text{D}_2^+$ ions, we use Eq. (\ref{D2_cont}), deriving the following boundary condition

\begin{equation}
    \partial_{\text{s}} n_{\text{D}_2^+} = \mp n_{\text{D}_2^+}/\sqrt{T_{\text{e}} F_T} \partial_{\text{s}} v_{\| D^+}.
    \label{nD2_bc}
\end{equation}

\noindent In order to derive the boundary conditions for $T_{\text{e}}$, $T_{\text{D}^+}$ and $T_{\text{D}_2^+}$, we notice that temperature gradients along the direction perpendicular to the wall are small compared to the gradients of the other physical quantities. In fact, \cite{Loizu2012,Mosetto2015} show that $\partial_{\text{s}} T_{\text{e}} \sim \partial_{\text{s}} T_{\text{D}^+} \simeq 0.1 \partial_{\text{s}} \phi$. In the present work, we follow this prescription and assume $\partial_{\text{s}} T_{\text{e}} = \partial_{\text{s}} T_{\text{D}^+} = \partial_{\text{s}} T_{\text{D}_2^+} = 0.1 \partial_{\text{s}} \phi$ (we note that our tests show that imposing $\partial_{\text{s}} T_{\text{e}} = \partial_{\text{s}} T_{\text{D}^+} = \partial_{\text{s}} T_{\text{D}_2^+} = 0$ does not affect the simulation results noticeably).

To obtain the boundary condition for $\Omega$ at the MP entrance, we start from its definition, $\Omega = \nabla\cdot\left[(n_{\text{D}^+}/B^2) \nabla_\perp\phi+(1/B^2)\nabla_\perp p_{\text{D}^+}\right] + \nabla\cdot\left[(n_{\text{D}_2^+}/B^2) \nabla_\perp\phi+(1/B^2)\nabla_\perp p_{\text{D}_2^+}\right]$. We write the second order derivatives in the directions perpendicular to the magnetic field retaining only derivatives along the $y$ direction, since $\partial^2_{\text{x}} \ll \partial^2_{\text{y}}$. Since $\partial_{\text{y}} B = 0$ at the limiter, the $1/B^2$ factor can be considered constant when the derivatives defining $\Omega$ are evaluated. We then write the derivatives along the $y$ direction in terms of derivatives along $s$ and consider $T_{\text{D}_2^+} = T_{\text{D}^+}$ (for simplicity). This yields

\begin{equation}
    \Omega = - \text{cos}\alpha \left[\partial_{\text{s}} (n_{\text{e}}+n_{\text{D}_2^+}) \partial_{\text{s}} \phi + T_{\text{D}^+} \partial^2_{\text{s}} (n_{\text{e}}+n_{\text{D}_2^+}) + (n_{\text{e}}+n_{\text{D}_2^+}) \partial^2_{\text{s}} \phi \right]. 
    \label{another}
\end{equation}

\noindent We now take advantage of Eqs. (\ref{ne_bc}) and (\ref{nD2_bc}) to express $\partial_{\text{s}} n_{\text{e}}$ and $\partial_{\text{s}} n_{\text{D}^+}$ in terms of $\partial_{\text{s}} \phi$ and use Eq. (\ref{phi_bc}) to obtain the final expression of the boundary condition for $\Omega$, that is

\begin{equation}
    \Omega = - (n_{\text{e}} + n_{\text{D}_2^+}) F_T \text{cos}^2\alpha \left[\pm \frac{\sqrt{T_{\text{e}}}}{\sqrt{F_T}} \partial^2_{\text{s}} v_{\| \text{D}^+} \mp \frac{1}{\sqrt{T_{\text{e}}} F_T} (\partial_{\text{s}} v_{\| \text{D}^+})^2 \right]. 
    \label{yet_another}
\end{equation}

Finally, the boundary condition for the electron parallel velocity is obtained from the analysis of the electron kinetic distribution function at the MP entrance. As discussed in \cite{Loizu2012}, this gives

\begin{equation}
    v_{\|\text{e}} = \sqrt{T_{\text{e}}} \left[\pm \text{exp}\left(\Lambda - \frac{\phi}{T_{\text{e}}}\right) \right],
    \label{vpare_bc}
\end{equation}

\noindent where $\Lambda = \text{log}\left[\sqrt{(1/2 \pi)(m_{\text{i}}/m_{\text{e}})}\right] \simeq 3$.

At the vessel outer wall and the core interface, $\it{ad}$ $\it{hoc}$ boundary conditions are considered, similarly to the approach used in previous models of GBS \cite{Loizu2012,Mosetto2015,Halpern2016}. In fact, a set of first-principles boundary conditions is yet to be derived for such boundaries. The impact of these $\it{ad}$ $\it{hoc}$ boundary conditions upon the simulation results is controlled by extending radially the simulation domain towards the wall and the core. The conditions we impose include homogeneous Neumann boundary conditions to $n_{\text{e}}$, $n_{\text{D}^+}$, $T_{\text{e}}$, $T_{\text{D}^+}$, $T_{\text{D}_2^+}$, $v_{\|\text{e}}$, $v_{\|\text{D}^+}$ and $v_{\|\text{D}_2^+}$. Since the density of $\text{D}_2^+$ ions is expected to be very low at the core-edge interface (no particles outflowing from the core), we use Dirichlet boundary conditions at the core interface for $n_{\text{D}_2^+}$, setting it to a residual value, while homogenous Neumann boundary conditions are considered at the vessel outer wall. We also use Dirichlet boundary conditions for the vorticity, setting $\Omega = 0$ at both the wall and the core interface. Regarding the $\phi$ boundary conditions, we follow the approach presented in Ref. \cite{Paruta2018}, where $\phi=\Lambda T_{\text{e}}$ is considered at the vessel wall. Finally, $\phi=\phi_{0}$ is considered at the core interface, where $\phi_{0}$ is a constant value chosen to prevent large gradients of $\phi$. 

\section{\label{sec:level4} The kinetic model for the neutral species and its formal solution}

In order to compute the neutral distribution functions of $\text{D}$ and $\text{D}_2$, $f_{\text{D}}$ and $f_{\text{D}_2}$, we consider a set of two coupled kinetic equations, that is

      \begin{flalign}\label{fD_char}
      &\begin{aligned}
      & \frac{\partial f_{\text{D}}}{\partial t} + \mathbf{v}\cdot\frac{\partial f_{\text{D}}}{\partial \mathbf{x}} = - \nu_{\text{iz,D}} f_{\text{D}} - \nu_{\text{cx,D}}\left(f_{\text{D}}-\frac{n_{\text{D}}}{n_{\text{D}^+}}f_{\text{D}^+}\right) + \nu_{\text{rec,D}^+} f_{\text{D}^+}\\
      & + \nu_{\text{cx,D}_2-\text{D}^+} \left(\frac{n_{\text{D}_2}}{n_{\text{D}^+}} f_{\text{D}^+}\right) - \nu_{\text{cx,D-D}_2^+} f_{\text{D}} + 2 \nu_{\text{diss,D}_2} f_{\text{D}_2} + \nu_{\text{diss-iz,D}_2} f_{\text{D}_2}\\
      & + \nu_{\text{diss,D}_2^+} f_{\text{D}_2^+} + 2 \nu_{\text{diss-rec,D}_2^+} f_{\text{D}_2^+},\\ 
      \end{aligned}&
      \end{flalign}
      
\noindent and
      
      \begin{flalign}\label{fD2_char}
      &\begin{aligned}
      & \frac{\partial f_{\text{D}_2}}{\partial t} + \mathbf{v}\cdot\frac{\partial f_{\text{D}_2}}{\partial \mathbf{x}} = - \nu_{\text{iz,D}_2} f_{\text{D}_2} - \nu_{\text{cx,D}_2}\left(f_{\text{D}_2}-\frac{n_{\text{D}_2}}{n_{\text{D}_2^+}}f_{\text{D}_2^+}\right) \\
      & + \nu_{\text{rec,D}_2^+} f_{\text{D}_2^+} - \nu_{\text{cx,D}_2-\text{D}^+} f_{\text{D}_2} + \nu_{\text{cx,D-D}_2^+} \left(\frac{n_{\text{D}}}{n_{\text{D}_2^+}} f_{\text{D}_2^+}\right) \\
      & - \nu_{\text{diss,D}_2} f_{\text{D}_2} - \nu_{\text{diss-iz,D}_2} f_{\text{D}_2}.
      \end{aligned}&
      \end{flalign}
      
The formal solution of Eqs. (\ref{fD_char}) and (\ref{fD2_char}) can be obtained by using the method of characteristics, assuming that the plasma quantities are known. This yields
      
      \begin{equation}\label{fD}
      \begin{aligned}
      & f_{\text{D}}(\mathbf{x},\mathbf{v},t) = \int_0^{r'_{\text{b}}} \left[\frac{S_{\text{D}}(\mathbf{x}',\mathbf{v},t')}{v}+\delta \left(r'-r'_{\text{b}}\right) f_{\text{D}} (\mathbf{x}'_{\text{b}},\mathbf{v},t'_{\text{b}})\right]\\
      & \times\exp\left[-\frac{1}{v}\int_0^{r'} \nu_{\text{eff}_{\text{D}}}(\mathbf{x}'',t'') dr''\right] \frac{J(\mathbf{x}')}{J(\mathbf{x})} dr' 
      \end{aligned}
      \end{equation}
      
\noindent and
      
      \begin{equation}\label{fD2}
      \begin{aligned}
      & f_{\text{D}_2}(\mathbf{x},\mathbf{v},t) = \int_0^{r'_{\text{b}}} \left[\frac{S_{\text{D}_2}(\mathbf{x}',\mathbf{v},t')}{v}+\delta \left(r'-r'_{\text{b}}\right) f_{\text{D}_2} (\mathbf{x}'_{\text{b}},\mathbf{v},t'_{\text{b}})\right]\\
      & \times\exp\left[-\frac{1}{v}\int_0^{r'} \nu_{\text{eff}_{\text{D}_2}}(\mathbf{x}'',t'') dr''\right] \frac{J(\mathbf{x}')}{J(\mathbf{x})} dr'. 
      \end{aligned}
      \end{equation}
      
\noindent The solutions presented in Eq. (\ref{fD}-\ref{fD2}) show that the distribution functions of $\text{D}$ and $\text{D}_2$ at position $\mathbf{x}$, velocity $\mathbf{v}$ and time $t$ result from the neutrals generated at a location $\mathbf{x'}=\mathbf{x}-r' \bm{\Omega}$, in the plasma volume or at the boundary, and at time $t'=t-r'/v$, with $\bm{\Omega} = \mathbf{v}/v$ the unit vector aligned with the neutral velocity and $r'$ the distance measured from $\mathbf{x'}$ to $\mathbf{x}$ (the subscript "$\text{b}$" denotes the intersection point between the domain boundary and the characteristic starting at $\mathbf{x}$ with direction $\bm{\Omega}$). Since the neutrals are solved on the $(R,Z)$ coordinate system, with $R$ the distance from the torus axis and $Z$ the vertical coordinate measured from the equatorial midplane, the integral includes the Jacobian corresponding to the coordinate system $J(\mathbf{x}) = R(\mathbf{x})$. The volumetric source due to the collisional processes in Eq. (\ref{fD}) is

      \begin{equation}\label{S_D}
      \begin{aligned}
      & S_{\text{D}}(\mathbf{x}',\mathbf{v},t')= \nu_{\text{cx,D}}(\mathbf{x}',t') n_{\text{D}}(\mathbf{x}',t') \Phi_{\left[\mathbf{v}_{\text{D}^+},T_{\text{D}^+}\right]}(\mathbf{x}',\mathbf{v},t') + \nu_{\text{cx,D}_2-\text{D}^+}(\mathbf{x}',t') n_{\text{D}_2}(\mathbf{x}',t') \Phi_{\left[\mathbf{v}_{\text{D}^+},T_{\text{D}^+}\right]}(\mathbf{x}',\mathbf{v},t')\\
      & + \nu_{\text{rec,D}^+}(\mathbf{x}',t') n_{\text{D}^+}(\mathbf{x}',\mathbf{v},t') \Phi_{\left[\mathbf{v}_{\text{D}^+},T_{\text{D}^+}\right]}(\mathbf{x}',\mathbf{v},t') + 2\nu_{\text{diss,D}_2}(\mathbf{x}',t')n_{\text{D}_2}(\mathbf{x}',t')\Phi_{\left[\mathbf{v}_{\text{D}_2},T_{\text{D,diss}\left(\text{D}_2\right)}\right]}(\mathbf{x}',\mathbf{v},t')\\
      & +\nu_{\text{diss-iz,D}_2}(\mathbf{x}',t') n_{\text{D}_2}(\mathbf{x}',t')\Phi_{\left[\mathbf{v}_{\text{D}_2},T_{\text{D,diss-iz}\left(\text{D}_2\right)}\right]}(\mathbf{x}',\mathbf{v},t') +\nu_{\text{diss,D}_2^+}(\mathbf{x}',t')n_{\text{D}_2^+}(\mathbf{x}',\mathbf{v},t') \Phi_{\left[\mathbf{v}_{\text{D}^+_2},T_{\text{D,diss}\left(D_2^+\right)}\right]}(\mathbf{x}',\mathbf{v},t')\\
      & +2\nu_{\text{diss-rec,D}_2^+}(\mathbf{x}',t') n_{\text{D}_2^+}(\mathbf{x}',\mathbf{v},t') \Phi_{\left[\mathbf{v}_{\text{D}^+_2},T_{\text{D,diss-rec}\left(\text{D}_2^+\right)}\right]}(\mathbf{x}',\mathbf{v},t')
      \end{aligned}
      \end{equation}
      
\noindent since $\text{D}$ ions can be generated in the plasma volume by $\text{D}-\text{D}^+$ and $\text{D}_2-\text{D}^+$ charge-exchange interactions, recombination of $\text{D}^+$ ions with electrons, dissociation of $\text{D}_2$ molecules into two $\text{D}$ atoms, dissociative ionization of $\text{D}_2$ into $\text{D}$ and $\text{D}^+$, dissociation of $\text{D}_2^+$ ions into $\text{D}$ and $\text{D}^+$, and dissociative recombination of $\text{D}_2^+$ into two $\text{D}$ atoms. 

Similarly, $\text{D}_2$ molecules can be generated in the plasma by $\text{D}_2-\text{D}_2^+$ and $\text{D}-\text{D}_2^+$ charge-exchange interactions or recombination of $\text{D}_2^+$ ions with electrons. Therefore, the volumetric source term in Eq. (\ref{fD2}) is

      \begin{equation}\label{S_D2}
      \begin{aligned}
      & S_{\text{D}_2}(\mathbf{x}',\mathbf{v},t')= \nu_{\text{cx,D}_2}(\mathbf{x}',t') n_{\text{D}_2}(\mathbf{x}',t') \Phi_{\left[\mathbf{v}_{\text{D}_2^+},T_{\text{D}_2^+}\right]}(\mathbf{x}',\mathbf{v},t')+\nu_{\text{rec,D}_2^+}(\mathbf{x}',t') n_{\text{D}_2^+}(\mathbf{x}',\mathbf{v},t') \Phi_{\left[\mathbf{v}_{\text{D}_2^+},T_{\text{D}_2^+}\right]}(\mathbf{x}',\mathbf{v},t')\\
      & + \nu_{\text{cx,D-D}_2^+}(\mathbf{x}',t') n_{\text{D}}(\mathbf{x}',t') \Phi_{\left[\mathbf{v}_{\text{D}_2^+},T_{\text{D}_2^+}\right]}(\mathbf{x}',\mathbf{v},t').
      \end{aligned}
      \end{equation}

We remark that $\Phi_{\left[\mathbf{v}_{\text{D}^+},T_{\text{D}^+}\right]}(\mathbf{x}',\mathbf{v},t') = \left[m_{\text{D}^+}/(2 \pi T_{\text{D}^+})\right]^{3/2} \exp\left[- m_{\text{D}^+} (\mathbf{v}-\mathbf{v}_{\text{D}^+})^2/(2 T_{\text{D}^+})\right]$ is a Maxwellian distribution function describing the $\text{D}^+$ ion population, centered at the ion velocity $\mathbf{v}_{\text{D}^+} (\mathbf{x}',t')$, which includes only the leading order components, i.e. $\mathbf{v}_{\text{D}^+} = v_{\|\text{D}^+} \mathbf{b} + \mathbf{v}_{\perp \text{D}^+ 0}$, and based on the $\text{D}^+$ temperature, $\mathbf{T}_{\text{D}^+} (\mathbf{x}',t')$. Similarly, $\Phi_{\left[\mathbf{v}_{\text{D}_2^+},T_{\text{D}_2^+}\right]} (\mathbf{x}',\mathbf{v},t')$ is a Maxwellian distribution that describes the $\text{D}_2^+$ ions and is defined analogously. We remark that, when evaluating the average velocity of the Maxwellian distributions describing neutrals generated from $\text{D}_2$ and $\text{D}_2^+$, we assume that $\mathbf{v}_{\text{D}_2}$ and $\mathbf{v}_{\text{D}_2^+}$ can be neglected, i.e. $\vert \mathbf{v}_{\text{D}_2} \vert \lesssim \vert \mathbf{v}_{\text{D}^+} \vert$ and $\vert \mathbf{v}_{\text{D}_2^+} \vert \lesssim \vert \mathbf{v}_{\text{D}^+} \vert$. Regarding the dissociative processes, we recall that the temperature $T_{\text{D,diss(D}_2\text{)}}$ is the average thermal energy of $\text{D}$ atoms generated by dissociation of $\text{D}_2$, presented in Table \ref{energies} and calculated in App. A. The energy of the neutral $\text{D}$ atoms generated by other dissociative processes is evaluated using a similar approach. 

The effective frequencies for depletion of neutral particles are given by

      \begin{equation}\label{nu_eff_D}
      \begin{aligned}
      & \nu_{\text{eff,D}}(\mathbf{x}'',t'')= \nu_{\text{iz,D}}(\mathbf{x}'',t'')+\nu_{\text{cx,D}}(\mathbf{x}'',t'')+ \nu_{\text{cx,D-D}_2^+}(\mathbf{x}'',t'')
      \end{aligned}
      \end{equation}
      
\noindent and
      
      \begin{equation}\label{nu_eff_D2}
      \begin{aligned}
      & \nu_{\text{eff,D}_2}(\mathbf{x}'',t'')= \nu_{\text{iz,D}_2}(\mathbf{x}'',t'') +\nu_{\text{cx,D}_2}(\mathbf{x}'',t'')+\nu_{\text{cx,D}_2-\text{D}^+}(\mathbf{x}'',t'')\\
      & +\nu_{\text{diss,D}_2}(\mathbf{x}'',t'')+\nu_{\text{diss-iz,D}_2}(\mathbf{x}'',t''),
      \end{aligned}
      \end{equation}

\noindent since the volumetric sinks of $\text{D}$ atoms are due to ionization or charge-exchange with $\text{D}^+$ or $\text{D}_2^+$, while $\text{D}_2$ are depleted by ionization, charge-exchange with $\text{D}_2^+$ or $\text{D}^+$, dissociation or dissociative ionization. 

A contribution to the neutral distribution functions in Eqs. (\ref{fD}) and (\ref{fD2}) is related to the neutral recycling at the boundary walls. Therefore, we now focus on the neutral processes that take place there. A fraction, $\alpha_{\text{refl}}(\mathbf{x'_{\text{b}}})$, of the $\text{D}_2^+$ ions that reach the boundary walls, after recombination with electrons and formation of $\text{D}_2$ neutrals is reflected back into the plasma. The remaining fraction, $1-\alpha_{\text{refl}}(\mathbf{x'_{\text{b}}})$, is absorbed and reemitted at wall temperature as $\text{D}_2$, also following a recombination process. Analogous considerations hold when describing $\text{D}_2$ neutrals that reach the boundary. The $\text{D}_2$ molecules can, in fact, be reflected or reemitted with the same probability as $\text{D}_2^+$. 

Regarding the atomic species, since the wall temperature is low, a fraction, $\beta_{\text{assoc}}$, of the $\text{D}^+$ and $\text{D}$ particles absorbed at the walls associate and are reemitted back into the plasma as $\text{D}_2$ molecules. The $\text{D}^+$ ions and $\text{D}$ neutrals reaching the boundaries that do not associate undergo reflection and reemission processes similar to the ones described for $\text{D}_2^+$ ions and $\text{D}_2$ particles, the probability of reflection, $\alpha_{\text{refl}}(\mathbf{x'_{\text{b}}})$, being the same. As a consequence, the distribution functions at the vessel, $f_{\text{D}} (\mathbf{x}'_{\text{b}},\mathbf{v},t')$ and $f_{\text{D}_2} (\mathbf{x}'_{\text{b}},\mathbf{v},t')$, for $v_p = \mathbf{v}\cdot\mathbf{\hat{n}}>0$ (with $\mathbf{\hat{n}}$ the unit vector normal to the boundary) yield

      \begin{equation}\label{f_D}
      \begin{aligned}
      & f_{\text{D}}(\mathbf{x}'_{\text{b}},\mathbf{v},t') = (1-\alpha_{\text{refl}}(\mathbf{x'_{\text{b}}})) \Gamma_{\text{\text{reem,D}}}(\mathbf{x'_b},t')\chi_{\text{in,D}}(\mathbf{x'_b},\mathbf{v}) \\
      & + \alpha_{\text{refl}}(\mathbf{x'_{\text{b}}}) \left[f_{\text{out,D}}(\mathbf{x'_b}, \mathbf{v}-2 \mathbf{v_p},t') + 
      \frac{\Gamma_{\text{out,D}^+}(\mathbf{x'_b},t')}{v_p} \Phi_{\left[\mathbf{v}_{\text{refl}\left(\text{D}^+\right)},T_{\text{D}^+}\right]}(\mathbf{x'_b},\mathbf{v},t')\right]
      \end{aligned}
      \end{equation}
      
\noindent and
      
      \begin{equation}\label{f_D2}
      \begin{aligned}
      & f_{\text{D}_2}(\mathbf{x}'_{\text{b}},\mathbf{v},t') = (1-\alpha_{\text{refl}}(\mathbf{x'_{\text{b}}})) \Gamma_{\text{reem},\text{D}_2}(\mathbf{x'_b},t')\chi_{\text{in,D}_2}(\mathbf{x'_b},\mathbf{v}) \\
      & + \alpha_{\text{refl}}(\mathbf{x'_{\text{b}}}) \left[f_{\text{out,D}_2}(\mathbf{x'_b}, \mathbf{v}-2 \mathbf{v_p},t') + \frac{\Gamma_{\text{out,D}_2^+}(\mathbf{x'_b},t')}{v_p} \Phi_{\left[\mathbf{v}_{\text{refl}\left(\text{D}_2^+\right)},T_{\text{D}_2^+}\right]}(\mathbf{x'_b},\mathbf{v},t')\right].
      \end{aligned}
      \end{equation}

\noindent We first analyse the contributions of reflected particles in Eqs. (\ref{f_D}-\ref{f_D2}). The reflected $\text{D}$ and $\text{D}_2$ are described by the distribution functions $f_{\text{out,D}}(\mathbf{x'_b}, \mathbf{v}-2 \mathbf{v_p},t')$ and $f_{\text{out,D}_2}(\mathbf{x'_b}, \mathbf{v}-2 \mathbf{v_p},t')$, since $\mathbf{v}-2 \mathbf{v_p}$ is the velocity of the reflected neutrals as they flow towards the wall, with $\mathbf{v_p} = v_p \mathbf{\hat{n}}$ the velocity along the direction normal to the wall surface. On the other hand, the contribution from the reflected $\text{D}^+$ and $\text{D}_2^+$ is modelled by considering the projection of the flux of outflowing $\text{D}^+$ and $\text{D}_2^+$ along the direction normal to the boundary surface, given respectively by $\Gamma_{\text{out,D}^+}(\mathbf{x'_b}) = -\mathbf{\Gamma_{\text{out,D}^+}}(\mathbf{x'_b}) \cdot \mathbf{\hat{n}}$ and $\Gamma_{\text{out,D}_2^+}(\mathbf{x'_b}) = -\mathbf{\Gamma_{\text{out,D}_2^+}}(\mathbf{x'_b}) \cdot \mathbf{\hat{n}}$. These fluxes include the contributions of the plasma parallel flow and the leading order perpendicular drifts, i.e. the $E \times B$ and diamagnetic drifts, yielding

      \begin{equation}\label{Gamma_out_Dp}
      \begin{aligned}
      & \mathbf{\Gamma_{\text{out,D}^+}}(\mathbf{x'_{\text{b}}}) = n_{\text{D}^+} v_{\| \text{D}^+} \mathbf{b} 
      + n_{\text{D}^+} \mathbf{v}_{\perp \text{D}^+ 0} \\
      \end{aligned}
      \end{equation}

\noindent and

      \begin{equation}\label{Gamma_out_D2p}
      \begin{aligned}
      & \mathbf{\Gamma_{\text{out,D}_2^+}}(\mathbf{x'_{\text{b}}}) = n_{\text{D}_2^+} v_{\| \text{D}_2^+} \mathbf{b} + n_{\text{D}_2^+} \mathbf{v}_{\perp \text{D}_2^+ 0}, \\
      \end{aligned}
      \end{equation} 
      
\noindent We assume that the velocity distribution of the $\text{D}$ neutrals generated by reflection of $\text{D}^+$ ions is described by a Maxwellian centered at the velocity, $\mathbf{v}_{\text{refl}\left(\text{D}^+\right)} = \mathbf{v}_{\text{D}^+}-2 \mathbf{v_p}_{\text{D}^+}$, with $\mathbf{v_p}_{\text{D}^+} = \left(\mathbf{v}_{\text{D}^+} \cdot \mathbf{\hat{n}}\right) \mathbf{\hat{n}}$, and with temperature of the incoming $\text{D}^+$ ions, $T_{\text{D}^+}$, given by $\Phi_{\left[\mathbf{v}_{\text{refl}\left(\text{D}^+\right)},T_{\text{D}^+}\right]}(\mathbf{x}',\mathbf{v},t')$. Analogously, the $\text{D}_2$ neutrals generated by reflection of $\text{D}_2^+$ ions are described by a Maxwellian distribution of velocities, $\Phi_{\left[\mathbf{v}_{\text{refl}\left(\text{D}_2^+\right)},T_{\text{D}^+}\right]}(\mathbf{x}',\mathbf{v},t')$, being $\mathbf{v}_{\text{refl}\left(\text{D}_2^+\right)} = \mathbf{v}_{\text{D}_2^+}-2 \mathbf{v_p}_{\text{D}_2^+}$, with $\mathbf{v_p}_{\text{D}_2^+} = \left(\mathbf{v}_{\text{D}_2^+} \cdot \mathbf{\hat{n}}\right) \mathbf{\hat{n}}$ and $T_{\text{D}_2^+}$ the temperature of the incoming $\text{D}_2^+$ ions. 

We now focus on the contributions in Eqs. (\ref{f_D}-\ref{f_D2}) accounting for reemission of neutrals from the boundary. These are written in terms of

      \begin{equation}\label{Gamma_D_reem}
      \begin{aligned}
      & \Gamma_{\text{reem,D}}(\mathbf{x'_{\text{b}}}) = (1-\beta_{\text{assoc}})\left[\Gamma_{\text{out,D}}(\mathbf{x'_{\text{b}}}) +\Gamma_{\text{out,D}^+}(\mathbf{x'_{\text{b}}})\right]
      \end{aligned}
      \end{equation}

\noindent and

      \begin{equation}\label{Gamma_D2_reem}
      \begin{aligned}
      & \Gamma_{\text{reem},\text{D}_2}(\mathbf{x'_{\text{b}}}) = \Gamma_{\text{out,D}_2}(\mathbf{x'_{\text{b}}}) + \Gamma_{\text{out,D}_2^+}(\mathbf{x'_{\text{b}}})+\frac{\beta_{\text{assoc}}}{2}\left[ \Gamma_{\text{\text{out,D}}}(\mathbf{x'_{\text{b}}}) + \Gamma_{\text{out,D}^+}(\mathbf{x'_{\text{b}}})\right].
      \end{aligned}
      \end{equation}

\noindent In addition to the projections of the boundary ion fluxes, $\Gamma_{\text{out,D}_2^+}$ and $\Gamma_{\text{out,D}_2^+}$, Eqs. (\ref{Gamma_D_reem}) and (\ref{Gamma_D2_reem}) take into account the projections along the direction normal to the boundary of the fluxes of $\text{D}$ and $\text{D}_2$ outflowing to the limiter and walls, $\Gamma_{\text{out,D}}$ and $\Gamma_{\text{out,D}_2}$. These are defined based on the neutral fluxes directed towards the boundary (i.e. for $v_p < 0$) as

      \begin{equation}\label{Gamma_D}
      \begin{aligned}
      \Gamma_{\text{out,D}}(\mathbf{x'_{\text{b}}}) = - \int_{v_p < 0} \left(\mathbf{v_p} \cdot \mathbf{n}\right) f_{\text{D}}(\mathbf{x'_{\text{b}}},\mathbf{v}) d\mathbf{v}
      \end{aligned}
      \end{equation}

\noindent and 

      \begin{equation}\label{Gamma_D2}
      \begin{aligned}
      \Gamma_{\text{out,D}_2}(\mathbf{x'_{\text{b}}}) = - \int_{v_p < 0} \left(\mathbf{v_p} \cdot \mathbf{n}\right) f_{\text{D}_2}(\mathbf{x'_{\text{b}}},\mathbf{v}) d\mathbf{v}.
      \end{aligned}
      \end{equation}

\noindent We assume that the velocity distribution of reemitted particles follows the Knudsen cosine law for a given wall temperature, $T_{\text{w}}$. This yields, for the $\text{D}$ neutrals,

      \begin{equation}\label{chi_D}
      \begin{aligned}
      & \chi_{\text{in,D}}(\mathbf{x'_{\text{b}}},\mathbf{v}) = \frac{3}{4 \pi}\frac{m_{\text{D}}^2}{T_{\text{w}}^2}\text{cos}(\theta)\exp\left(-\frac{m_{\text{D}} v^2}{2 T_{\text{w}}}\right),
      \end{aligned}
      \end{equation}

\noindent while the expression for $\text{D}_2$ molecules is analogously given by

      \begin{equation}\label{chi_D2}
      \begin{aligned}
      & \chi_{\text{in,D}_2}(\mathbf{x'_{\text{b}}},\mathbf{v}) = \frac{3}{4 \pi}\frac{m_{\text{D}_2}^2}{T_{\text{w}}^2}\text{cos}(\theta)\exp\left(-\frac{m_{\text{D}_2} v^2}{2 T_{\text{w}}}\right).
      \end{aligned}
      \end{equation}
      
We now follow the same approach described in \cite{Wersal2015} to obtain a set of time-independent two-dimensional integral equations for the $\text{D}$ and $\text{D}_2$ densities, making the numerical implementation of the formal solution in Eqs. (\ref{fD}) and (\ref{fD2}) feasible. More precisely, we first make use of the fact that the neutral time of flight is typically shorter than the characteristic timescales of turbulence, $\tau_{\text{n}} \ll \tau_{\text{turb}}$, a condition that we denote as the neutral adiabatic regime. This allows us to approximate $t'=t$ in Eqs. (\ref{fD}-\ref{f_D2}) or, equivalently, $\partial_t f_{\text{D}} = 0$ and $\partial_t f_{\text{D}_2} = 0$ in Eqs. (\ref{fD_char}-\ref{fD2_char}). Second, we note that the neutral mean free path is typically smaller than the characteristic elongation of turbulence structures along the magnetic field, $\lambda_{\text{mfp,n}} k_{\|} \ll 1$. Therefore, our description of neutral motion is reduced to the analysis of a set of independent two-dimensional planes perpendicular to the magnetic field, approximately coincident with the poloidal planes. Then, integrating Eqs. (\ref{fD}-\ref{fD2}) over the velocity space, a system of two coupled equations for the densities of $\text{D}$ and $\text{D}_2$ is obtained,

      \begin{equation}\label{before_nD}
      \begin{aligned}
      & n_{\text{D}}(\mathbf{x}_\perp) = \int_{\text{D}} dA' \frac{1}{r'_\perp} \int_0^{\infty} dv_{\perp} v_{\perp} \int_0^{\infty} dv_{\|} \left\{ \frac{S_D(\mathbf{x}'_\perp,\mathbf{v})}{v_{\perp}} \text{exp}\left[-\frac{1}{v_\perp}\int_0^{r'_{\perp}}\nu_{\text{eff,D}}(\mathbf{x}''_\perp)dr''_\perp\right]\right\} \\
      & + \int_{\partial \text{D}} da'_{\text{b}} \frac{\text{cos}\theta'}{r'_{\perp \text{b}}} \int_0^{\infty} dv_{\perp} v_{\perp} \int_0^{\infty} dv_{\|} \left\{ f_{\text{D}}(\mathbf{x'_{\perp \text{b}}},\mathbf{v}) \text{exp}\left[-\frac{1}{v_\perp}\int_0^{r'_{\perp}}\nu_{\text{eff,D}}(\mathbf{x}''_\perp)dr''_\perp\right]\right\},
      \end{aligned}
      \end{equation}

\noindent and

      \begin{equation}\label{before_nD2}
      \begin{aligned}
      & n_{\text{D}_2}(\mathbf{x}_\perp) = \int_{\text{D}} dA' \frac{1}{r'_\perp} \int_0^{\infty} dv_{\perp} v_{\perp} \int_0^{\infty} dv_{\|} \left\{ \frac{S_{\text{D}_2}(\mathbf{x}'_\perp,\mathbf{v})}{v_{\perp}} \text{exp}\left[-\frac{1}{v_\perp}\int_0^{r'_{\perp}}\nu_{\text{eff,D}_2}(\mathbf{x}''_\perp)dr''_\perp\right]\right\} \\
      & + \int_{\partial \text{D}} da'_{\text{b}} \frac{\text{cos}\theta'}{r'_{\perp \text{b}}} \int_0^{\infty} dv_{\perp} v_{\perp} \int_0^{\infty} dv_{\|} \left\{ f_{\text{D}_2}(\mathbf{x'_{\perp \text{b}}},\mathbf{v}) \text{exp}\left[-\frac{1}{v_\perp}\int_0^{r'_{\perp}}\nu_{\text{eff,D}_2}(\mathbf{x}''_\perp)dr''_\perp\right]\right\}.
      \end{aligned}
      \end{equation}
      
\noindent where the same geometrical arguments presented in \cite{Wersal2015} is used when considering the integral along the neutral path and the integral along the perpendicular velocity angle, that is 

      \begin{equation}\label{first}
      \begin{aligned}
      \int_0^{r_{\perp,\text{b}}} dr'_{\perp} \int_0^{2 \pi} d\vartheta F(\mathbf{x}_{\perp},\mathbf{x}'_{\perp}) =      
      \int_{\text{D}} dA' \frac{1}{r'_{\perp}} F(\mathbf{x}_{\perp},\mathbf{x}'_{\perp}),
      \end{aligned}
      \end{equation}

\noindent where $dA'$ is the area element in the 2D poloidal plane and $F(\mathbf{x}_{\perp},\mathbf{x}'_{\perp})$ is a generic function. In addition, we use

      \begin{equation}\label{second}
      \begin{aligned}
      \int_0^{r_{\perp,\text{b}}} dr'_{\perp} \int_0^{2 \pi} d\vartheta \delta(r'_{\perp} - r'_{\perp\text{b}}) F(\mathbf{x}_{\perp},\mathbf{x}'_{\perp}) =      
      \int_{\partial \text{D}} da'_{\text{b}} \frac{\text{cos}\theta'}{r'_{\perp\text{b}}} F(\mathbf{x}_{\perp},\mathbf{x}'_{\perp\text{b}}),
      \end{aligned}
      \end{equation}

\noindent with $da'_{\text{b}}$ being a line element along the boundary of $\text{D}$, denoted as $\partial \text{D}$, and $\theta'=\text{arccos}(\mathbf{\Omega}_{\perp} \cdot \mathbf{\hat{n}})$.
      
We now express the volumetric source terms appearing in Eqs. (\ref{S_D}) and (\ref{S_D2}), $S_{\text{D}}(\mathbf{x}',\mathbf{v})$ and $S_{\text{D}_2}(\mathbf{x}',\mathbf{v})$, in terms of $n_{\text{D}}$ and $n_{\text{D}_2}$, and the distribution functions of the neutral species at the boundary appearing in Eqs. (\ref{f_D}) and (\ref{S_D2}), $f_{\text{D}}$ and $f_{\text{D}_2}$, in terms of $\Gamma_{\text{out,D}^+}$, $\Gamma_{\text{out,D}_2^+}$, $\Gamma_{\text{out,D}}$ and $\Gamma_{\text{out,D}_2}$. For $n_{\text{D}_2}$, this yields

      \begin{equation}\label{nD2}
      \begin{aligned}
      & n_{\text{D}_2}(\mathbf{x}_\perp)= \int_{\text{D}} n_{\text{D}_2}(\mathbf{x}'_\perp) \nu_{\text{cx,D}_2}(\mathbf{x}'_\perp) K_{p\rightarrow p}^{\text{D}_2,\text{D}_2^+}(\mathbf{x}_\perp,\mathbf{x}'_\perp) dA'\\
      & + \int_{\partial \text{D}} (1-\alpha_{\text{refl}}(\mathbf{x}'_{\perp,\text{b}})) \Gamma_{\text{out,D}_2}(\mathbf{x}'_{\perp,\text{b}})K_{b\rightarrow p}^{\text{D}_2}(\mathbf{x}_\perp,\mathbf{x}'_{\perp,\text{b}}) da'_{\text{b}}\\
      & + \int_{\partial \text{D}} (1-\alpha_{\text{refl}}(\mathbf{x}'_{\perp,\text{b}})) \frac{\beta_{\text{assoc}}}{2}
      \Gamma_{\text{out,D}}(\mathbf{x}'_{\perp,\text{b}})K_{b\rightarrow p}^{\text{D}_2}(\mathbf{x}_\perp,\mathbf{x}'_{\perp,\text{b}}) da'_{\text{b}}\\
      & + \int_{\text{D}} n_{\text{D}}(\mathbf{x}'_\perp) \nu_{\text{cx,D-D}_2^+}(\mathbf{x'_\perp}) K_{p\rightarrow p}^{\text{D}_2,\text{D}_2^+}(\mathbf{x}_\perp,\mathbf{x}'_\perp) dA' + n_{\text{D}_2[\text{rec}(\text{D}_2^+)]}(\mathbf{x}_\perp) \\
      & + n_{\text{D}_2[\text{out}(\text{D}_2^+)]}(\mathbf{x}_\perp) + n_{\text{D}_2[\text{out}(\text{D}^+)]}(\mathbf{x}_\perp),
      \end{aligned}
      \end{equation}

\noindent while for $n_{\text{D}}$ one has

      \begin{equation}\label{nD}
      \begin{aligned}
      & n_{\text{D}}(\mathbf{x}_\perp)= \int_{\text{D}} n_{\text{D}}(\mathbf{x}'_\perp) \nu_{\text{cx,D}}(\mathbf{x}'_\perp) K_{p\rightarrow p}^{\text{D,D}^+}(\mathbf{x}_\perp,\mathbf{x}'_\perp) dA'\\
      & + \int_{\text{D}} n_{\text{D}_2}(\mathbf{x}'_\perp) \nu_{\text{cx,D}_2-\text{D}^+}(\mathbf{x}'_\perp) K_{p\rightarrow p}^{\text{D,D}^+}(\mathbf{x}_\perp,\mathbf{x}'_\perp) dA'\\
      & + \int_{\text{D}} 2 n_{\text{D}_2}(\mathbf{x}'_\perp) \nu_{\text{diss,D}_2^+}(\mathbf{x}'_\perp) K_{p\rightarrow p}^{\text{D,diss}\left(\text{D}_2^+\right)}(\mathbf{x}_\perp,\mathbf{x}'_\perp) dA'\\
      & + \int_{\text{D}} n_{\text{D}_2}(\mathbf{x}'_\perp) \nu_{\text{diss-iz,D}_2^+}(\mathbf{x}'_\perp) K_{p\rightarrow p}^{\text{D,diss-iz}\left(\text{D}_2^+\right)}(\mathbf{x}_\perp,\mathbf{x}'_\perp) dA'\\
      & + \int_{\partial \text{D}} (1-\alpha_{\text{refl}}(\mathbf{x}'_{\perp,\text{b}})) (1-\beta_{\text{assoc}}) \Gamma_{\text{out,D}}(\mathbf{x}'_{\perp,\text{b}})K_{b\rightarrow p}^{\text{D,reem}} (\mathbf{x}_\perp,\mathbf{x}'_{\perp,\text{b}}) da'_{\text{b}}\\
      & + n_{\text{D}[\text{rec}(\text{D}^+)]}(\mathbf{x}_\perp) 
      + n_{\text{D}[\text{out}(\text{D}^+)]}(\mathbf{x}_\perp) + n_{\text{D}[\text{diss}(\text{D}_2^+)]}(\mathbf{x}_\perp).
      \end{aligned}
      \end{equation}

\noindent Replacing $v_p$ in Eqs. (\ref{Gamma_D2}) and (\ref{Gamma_D}), the normal projections of the fluxes of $\text{D}_2$ and $\text{D}$ can be written respectively as $\Gamma_{\text{out,D}_2}(\mathbf{x}'_{\perp,\text{b}}) = - \int_{\text{cos}(\theta) < 0} v_{\perp} \text{cos}\theta f_{\text{D}_2}(\mathbf{x}'_{\perp,\text{b}},\mathbf{v}_{\perp}) d\mathbf{v}_{\perp}$ and $\Gamma_{\text{out,D}}(\mathbf{x}'_{\perp,\text{b}}) = - \int_{\text{cos}(\theta) < 0} v_{\perp} \text{cos}\theta f_{\text{D}}(\mathbf{x}'_{\perp,\text{b}},\mathbf{v}_{\perp}) d\mathbf{v}_{\perp}$. By replacing $f_{\text{D}_2}(\mathbf{x}'_{\perp,\text{b}},\mathbf{v}_{\perp})$ and $f_{\text{D}}(\mathbf{x}'_{\perp,\text{b}},\mathbf{v}_{\perp})$ by their expressions as given in Eqs. (\ref{f_D}) and (\ref{f_D2}), these fluxes can be rewritten in terms of $n_{\text{D}}$, $n_{\text{D}_2}$, $\Gamma_{\text{out,D}^+}$, $\Gamma_{\text{out,D}_2^+}$, $\Gamma_{\text{out,D}}$ and $\Gamma_{\text{out,D}_2}$ as 

      \begin{equation}\label{GammaD2}
      \begin{aligned}
      & \Gamma_{\text{out,D}_2}(\mathbf{x}_{\perp,\text{b}}) = \int_{\text{D}} n_{\text{D}_2}(\mathbf{x}'_\perp) \nu_{\text{cx,D}_2}(\mathbf{x}'_\perp) K_{p\rightarrow b}^{\text{D}_2,\text{D}_2^+}(\mathbf{x}_\perp,\mathbf{x}'_\perp) dA'\\
      & + \int_{\partial \text{D}} (1-\alpha_{\text{refl}}(\mathbf{x}'_{\perp,\text{b}})) \Gamma_{\text{out,D}_2}(\mathbf{x}'_{\perp,\text{b}})K_{b\rightarrow b}^{\text{D}_2}(\mathbf{x}_\perp,\mathbf{x}'_{\perp,\text{b}}) da'_{\text{b}}\\
      & + \int_{\partial \text{D}} (1-\alpha_{\text{refl}}(\mathbf{x}'_{\perp,\text{b}})) \frac{\beta_{\text{assoc}}}{2}
      \Gamma_{\text{out,D}}(\mathbf{x}'_{\perp,\text{b}})K_{b\rightarrow b}^{\text{D}_2}(\mathbf{x}_\perp,\mathbf{x}'_{\perp,\text{b}}) da'_{\text{b}}\\
      & + \int_{\text{D}} n_{\text{D}}(\mathbf{x}'_\perp) \nu_{\text{cx,D-D}_2^+}(\mathbf{x'_\perp}) K_{p\rightarrow b}^{\text{D}_2,\text{D}_2^+}(\mathbf{x}_\perp,\mathbf{x}'_\perp) dA'\\
      & + \Gamma_{\text{out,D}_2[\text{rec}(\text{D}_2^+)]}(\mathbf{x}_\perp) + \Gamma_{\text{out,D}_2[\text{out}(\text{D}_2^+)]}(\mathbf{x}_\perp) + \Gamma_{\text{out,D}_2[\text{out}(\text{D}^+)]}(\mathbf{x}_\perp),\\
      \end{aligned}
      \end{equation}

\noindent and

      \begin{equation}\label{GammaD}
      \begin{aligned}
      & \Gamma_\text{out,D}(\mathbf{x}_{\perp,\text{b}})= \int_{\text{D}} n_{\text{D}}(\mathbf{x}'_\perp) \nu_{\text{cx,D}}(\mathbf{x}'_\perp) K_{p\rightarrow b}^{\text{D,D}^+}(\mathbf{x}_\perp,\mathbf{x}'_\perp) dA'\\
      & + \int_{\text{D}} n_{\text{D}_2}(\mathbf{x}'_\perp) \nu_{\text{cx,D}_2-\text{D}^+}(\mathbf{x}'_\perp) K_{p\rightarrow b}^{\text{D,D}^+}(\mathbf{x}_\perp,\mathbf{x}'_\perp) dA'\\
      & + \int_{\text{D}} 2 n_{\text{D}_2}(\mathbf{x}'_\perp) \nu_{\text{diss,D}_2^+}(\mathbf{x}'_\perp) K_{p\rightarrow b}^{\text{D,diss}\left(\text{D}_2^+\right)}(\mathbf{x}_\perp,\mathbf{x}'_\perp) dA'\\
      & + \int_{\text{D}} n_{\text{D}_2}(\mathbf{x}'_\perp) \nu_{\text{diss-iz,D}_2^+}(\mathbf{x}'_\perp) K_{p\rightarrow b}^{\text{D,diss-iz}\left(\text{D}_2^+\right)}(\mathbf{x}_\perp,\mathbf{x}'_\perp) dA'\\
      & + \int_{\partial \text{D}} (1-\alpha_{\text{refl}}(\mathbf{x}'_{\perp,\text{b}})) (1-\beta_{\text{assoc}}) \Gamma_{\text{out,D}}(\mathbf{x}'_{\perp,\text{b}})K_{b\rightarrow b}^{\text{D,reem}} (\mathbf{x}_\perp,\mathbf{x}'_{\perp,\text{b}}) da'_{\text{b}}\\
      & + \Gamma_{\text{D}[\text{rec}(\text{D}^+)]}(\mathbf{x}_\perp) 
      + \Gamma_{\text{D}[\text{out}(\text{D}^+)]}(\mathbf{x}_\perp) + \Gamma_{\text{D}[\text{diss}(\text{D}_2^+)]}(\mathbf{x}_\perp),
      \end{aligned}
      \end{equation}
      
\noindent We note that that the neutral particle densities and fluxes in Eqs. (\ref{nD2}-\ref{GammaD}) are multiplied by a factor $1-\alpha_{\text{refl}}(\mathbf{x}'_{\perp,\text{b}})$ in order to account only for the contribution of particles that are reemitted at the boundary, hence excluding reflection. Neutral reflection is included, in the definition of the kernel functions that appear in Eqs. (\ref{before_nD}-\ref{before_nD2}). 

We now turn to the definition of the kernel functions appearing in Eqs. (\ref{nD2}-\ref{GammaD}). These are defined as integrals over velocity space. For instance, $K_{p\rightarrow p}^{\text{D}_2,\text{D}_2^+}(\mathbf{x}_\perp,\mathbf{x}'_\perp)$ quantifies the amount of $\text{D}_2$ neutrals found at a location $\mathbf{x}_\perp$ in the plasma volume ($p$) being generated from collisions involving neutralization of $\text{D}_2^+$ ions at a location $\mathbf{x}'_\perp$ inside the plasma volume ($p$). Its expression is given by

      \begin{flalign}\label{kernels_dr}
      &\begin{aligned}
      &K_{p\rightarrow p}^{\text{D}_2,\text{D}_2^+}(\mathbf{x}_\perp,\mathbf{x}'_\perp) = K_{p\rightarrow p, \text{dir}}^{\text{D}_2,\text{D}_2^+}(\mathbf{x}_\perp,\mathbf{x}'_\perp) + \alpha_{\text{refl}} K_{p\rightarrow p, \text{refl}}^{\text{D}_2,\text{D}_2^+}(\mathbf{x}_\perp,\mathbf{x}'_\perp).
      \end{aligned}&
      \end{flalign}
      
\noindent which separates the contributions to $n_{\text{D}_2}$ arising from the direct path of length $r'_{\perp,\text{dir}}$ connecting $\mathbf{x}_\perp$ and $\mathbf{x}'_\perp$, $K_{p\rightarrow p, \text{dir}}^{\text{D}_2,\text{D}_2^+}(\mathbf{x}_\perp,\mathbf{x}'_\perp)$, and the path corresponding to the trajectory of neutrals that are reflected at the boundary, $K_{p\rightarrow p, \text{refl}}^{\text{D}_2,\text{D}_2^+}(\mathbf{x}_\perp,\mathbf{x}'_\perp)$. Both $K_{p\rightarrow p, \text{dir}}^{\text{D}_2,\text{D}_2^+}$ and $K_{p\rightarrow p, \text{refl}}^{\text{D}_2,\text{D}_2^+}$ have the same expression,

      \begin{flalign}\label{dir}
      &\begin{aligned}
      &K_{p\rightarrow p,\text{path}}^{\text{D}_2,\text{D}_2^+}(\mathbf{x}_\perp,\mathbf{x}'_\perp) = \int_0^\infty \frac{1}{r'_{\perp,\text{path}}} \Phi_{\perp\left[\mathbf{v_{\perp}}_{\text{D}_2^+},T_{\text{D}_2^+}\right]}(\mathbf{x'_{\perp}},\mathbf{v_{\perp}})
      \text{exp}\left[-\frac{1}{v_\perp}\int_0^{r'_{\perp,\text{path}}}\nu_{\text{eff,D}_2}(\mathbf{x}''_\perp)dr''_\perp\right]dv_\perp,
      \end{aligned}&
      \end{flalign}  

\noindent where $\text{path} = \{ \text{dir}, \text{refl} \}$ and $r'_{\perp,\text{path}}$ is the distance between $\mathbf{x}_\perp$ and $\mathbf{x}'_\perp$ measured along the path (for the direct trajectory $r'_{\perp,\text{dir}}$ is given by the distance between the two points along a straight line, while for the reflected trajectory $r'_{\perp,\text{refl}}$ is the sum of the distances between $\mathbf{x}'_\perp$ and the boundary and the distance from the boundary to $\mathbf{x}_\perp$). We remark that $\Phi_{\perp\left[\mathbf{v_{\perp}}_{\text{D}_2^+},T_{\text{D}_2^+}\right]}(\mathbf{x'_{\perp}},\mathbf{v_{\perp}})$ is the integral along the parallel velocity of the $\text{D}_2^+$ Maxwellian distribution function, $\Phi_{\perp\left[\mathbf{v_{\perp}}_{\text{D}_2^+},T_{\text{D}_2^+}\right]}(\mathbf{x'_{\perp}},\mathbf{v_{\perp}}) = \int_{-\infty}^\infty \Phi_{\left[\mathbf{v_{\perp}}_{\text{D}_2^+},T_{\text{D}_2^+}\right]}(\mathbf{x}',\mathbf{v_{\perp}}) dv_{\|}$. We also remark that $K_{p\rightarrow p, \text{dir}}^{\text{D}_2,\text{D}_2^+}$ in Eq. (\ref{dir}) is valid in case the points are optically connected, i.e. if the straight line connecting the two points does not cross the core region nor the limiter plates. Otherwise, if the points are not connected, one has $K_{p\rightarrow p, \text{dir}}^{\text{D}_2,\text{D}_2^+} = 0$. As for $K_{p\rightarrow p, \text{refl}}^{\text{D}_2,\text{D}_2^+}$, in the present work we assume no reflection at the outer walls, while reflection of ions and neutrals may take place at the limiter plates. The other kernels appearing in Eqs. (\ref{before_nD}-\ref{before_nD2}) have the same structure as $K_{p\rightarrow p}^{\text{D}_2,\text{D}_2^+}$, and they take into account possible direct and reflected paths connecting the two points. These kernels are presented in detail in App. C.

We now turn to the evaluation of the non-homogeneous terms appearing in Eqs. (\ref{nD2}-\ref{GammaD}), i.e. the terms that are not proportional to $n_{\text{D}}$ nor $n_{\text{D}_2}$. For instance, these terms include the contribution of the ions recycled at the wall. In fact, the reflection and reemission of $\text{D}^+$ ions that outflow to the boundary and recombine with electrons contribute to the density of neutral $\text{D}$ atoms, through the term

      \begin{flalign}
      &\begin{aligned}
      & n_{\text{D}[\text{out,D}^+]}(\mathbf{x}_\perp)=\int_{\partial \text{D}} \Gamma_{\text{out,D}^+}(\mathbf{x}'_{\perp,\text{b}}) \left[(1-\alpha_{\text{refl}}(\mathbf{x}'_{\perp,\text{b}}))\left(1-\beta_{\text{assoc}}\right)K_{b\rightarrow p}^{\text{D,reem}}(\mathbf{x}_\perp,\mathbf{x}'_{\perp,\text{b}}) \right. \\
      & \left. + \alpha_{\text{refl}}(\mathbf{x}'_{\perp,\text{b}}) K_{b\rightarrow p}^{\text{D,refl}}(\mathbf{x}_\perp,\mathbf{x}'_{\perp,\text{b}}) \right] da'_{\text{b}}, 
      \end{aligned}&
      \end{flalign}

\noindent where $\Gamma_{\text{out,D}^+}$ is defined in Eq. (\ref{Gamma_out_Dp}). Similarly, the recombination of $\text{D}_2^+$ ions with electrons at the walls that are then either reflected or reemitted as $\text{D}_2$, and the recombination of $\text{D}^+$ ions with electrons at the walls and the following association into $\text{D}_2$ molecules contribute to the density of the $\text{D}_2$ species. These contributions can be expressed as   
      
      \begin{flalign}
      &\begin{aligned}
      & n_{\text{D}_2[\text{out,D}_2^+]}(\mathbf{x}_\perp)=\int_{\partial \text{D}} \Gamma_{\text{out,D}_2^+}(\mathbf{x}'_{\perp,\text{b}}) \left[(1-\alpha_{\text{refl}}(\mathbf{x}'_{\perp,\text{b}}))K_{b\rightarrow p}^{\text{D}_2\text{,reem}}(\mathbf{x}_\perp,\mathbf{x}'_{\perp,\text{b}}) \right. \\
      & \left. + \alpha_{\text{refl}}(\mathbf{x}'_{\perp,\text{b}}) K_{b\rightarrow p}^{\text{D}_2\text{,refl}}(\mathbf{x}_\perp,\mathbf{x}'_{\perp,\text{b}}) \right] da'_{\text{b}},
      \end{aligned}&
      \end{flalign}

\noindent and

      \begin{flalign}
      &\begin{aligned}
      & n_{\text{D}_2[\text{out,D}^+]}(\mathbf{x}_\perp)=\int_{\partial \text{D}} \Gamma_{\text{out,D}^+}(\mathbf{x}'_{\perp,\text{b}}) \left[(1-\alpha_{\text{refl}}(\mathbf{x}'_{\perp,\text{b}}))\frac{\beta_{\text{assoc}}}{2}
      K_{b\rightarrow p}^{\text{D}_2\text{,reem}}(\mathbf{x}_\perp,\mathbf{x}'_{\perp,\text{b}})\right] da'_{\text{b}}.
      \end{aligned}&
      \end{flalign}
      
We also define the non-homogeneous terms appearing in Eqs. (\ref{GammaD2}-\ref{GammaD}), that provide the contributions to the flux of neutrals at the boundary, $\Gamma_{\text{out,D}}$ and $\Gamma_{\text{out,D}_2}$, given by the ions outflowing to the wall. Following a similar approach to the one presented above, this can be expressed as  
      
      \begin{flalign}
      &\begin{aligned}
      & \Gamma_{\text{out,D}_2[\text{out,D}_2^+]}(\mathbf{x}_{\perp,\text{b}})=\int_{\partial \text{D}} \Gamma_{\text{out,D}_2^+}(\mathbf{x}'_{\perp,\text{b}}) \left[(1-\alpha_{\text{refl}}(\mathbf{x}'_{\perp,\text{b}}))K_{b\rightarrow b}^{\text{D}_2\text{,reem}}(\mathbf{x}_{\perp,\text{b}},\mathbf{x}'_{\perp,\text{b}}) \right. \\
      & \left. + \alpha_{\text{refl}}(\mathbf{x}'_{\perp,\text{b}}) K_{b\rightarrow b}^{\text{D}_2\text{,refl}}(\mathbf{x}_{\perp,\text{b}},\mathbf{x}'_{\perp,\text{b}}) \right] da'_{\text{b}},
      \end{aligned}&
      \end{flalign}
      
      \begin{flalign}
      &\begin{aligned}
      & \Gamma_{\text{out,D}_2[\text{out,D}^+]}(\mathbf{x}_{\perp,\text{b}})=\int_{\partial \text{D}} \Gamma_{\text{out,D}^+}(\mathbf{x}'_{\perp,\text{b}}) \left[(1-\alpha_{\text{refl}}(\mathbf{x}'_{\perp,\text{b}}))\frac{\beta_{\text{assoc}}}{2}
      K_{b\rightarrow b}^{\text{D}_2\text{,reem}}(\mathbf{x}_{\perp,\text{b}},\mathbf{x}'_{\perp,\text{b}})\right] da'_{\text{b}},
      \end{aligned}&
      \end{flalign}
      
\noindent and
      
      \begin{flalign}
      &\begin{aligned}
      & \Gamma_{\text{out,D}[\text{out,D}^+]}(\mathbf{x}_{\perp,\text{b}})=\int_{\partial \text{D}} \Gamma_{\text{out,D}^+}(\mathbf{x}'_{\perp,\text{b}}) \left[(1-\alpha_{\text{refl}}(\mathbf{x}'_{\perp,\text{b}}))\left(1-\beta_{\text{assoc}}\right)K_{b\rightarrow b}^{\text{D,reem}}(\mathbf{x}_{\perp,\text{b}},\mathbf{x}'_{\perp,\text{b}}) \right. \\
      & \left. + \alpha_{\text{refl}}(\mathbf{x}'_{\perp,\text{b}}) K_{b\rightarrow b}^{\text{D,refl}}(\mathbf{x}_{\perp,\text{b}},\mathbf{x}'_{\perp,\text{b}}) \right] da'_{\text{b}}. \\
      \end{aligned}&
      \end{flalign}

We now turn to the evaluation of the contributions to the neutral particles appearing in Eqs. (\ref{nD2}-\ref{GammaD}) caused by volumetric processes that involve the ion species $\text{D}^+$ and $\text{D}_2^+$. The contribution to the $\text{D}_2$ density as a result of $\text{D}_2^+$ recombination processes is given by

      \begin{flalign}\label{nD2_rec}
      &\begin{aligned}
      n_{\text{D}_2[\text{rec,D}_2^+]}(\mathbf{x}_\perp) = \int_{\text{D}} n_{\text{D}_2^+}(\mathbf{x}'_\perp) \nu_{\text{rec,D}_2^+}(\mathbf{x}'_\perp) K_{p\rightarrow   p}^{\text{D}_2,\text{D}_2^+}(\mathbf{x}_\perp,\mathbf{x}'_\perp) dA',
      \end{aligned}&
      \end{flalign}
    
\noindent and the contribution to the flux of $\text{D}_2$ to the boundary, also associated to $\text{D}_2^+$ recombination events, is expressed as
      
      \begin{flalign}\label{GammaD2_rec}
      &\begin{aligned}
      \Gamma_{\text{out,D}_2[\text{rec,D}_2^+]}(\mathbf{x}_\perp) = \int_{\text{D}} n_{\text{D}_2^+}(\mathbf{x}'_\perp) \nu_{\text{rec,D}_2^+}(\mathbf{x}'_\perp) K_{p\rightarrow   b}^{\text{D}_2,\text{D}_2^+}(\mathbf{x}_\perp,\mathbf{x}'_\perp) dA'.
      \end{aligned}&
      \end{flalign}
      
\noindent Similar contributions from volumetric recombination processes are considered for the $\text{D}$ neutral species. The contribution to the $\text{D}$ density as a result of $\text{D}^+$ recombination yields
    
      \begin{flalign}\label{nD_rec}
      &\begin{aligned}
      n_{\text{D}[\text{rec,D}^+]}(\mathbf{x}_\perp)=\int_{\text{D}} n_{\text{D}^+}(\mathbf{x}'_\perp) \nu_{\text{rec,D}^+}(\mathbf{x}'_\perp) K_{p\rightarrow p}^{\text{D,D}^+}(\mathbf{x}_\perp,\mathbf{x}'_\perp) dA',
      \end{aligned}&
      \end{flalign}
      
\noindent while an analogous definition is used for the flux of $\text{D}$,

      \begin{flalign}\label{GammaD_rec}
      &\begin{aligned}
      \Gamma_{\text{out,D}[\text{rec,D}^+]}(\mathbf{x}_\perp)=\int_{\text{D}} n_{\text{D}^+}(\mathbf{x}'_\perp) \nu_{\text{rec,D}^+}(\mathbf{x}'_\perp) K_{p\rightarrow b}^{\text{D,D}^+}(\mathbf{x}_\perp,\mathbf{x}'_\perp) dA'.
      \end{aligned}&
      \end{flalign}
      
Finally, the contribution of the dissociation of $\text{D}_2^+$ ions to $n_{\text{D}}$ appearing in Eq. (\ref{nD}) is evaluated as
      
      \begin{flalign}
      &\begin{aligned}
      & n_{\text{D}[\text{diss}(\text{D}_2^+)]}(\mathbf{x_{\perp}})=\int_{\text{D}} n_{\text{D}_2^+}(\mathbf{x}'_\perp) \nu_{\text{diss,D}_2^+}(\mathbf{x}'_\perp) K_{p\rightarrow p}^{\text{D,diss}\left(\text{D}_2^+\right)}(\mathbf{x_{\perp}},\mathbf{x}'_\perp) dA' \\
      & + \int_{\text{D}} 2 n_{\text{D}_2^+}(\mathbf{x}'_\perp) \nu_{\text{diss-rec,D}_2^+}(\mathbf{x}'_\perp) K_{p\rightarrow p}^{\text{D,diss-rec}\left(\text{D}_2^+\right)}(\mathbf{x_{\perp}},\mathbf{x}'_\perp) dA'.
      \end{aligned}&
      \end{flalign}

Similarly, a dissociation process of $\text{D}_2^+$ ions results in a contribution to $\Gamma_{\text{out,D}}$ in Eq. (\ref{GammaD}) given by

      \begin{flalign}
      &\begin{aligned}
      & \Gamma_{\text{out,D}[\text{diss}(\text{D}_2^+)]}(\mathbf{x}_\perp)=\int_{\text{D}} n_{\text{D}_2^+}(\mathbf{x}'_\perp) \nu_{\text{diss,D}_2^+}(\mathbf{x}'_\perp) K_{p\rightarrow b}^{\text{D,diss}\left(\text{D}_2^+\right)}(\mathbf{x_{\perp,b}},\mathbf{x}'_\perp) dA' \\
      & + \int_{\text{D}} 2 n_{\text{D}_2^+}(\mathbf{x}'_\perp) \nu_{\text{diss-rec,D}_2^+}(\mathbf{x}'_\perp) K_{p\rightarrow b}^{\text{D,diss-rec}\left(\text{D}_2^+\right)}(\mathbf{x_{\perp,b}},\mathbf{x}'_\perp) dA'.
      \end{aligned}&
      \end{flalign}
      
For their numerical solution, the system of kinetic equations for the neutral species is discretized on a regular cartesian grid in the $(R,Z)$ plasma and then written in matrix form. The details of the numerical implementation of the neutral model are discussed in App. D.

\section{\label{sec:level5} First simulation of a multi-component plasma with the GBS code}

We present the first results from simulations of turbulence in the tokamak boundary carried out by using the multi-component plasma model described in Secs. II-IV and implemented in the GBS code. Similarly to \cite{Ricci2012,Wersal2015}, we consider a tokamak with an infinitesimally thin toroidal limiter at the HFS equatorial midplane, with major radius $R_0/\rho_{\text{s}0} = 500$, and we simulate a three-dimensional domain with an annular cross section that includes the edge and the open-field line region of the device. The radial size of the domain is $S_{\text{rad}} = 150 \rho_{\text{s}0}$ and a the poloidal size is $S_{\text{pol}} = 800 \rho_{\text{s}0}$ at the core interface. Since the limiter has a radial width of $75 \rho_{\text{s}0}$), both the open and closed field-line regions have a radial extension of $75 \rho_{\text{s}0}$, corresponding to half the size along the radial direction. The parameters chosen for the present simulation are $q = 3.992$, $n_0 = 2 \times 10^{13} \text{cm}^{-3}$, $T_0 = 20.0 \text{eV}$, $\Omega_{\text{ci}} = 5.0 \times 10^7 \text{s}^{-1}$, $T_{\text{W}} = 0.3 \text{eV}$, $\nu = 0.1$, $\eta_{0\text{e}} = \eta_{0\text{D}^+} = 1.0$, $\eta_{0\Omega} = 4.0$, $\chi_{\|0,\text{e}} = 0.5$, $\chi_{\|0,\text{D}^+} = 0.05$, $\chi_{\|0,\text{D}_2^+} = 0.05$, $D_{\| \text{n}_e} = 0.5$, $D_{\| \text{n}_{\text{D}_2^+}} = 0.0$,  $D_{\| v_{\|\text{e}}} = 0.5$, $D_{\| v_{\|\text{D}^+}} = 0.0$, $D_{\| v_{\|\text{D}_2^+}} = 0.5$, $D_{\perp \text{n}_{\text{e}}} = 21.0$, $D_{\| v_{\|\text{e}}} = 0.5$, $D_{\| v_{\|\text{D}^+}} = 0.0$, $D_{\| v_{\|\text{D}_2^+}} = 0.5$ and $D_{\perp \text{n}_{\text{e}}} = 21.0$, $D_{\perp \text{n}_{\text{D}_2^+}} = D_{\perp \Omega} = D_{\perp v_{\|\text{e}}} = D_{\perp v_{\|\text{D}^+}} = D_{\perp v_{\|\text{D}_2^+}} = D_{\perp T_{\text{e}}} = D_{\perp T_{\text{D}^+}} = D_{\perp T_{\text{D}_2^+}} = 7.0$. Regarding the reflection probability at the limiter, we remark that it depends strongly on the particle energy and the wall material (see \cite{Stangeby2000}).  In this simulation, reflection of ions and neutrals takes place at the limiter plates with a given probability $\alpha_{\text{refl,lim}}$, constant along the limiter surface. The fraction of reflection at the boundary is therefore defined as 

      \begin{flalign}
      &\begin{aligned}
      \alpha_{\text{refl}}(\mathbf{x}'_{\perp,\text{b}}) = \left\{
      \begin{array}{ll}
      \alpha_{\text{refl,lim}} \neq 0 & \text{if\ } \mathbf{x}'_{\perp,\text{b}} \text{is\ located\ at\ limiter\ walls} \\
      0 & \text{if\ } \mathbf{x}'_{\perp,\text{b}} \text{is\ located\ at\ the\ outer\ and\ inner\ boundary.}
      \end{array}
      \right.
      \end{aligned}&
      \end{flalign}
      
\noindent We choose to consider metallic boundaries and hence we assume $\alpha_{\text{refl,lim}} = 0.8$, a value similar to the one adopted in \cite{Wersal2015}. We also assume $\beta_{\text{assoc}} = 0.95$, which is consistent with the usual assumption that most $\text{D}$ atoms associate into $\text{D}_2$ molecules at the boundary (see e.g. \cite{Zhang2019,Galassi2020}). 

Regarding the numerical parameters, we note that the plasma grid resolution is $n_{\text{x,p}} \times n_{\text{y,p}}\times n_{\text{z,p}} = 255\times 511\times 64$ while neutral grid resolution is $n_{\text{x,n}}\times n_{\text{y,n}}\times n_{\text{z,n}} = 24\times 138\times 64$. The time step is $3.75 \times 10^{-5} R_0/c_{\text{s}}$ and the neutral quantities are evaluated every $\Delta t = 0.1 R_0/c_{\text{s}}$. Although we have not carried out convergence studies with the multispecies model presented in this paper, convergence on plasma and neutral grid refinement has been studied within the single component framework. The conclusions presented in \cite{Giacomin2021}, which we expect to remain valid in the multispecies model presented here, show that our results are converged with respect to the frequency of neutral calculation. 

\begin{figure}[H]
\centering
\includegraphics[keepaspectratio=t,width=0.65\textwidth]{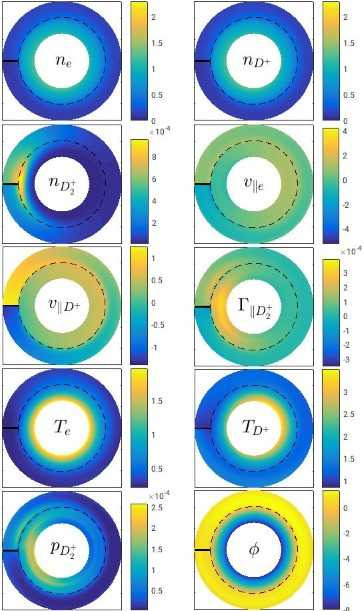}
\caption{Cross section plots of the electron density ($n_\text{e}$), $\text{D}^+$ density ($n_{\text{D}^+}$), $\text{D}_2^+$ density ($n_{\text{D}_2^+}$), electron parallel velocity ($v_{\| \text{e}}$), $\text{D}^+$ parallel velocity ($v_{\| \text{D}^+}$), $\text{D}_2^+$ parallel velocity ($v_{\| \text{D}_2^+}$), electron temperature ($T_{\text{e}}$), $\text{D}^+$ temperature ($T_{\text{D}^+}$), $\text{D}_2^+$ temperature ($T_{\text{D}_2^+}$) and electrostatic potential ($\phi$), toroidal and time-averaged over an interval of $\Delta t = 10.1 R_0/c_{\text{s}0}$ from the quasi-steady state of the multi-component plasma simulation described in Sec. 6.}
\label{plasma}
\end{figure}

\begin{figure}[H]
\centering
\includegraphics[keepaspectratio=t,width=0.85\textwidth]{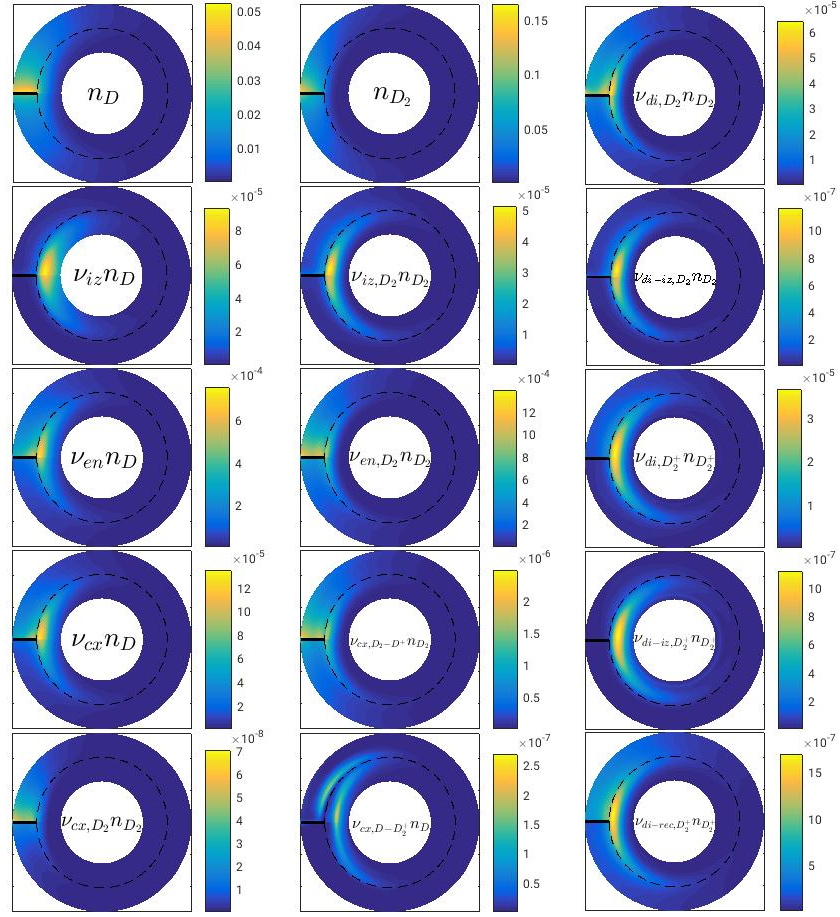}
\caption{Cross section plots of the neutral species densities and source terms resulting from the neutral-plasma interaction, toroidal and time-averaged over an interval of $\Delta t = 10.1 R_0/c_{\text{s}0}$ from the quasi-steady state of the multi-component plasma simulation described in Sec. 6.}
\label{neutral}
\end{figure}

\begin{figure}[ht]
\centering
\includegraphics[keepaspectratio=t,width=0.85\textwidth]{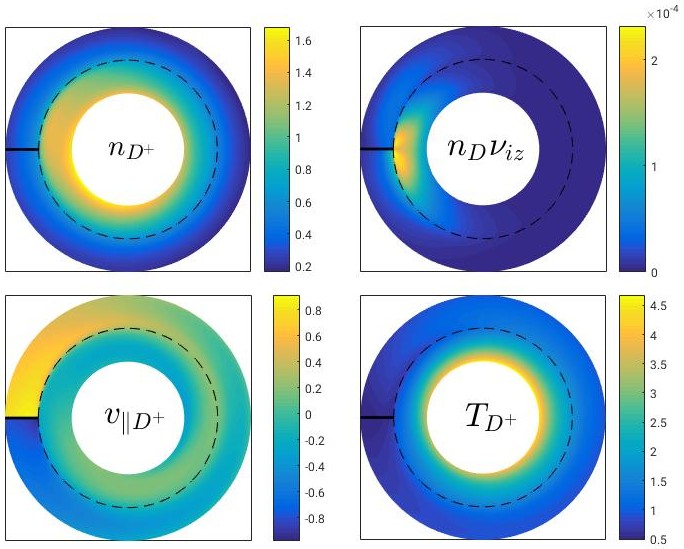}
\caption{Cross section plots of plasma density $n = n_{\text{e}} = n_{\text{D}^+}$, ion parallel velocity $v_{\|\text{D}^+}$, ion temperature $T_{\text{D}^+}$ and ionization source term $n_{\text{D}^+} \nu_{\text{iz}}$, toroidal and time-averaged over an interval of $\Delta t = 10.1 R_0/c_{\text{s}0}$ from a quasi-steady state single-component plasma simulation. The grid sizes and simulation parameters are the same as the ones considered in the multi-component simulations, except for the wall re-emission temperature, which is set to $T_{\text{W}} = 3.0 \text{eV}$, to mimick Franck-Condon dissociation processes, and $D_{\perp \text{n}_{\text{e}}} = 7.0$.}
\label{single_cs}
\end{figure}

For the description of the simulation results we focus on the quasi-steady state regime, established after a transient, when the plasma and neutral profiles fluctuate around constant values. We take toroidal and time averages of the plasma quantities evolved by Eqs. (\ref{n_e}-\ref{TD2}) and (\ref{poisson}) over a time interval of $\Delta t \simeq 10 R_0/c_{\text{s}0}$. These quantities are shown in Fig. \ref{plasma} on a poloidal cross section. In Fig. \ref{neutral}, we present the density of the neutral species, $n_{\text{D}}$ and $n_{\text{D}_2}$, and the neutral-plasma collisional interaction terms taken into account in our model. The results of the multispecies simulations are compared with the one of a single-component plasma, with corresponding parameters. The time and toroidal averages of the plasma and neutral main quantities for the single species simulation are shown in Fig. \ref{single_cs}. 

We first focus on some general considerations on the plasma and neutral densities. The plots in Figs. \ref{plasma} reveal that the density of the molecular ion species $\text{D}_2^+$ is three to four orders of magnitude smaller than the density of the main ion species $\text{D}^+$, a result in agreement with the assumption $n_{\text{D}_2^+}/n_{\text{D}^+} \ll 1$ used in Eqs. (\ref{vpare}-\ref{TD2}) for the derivation of the parallel friction and heat flux terms and in Eqs. (\ref{D_cont}-\ref{e_mom}) to obtain the boundary conditions at the limiter. We highlight that the density of $\text{D}_2^+$ peaks just inside the LCFS next to the limiter, since most of the $\text{D}_2$ molecules cross the open-field line region without interacting and are then dissociated and/or ionized by the denser and warmer plasma inside the LCFS. As a matter of fact, $n_{\text{D}_2^+}$ exhibits a similar behavior to the profile of the molecular ionization source $n_{\text{D}_2} \nu_{\text{iz},\text{D}_2}$ presented in Fig. \ref{neutral}, which also peaks in the edge near the limiter. On the other hand, Fig. 2 shows that $n_{\text{D}}$ and $n_{\text{D}_2}$ are comparable to $n_{\text{D}^+}$ near the limiter plates, while they are about one order of magnitude smaller than $n_{\text{D}^+}$ in the rest of the SOL and up to two orders of magnitude smaller inside the LCFS. Furthermore, regarding the relative importance of $\text{D}$ and $\text{D}_2$, Fig. \ref{neutral} shows that $n_{\text{D}_2}$ is larger than $n_{\text{D}}$ by a factor between two and three in the open-field line region around the limiter, while $n_{\text{D}}$ is larger than $n_{\text{D}_2}$ inside the LCFS at the HFS, as a consequence of the higher plasma densities and temperatures that lead to the dissociation of $\text{D}_2$ molecules in that region.

As a second set of observations, we focus on the asymmetry of the plasma density and flow. An up-down asymmetry in the edge region is shown by the profiles of $n_{\text{e}}$ and $n_{\text{D}^+}$, which are noticeably larger below the equatorial midplane than above it. The underlying reason of this asymmetry can be inferred from the $v_{\| \text{e}}$, $v_{\| \text{D}^+}$ and $\Gamma_{\| \text{D}_2^+}$ profiles. In fact, the $\text{e}^-$ and $\text{D}^+$ parallel flows are directed in the counterclockwise direction in the edge region. Therefore, the ionization of neutrals inside the LCFS, which occurs mostly in the proximity of the limiter at the HFS, leads to plasma particles subject to a downward flow. Albeit being small, this flow leads to a slightly larger density of $\text{e}^-$ and $\text{D}^+$ below the equatorial midplane of the device. The parallel flux of $\text{D}_2^+$ ions is also directed counterclockwise in the edge at the HFS, which further enhances this mechanism, even though $\text{D}_2^+$ densities are small compared to the other species. We highlight that the $n_{\text{D}^+}$ and $v_{\|\text{D}^+}$ profiles are slightly different when the single-component model of GBS is considered, as illustrated in Fig. \ref{single_cs}. In this case, although it is also observed an up-down asymmetry in the $n_{\text{D}^+}$ profile, this is related to the fact that the ionization source, $n_{\text{D}} \nu_{\text{iz}}$, is larger in the edge region below the limiter than above it, due to larger recycling rates at the lower limiter plate. In fact, contrary to the multispecies case, the $v_{\|\text{D}^+}$ is characterized by a counterclockwise parallel flow of $\text{D}^+$ ions in the edge below the midplane, while above it the parallel flow is directed clockwise.

In the multi-component plasma simulation we also observe a larger parallel flow of plasma in the open-field line region towards the upper side of the limiter when compared to the lower side. This can be observed in Fig. \ref{plasma}, that shows larger $n_{\text{D}^+}$ and $v_{\| \text{D}^+}$ above the limiter plates than below it, ultimately leading to higher recycling rates and hence larger $n_{\text{D}}$ and $n_{\text{D}_2}$ densities in the region above the limiter, as shown in Fig. \ref{neutral}. The reason behind this behaviour is again related to the $v_{\|\text{D}^+}$ profile. In fact, while the $\text{D}^+$ ions flow counterclockwise along the magnetic field lines in the edge, they undergo cross-field transport towards the SOL. As a result of the counterclockwise parallel flow and related asymmetry of $n_{\text{D}^+}$ in the edge region, most ions cross the LCFS above the equatorial midplane, while flowing along the magnetic field lines towards the upper side of the limiter. 

It is also observed that $n_{\text{D}^+}$ is slightly larger in the HFS compared to the LFS, which is due to the existence of $\text{D}^+$ sources in the HFS around the midplane. Similarly, also in the single-species simulation, $n_{\text{D}^+}$ is larger in the HFS as a consequence of the ionization source, $n_{\text{D}} \nu_{\text{iz}}$. 

Focusing on the temperature of the plasma components, we observe that the $T_{\text{e}}$ profile presents a similar behaviour to the one observed in single-component plasma simulations. A clear asymmetry between the HFS and the LFS is observed for $T_{\text{D}^+}$, which is qualitatively similar to the results for a single-component simulation in Fig. \ref{single_cs}. As a matter of fact, the temperature is considerably lower on the HFS compared to the LFS, which is related to the generation of cold $\text{D}^+$ ions inside the LCFS due to ionization of $\text{D}$ atoms, dissociative processes and charge-exchange interactions. This effect is particularly important above the limiter, where the recycling rates are larger. On the other hand, the profile of $p_{\text{D}_2^+}$ exhibits a maximum inside the LCFS at the HFS, where the majority of the $\text{D}_2^+$ ions are generated by ionization of $\text{D}_2$ molecules coming from the limiter. The up-down asymmetry of the $\text{D}_2^+$ pressure around the limiter is also due to the asymmetry of the recycling rates. As an aside note, we remark that, since it is strongly related to the $T_{\text{e}}$ profile \cite{Loizu2013}, the electrostatic potential profile revealed by the multi-fluid simulations is similar to the one observed in the single-component plasma model.

Analyzing the neutral-plasma interaction terms presented in Fig. \ref{neutral}, we first notice that ionization processes tend to be more important in the edge region at the HFS, with atomic and molecular ionization rates exhibiting similar profiles. However, $n_{\text{D}_2} \nu_{\text{iz},\text{D}_2}$ peaks in the vicinity of the LCFS, while $n_{\text{D}} \nu_{\text{iz},\text{D}}$ peaks further inside the LCFS and has a larger radial spread. In fact, $\text{D}_2$ molecules generated in the open-field line region and are dissociated and/or ionized in the proximity of the LCFS, where the plasma is warmer and denser. In contrast, although most $\text{D}$ atoms are generated in the open-field line region, they are also created by dissociation of $\text{D}_2$ molecules in the edge. This shifts the maximum of $n_{\text{D}} \nu_{\text{iz},\text{D}}$ radially inwards and makes the ionization source spread across a wider area. As a result of $n_{\text{D}}$ being larger than $n_{\text{D}_2}$ in the edge, the maximum of $n_{\text{D}} \nu_{\text{iz},\text{D}}$ is also almost two times larger than $n_{\text{D}_2} \nu_{\text{iz},\text{D}_2}$.

Focusing on the electron-neutral collisions, we note that the reactions involving $\text{D}_2$ occur more often in the open-field line region, mainly in the area surrounding the limiter plates where the majority of the neutral molecules are generated. Reactions with $\text{D}_2$ become less important in the edge, since most molecules are dissociated and/or ionized due to the higher densities and temperatures. On the other hand, electron-atom collision reactions involving the $\text{D}$ species peak inside the LCFS, because the cross sections of these reactions are larger in the edge region due to the higher plasma density and temperature and because of the presence of $\text{D}$ atom resulting from dissociative processes. We also highlight that elastic collisions and charge-exchange reactions are more frequent on the upper side of the limiter, in agreement with the strong up-down asymmetry discussed above. Regarding charge-exchange reactions, we observe that they are spatially localized similarly to the electron-neutral collisions. The reactions between the two molecular species ($\text{D}_2 - \text{D}_2^+$ collisions) occur less often than the charge-exchange between mono-atomic species ($\text{D} - \text{D}^+$ collisions) by three to four orders of magnitude, which is a result of the $n_{\text{D}_2^+}$ to $n_{\text{D}^+}$ ratio. In addition, the terms arising from charge-exchange interactions between $\text{D}_2$ molecules and $\text{D}^+$ ions ($\text{D}_2 - \text{D}^+$ collisions) are found to be two orders of magnitude smaller than the ones between the atomic species ($\text{D} - \text{D}^+$ collisions) in the region of the domain where these interactions are important. In turn, charge-exchange between $\text{D}_2^+$ ions and $\text{D}$ atoms is three orders of magnitude smaller than $\text{D} - \text{D}^+$ charge-exchange, which is due to the fact that $n_{\text{D}} \nu_{\text{cx},\text{D}-\text{D}_2^+}$ is proportional to $n_{\text{D}_2^+}$.

Finally, we analyse the dissociative processes, which represent a sink of molecular species $\text{D}_2$ and $\text{D}_2^+$ and sources of $\text{D}$ atoms and $\text{D}^+$ ions. Simple dissociation of $\text{D}_2$ and $\text{D}_2^+$, described by the terms $n_{\text{D}_2} \nu_{\text{di}, \text{D}_2}$ and $n_{\text{D}_2^+} \nu_{\text{di}, \text{D}_2^+}$ respectively, which do not involve ionization nor recombination processes, are found to be dominant dissociation processes, and occur with a frequency similar to that of the ionization of $\text{D}$ and $\text{D}_2$. We remark that dissociation of $\text{D}_2$ molecules peaks just above the limiter plate (where most $\text{D}_2$ molecules are generated) and in the edge region, in the vicinity of the LCFS, and then it is significantly smaller in the core, since $n_{\text{D}_2}$ drops rapidly across the edge. In contrast, dissociation of $\text{D}_2^+$ ions is very small in the open-field line region, where the density of $\text{D}_2^+$ is negligible (at the typical electron temperature of the SOL, the $\text{D}_2$ ionization cross section is small), and is important only inside the LCFS, where $\text{D}_2^+$ ions are generated. The $n_{\text{D}_2^+} \nu_{\text{di}, \text{D}_2^+}$ profile therefore closely follows the $n_{\text{D}_2^+}$ profile, with a larger radial spread when compared with the dissociation of $\text{D}_2$. As for dissociative ionization of $\text{D}_2$ and $\text{D}_2^+$, $n_{\text{D}_2} \nu_{\text{di}-{\text{iz}}, \text{D}_2}$ and $n_{\text{D}_2^+} \nu_{\text{di}-{\text{iz}}, \text{D}_2^+}$ respectively, we observe that the rates are smaller by one to two orders of magnitude with respect to the simple dissociation of $\text{D}_2$ and $\text{D}_2^+$ and peak in the edge region a bit further inside. This is due to the fact that the energy required to trigger dissociative ionization processes is considerably larger than the one needed to dissociate the particles without triggering an ionization process, as shown in Table II. Hence, these processes are only relevant in the edge region, where densities and temperatures are sufficiently high to make these cross sections significant. This is particularly the case of $n_{\text{D}_2^+} \nu_{\text{di}-{\text{iz}}, \text{D}_2^+}$, since this term is also proportional to the density of $\text{D}_2^+$ ions, which is relevant only inside the LCFS. Nevertheless, we highlight that these reactions become considerably less important towards the core, as very few $\text{D}_2$ and $\text{D}_2^+$ cross the edge region without being dissociated. As for dissociative-recombination of $\text{D}_2^+$ particles, $n_{\text{D}_2^+} \nu_{\text{di}-{\text{rec}}, \text{D}_2^+}$, its amplitude is also smaller than that of simple dissociation by one to two orders of magnitude and follows very closely the $n_{\text{D}_2^+}$ profile, since there is no energy threshold to trigger the reaction, unlike dissociative ionization processes.

These results allow us to draw a global picture of the main processes determining the dynamics of $\text{D}_2$ neutrals in the boundary. Although some $\text{D}_2$ molecules are dissociated in the SOL region, most of them cross the LCFS and are dissociated  into $\text{D}$ atoms within a short distance as they get in contact with the warmer and denser plasma of the edge. The remaining $\text{D}_2$ molecules penetrate further towards the core and are ionized by the increasingly warmer and denser plasma, giving rise to $\text{D}_2^+$ ions, which in turn are quickly dissociated into $\text{D}^+$ ions and $\text{D}$ atoms. 

We remark that, in the multi-component as well as in the single-component simulations, given the low plasma density of the SOL, a significant amount of $\text{D}$ atoms generated in the open-field line region (emitted at the limiter or created by dissociation of $\text{D}_2$ molecules) penetrate in the edge, where ionization takes place due to the higher plasma density and temperature. However, the presence of the $\text{D}$ sources inside the LCFS in the multi-component simulations shifts the ionization processes, $n_{\text{D}} \nu_{\text{iz}}$, towards the core with respect to the results with respect to single-component simulations, as shown in Fig. \ref{single_cs}.

\begin{figure}[ht]
\includegraphics[keepaspectratio=t,width=0.95\textwidth]{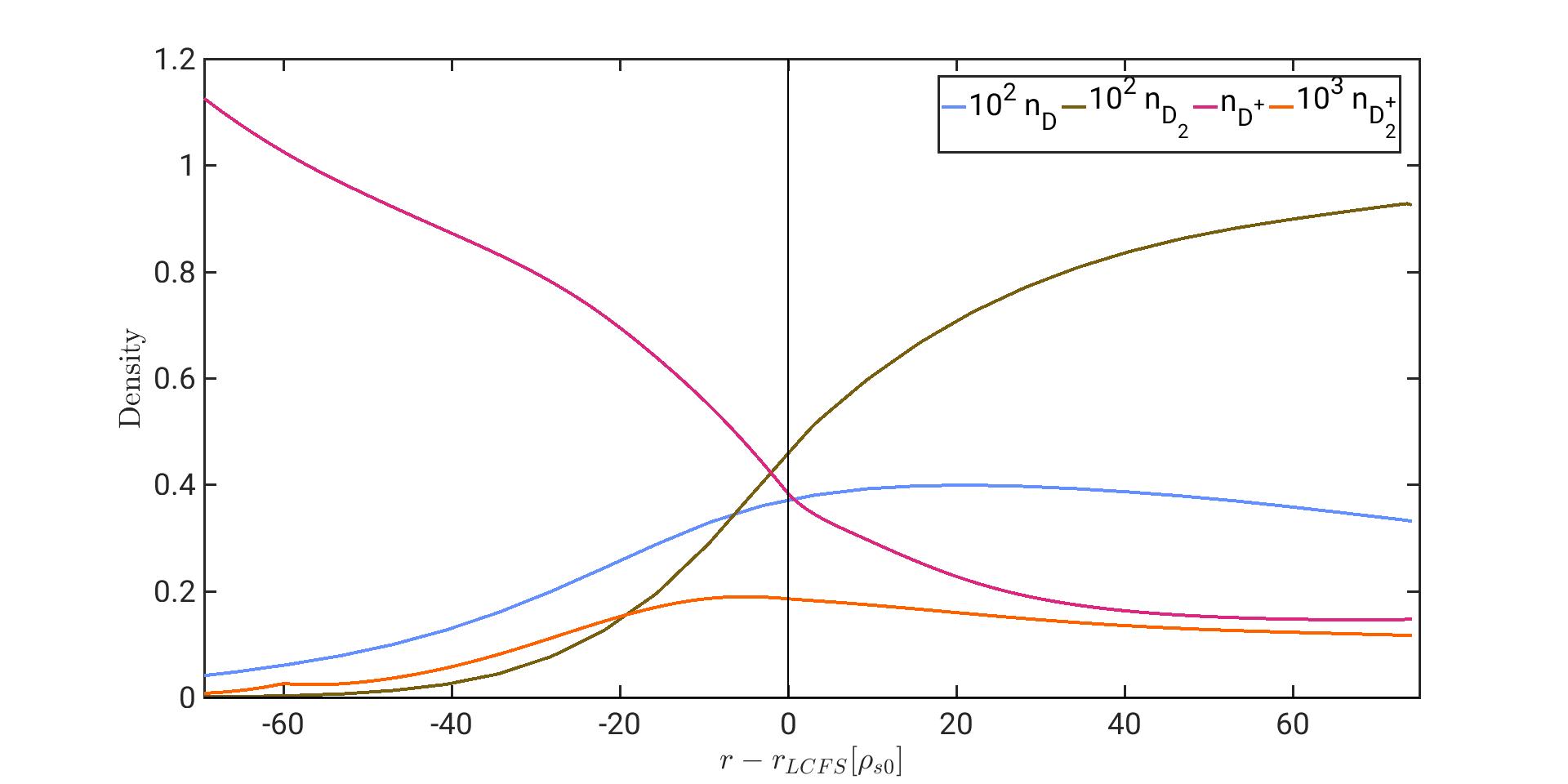}
\caption{Radial profiles of the ions and neutrals species densities, averaged over the toroidal and poloidal directions, evaluated over an interval of $\Delta t = 10.1 R_0/c_{\text{s}0}$ from a quasi-steady state simulation described in Sec. 6.}
\label{densities}
\end{figure}

\begin{figure}[ht]
\includegraphics[keepaspectratio=t,width=0.95\textwidth]{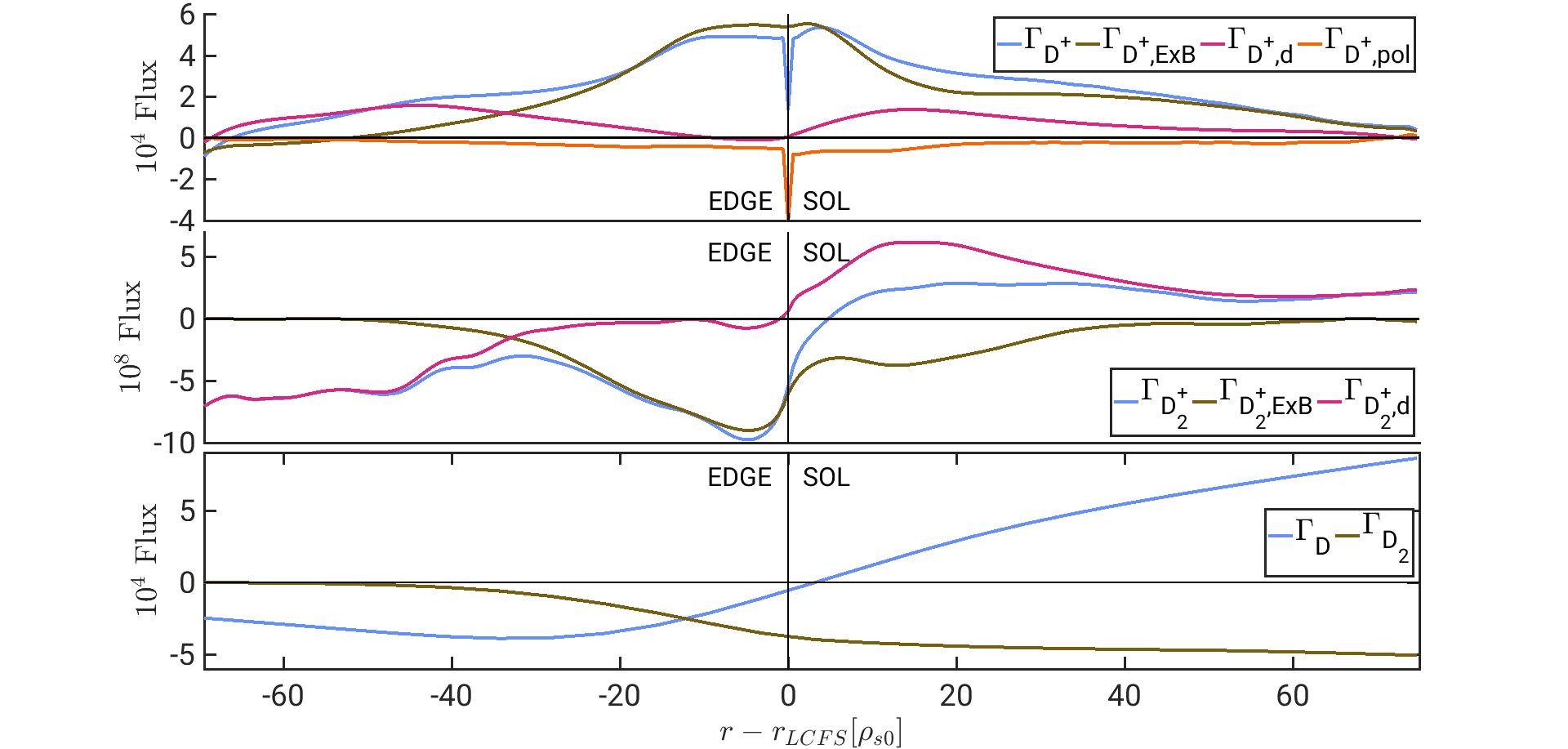}
\caption{Radial profiles of the radial flux for $\text{D}^+$ ions (top), $\text{D}_2^+$ ions (middle) and neutral species $\text{D}$ and $\text{D}_2$ (bottom), averaged over the toroidal and poloidal directions, evaluated over an interval of $\Delta t = 10.1 R_0/c_{\text{s}0}$ from the quasi-steady state multi-component plasma simulation described in Sec. 6. The components of the $\text{D}^+$ and $\text{D}_2^+$ radial flux are discriminated.}
\label{fluxes}
\end{figure}

\begin{figure}[ht]
\includegraphics[keepaspectratio=t,width=0.95\textwidth]{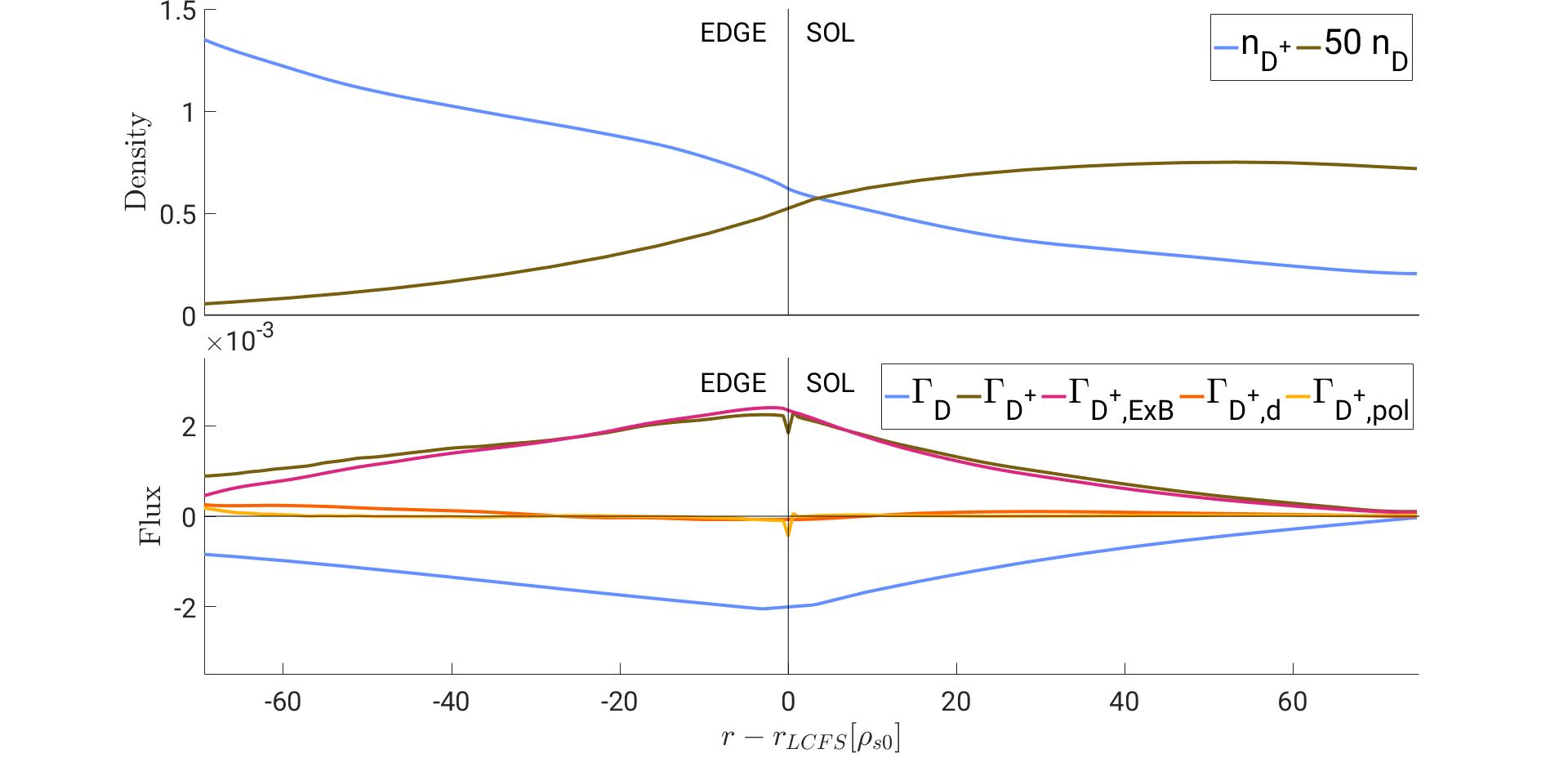}
\caption{Radial profiles of density (top) and radial flux (bottom) for the $\text{D}^+$ and $\text{D}$ species, averaged over the toroidal and poloidal directions, evaluated over an interval of $\Delta t = 10.1 R_0/c_{\text{s}0}$ from a quasi-steady state single-component plasma situation. The components behind the radial ion flux are discriminated. Plasma and neutral grid resolution, as well as simulation parameters, are the same considered in Fig. \ref{single_cs}.}
\label{single_fd}
\end{figure}

To conclude, we present radial plots of the particle densities (Fig. \ref{densities}) and radial fluxes (Fig. \ref{fluxes}), obtained by evaluating the time, toroidal and poloidal average of these quantities. In Fig. \ref{fluxes}, we discriminate the contributions of the $\text{E} \times \text{B}$, diamagnetic and polarization drifts to the flux of the plasma ion species, $\text{D}^+$ and $\text{D}_2^+$. The results from the single-component simulations are shown in Fig. \ref{single_fd}. The $n_{\text{D}^+}$ profile in Fig. \ref{densities} is similar to the one observed within the single-component plasma simulation in Fig. \ref{single_fd}, with a large density gradient region near the LCFS and a density shoulder appearing in the far SOL. In turn, the density of $\text{D}_2^+$ is small in the whole domain and peaks in the edge, across the LCFS, where most $\text{D}_2$ molecules are ionized and decreases rapidly towards the core, due to the small penetration of $\text{D}_2$ molecules in the warmer and denser plasma of the region. On the other hand, the $\text{D}_2^+$ ions observed in the open-field line region result from charge-exchange interactions between $\text{D}_2$ and $\text{D}^+$ (see Fig. \ref{neutral}) and the ionization of $\text{D}_2$ molecules reemitted from the limiter and vessel wall. 

Focusing on the neutral species, we note that $n_{\text{D}}$ peaks in the open-field line region, in contrast to the single-component plasma simulation. This is the result of the $\text{D}_2$ molecules dissociated into $\text{D}$ atoms in the edge and near SOL. On the other hand, we observe that $n_{\text{D}_2}$ decreases monotonically from the outer wall to the core interface, since $\text{D}_2$ molecules are generated in the open-field line region as the result of recycling processes are lost due to dissociation and ionization processes which take place mostly in the edge and near SOL.

The dissociation of $\text{D}_2$ molecules also impacts $\Gamma_{\text{D}}$, the radial flux of $\text{D}$, presented in Fig. \ref{fluxes}. In contrast with the single-component plasma simulation presented in Fig. \ref{single_fd}, $\Gamma_{\text{D}}$ points radially inwards in the edge, but reverses sign in the SOL region, a consequence of the release of $\text{D}$ atoms because of the dissociation of $\text{D}_2$ molecules, particularly important close to the LCFS. In addition, the $\text{D}$ atoms reaching the outer wall associate and are reemitted as $\text{D}_2$ molecules, thus contributing to the outward flux of $\text{D}$. The multi-component simulation shows that $\Gamma_{\text{D}}$ peaks in the edge region, while for a single-component model $\Gamma_{\text{D}}$ is maximum at the LCFS. This is due to the $\text{D}$ atoms that are generated in the edge region close to the LCFS in a multi-component model, compensating their ionization. At the same time, we note that $\Gamma_{\text{D}_2}$, the radial flux of $\text{D}_2$ molecules, points radially inwards in the whole domain (see Fig. \ref{fluxes}). More precisely, $\Gamma_{\text{D}_2}$ is approximately constant in the SOL, because the loss of $\text{D}_2$ molecules due to dissociation is compensated by the $\text{D}_2$ molecules recycled at the limiter. Then, $\Gamma_{\text{D}_2}$ decreases in the edge as a consequence of the molecules being dissociated and/or ionized because of the larger temperatures and densities in this region and becomes negligible towards the core.

Turning to the dynamics of the ion species, we note that the radial flux of $\text{D}^+$ ions points radially outwards across the whole domain and is mostly determined by the dominant $\text{E} \times \text{B}$ flux except near the core, where the diamagnetic flux, dominates over the $\text{E} \times \text{B}$ flux. The polarization drift contribution is negligible in the whole domain. We also remark that the flux increases across the edge region from the core to the separatrix, having a maximum in the near SOL, and then decreases gradually across the open-field line region. This contrasts with the behavior of the ion flux in the single-component plasma simulation (see Fig. \ref{single_fd}), where the flux peaks at the LCFS. This difference is related to the location of the ionization source $n_{\text{D}} \nu_{\text{iz}}$. Indeed, while the source has a smooth profile and peaks at the LCFS in the single-component model, the ionization source peaks further inside the edge in the multi-component model, accounting for a sharp increase of the $\text{D}^+$ flux in the edge close to the LCFS.

Fig. \ref{fluxes} shows that the radial flux of $\text{D}_2^+$ ions points radially outwards in the SOL and radially inwards in the edge. This is a consequence of the fact that most $\text{D}_2^+$ are generated in the vicinity of the LCFS, where the $\text{D}_2$ molecules are ionized by the warmer and denser plasma. The $\text{D}_2^+$ radial flux is determined by the balance between the inward pointing $\text{E} \times \text{B}$ and outward pointing diamagnetic drift components in the SOL, by the $\text{E} \times \text{B}$ flux in the edge close to the LCFS, and by the diamagnetic component towards the core.

We also note that the inward pointing $\Gamma_{\text{D}_2^+}$ is sharply peaked in the edge, close to the LCFS. This is because most $\text{D}_2^+$ ions are generated by ionization of $\text{D}_2$ molecules in that region and are then dissociated after traveling a short distance. Indeed, the location of the peak of $\Gamma_{\text{D}_2^+}$ corresponds to the one of the $n_{\text{D}_2^+}$ profile in Fig. \ref{densities}. The flux of $\text{D}_2^+$ associated with the polarization drift is not represented in Fig. \ref{fluxes} because it is neglected in our model. We note that $\Gamma_{\text{D}_2^+}$ is three to four orders of magnitude smaller than $\Gamma_{\text{D}^+}$, which is a consequence of the ratio $n_{\text{D}_2^+}/n_{\text{D}^+}$. Since the polarization drift component is expected to be small compared to the total molecular ion flux, $\Gamma_{\text{D}_2^+}$, we conclude that neglecting the polarization drift terms in Eqs. (\ref{n_e}-\ref{TD2}) has indeed a negligible impact on the simulation results.

\section{\label{sec:level6} Conclusions}

In this work we present a multi-component model for the self-consistent description of the neutral and plasma dynamics in the tokamak boundary. This model is implemented in the GBS code, allowing for the simulation of a deuterium plasma in the edge and SOL regions of a tokamak, including electrons, $\text{D}^+$ and $\text{D}_2^+$ ions, $\text{D}$ atoms and $\text{D}_2$ molecules. The neutral and the plasma models are coupled through a number of collisional processes, which give rise to neutral-plasma interaction terms in the plasma and neutral equations. The reactions considered include ionization, electron-neutral elastic collisions, charge-exchange and dissociative processes. The multi-component plasma model relies on the Braginskii fluid equations derived in the drift limit, being an extension of the single ion species model to account for $\text{D}_2^+$ ions and closed by following Zhdanov approach. As for the neutral species, we extend the approach considered in the single neutral species model of GBS \cite{Wersal2015} to include the molecular species, $\text{D}_2$. The neutrals are computed by solving two coupled kinetic equations for the $\text{D}$ and $\text{D}_2$ species, which is carried out by using the method of characteristics. The resulting system of linear integral equations are then discretized and solved for the $n_{\text{D}}$ and $n_{\text{D}_2}$ densities. 

The results from the first simulation carried out using the multi-component model are described in the sheath-limited regime in a toroidally limited plasma. The results exhibit some noticeable differences with respect to the single-ion component implemented in GBS. We observe an up-down asymmetry in the $n_{\text{e}}$ and $n_{\text{D}^+}$ density, which are larger below the equatorial midplane. This is related to the counterclockwise parallel flow of the plasma in the edge, observed in the profiles of $v_{\| \text{e}}$, $v_{\| \text{D}^+}$ and $v_{\| \text{D}_2^+}$. This feature also leads to larger recycling rates and a higher density of neutral particles in the upper side of the limiter, compared to the lower side. Moreover, the simulation shows that the density of the neutral species, $n_{\text{D}}$, is about one order of magnitude smaller than $n_{\text{D}^+}$ in the open-field line region and two orders of magnitude smaller in the edge, while $n_{\text{D}_2^+}$ is about three to four orders of magnitude smaller than $n_{\text{D}^+}$, even in the edge close to the LCFS, where $n_{\text{D}_2^+}$ peaks. 

By taking into account the molecular dynamics, the first simulations based upon the multi-component model also shed some light on the role played by molecules on the plasma fuelling. As a matter of fact, $\text{D}_2$ particles are generated close to the LCFS. A large fraction of $\text{D}_2$ molecules reach the closed field line region, where they are most often dissociated into atomic $\text{D}$ by the warmer and denser plasma. The resulting $\text{D}$ atoms and the remaining $\text{D}_2$ molecules are then ionized inside the edge, with the $\text{D}_2^+$ ions being quickly dissociated as a consequence of the high electron densities and temperatures. The simulation results therefore show that the peak of the ionization of $\text{D}$ atoms is shifted radially inwards with respect to the results from the single-species simulations. 

The radial profiles of the densities and radial fluxes are also impacted by the presence of molecular species. We observe that the radial flux of $\text{D}^+$ increases sharply in the edge close to the LCFS as a result of the peak of the ionization source observed in that region. The flux of $\text{D}^+$ then remains high in the vicinity of the LCFS, and decreases sharply again in the near SOL, where the sources of $\text{D}^+$ are outweighed by the sinks at the limiter. This is a major difference with respect to the $\text{D}^+$ flux observed in the single-ion species simulation, which is maximum at the LCFS. On the other hand, the $\text{D}$ density peaks in the SOL due to the $\text{D}_2^+$ ions dissociated there. This also explains why the $\text{D}$ radial flux reverses sign, pointing radially outwards in the far SOL. On the other hand, the inward flux of $\text{D}$ atoms in the edge increases radially inwards in the vicinity of the LCFS, since $\text{D}$ atoms are also generated in that region as a result of dissociation of $\text{D}_2$ molecules.

Ultimately, our results show that the multi-component model for the self-consistent description of the neutral-plasma interaction can provide a description of a deuterium plasma that captures the main features of the molecular dynamics and its overall impact. While describing the turbulent phenomena that lead to cross-field transport, it is possible to address a multi-component plasma and more than one neutral species at a kinetic level. The procedure described here can be extended to include additional plasma and neutral species, as well as additional collisional processes.

\section*{Appendix A: Evaluation of average electron energy loss and reaction product energies in collisional processes}

The Franck-Condon principle \cite{Franck1926,Condon1928} states that electronic excitation occurs over a timescale considerably shorter than the characteristic timescale associated with vibration or dissociation of the diatomic species. In turn, the vibration or dissociation timescales are much shorter than the electron deexcitation timescale. As a result, when an electron impacts a $\text{D}_2$ molecule or a $\text{D}_2^+$ ion, an electronic excitation is observed with no significant change in the inter-atomic distance (vertical transition). If the excited state is not stable, the molecule dissociates before deexcitation takes place. In this case, the difference between the excitation energy and the dissociation energy is converted into kinetic energy of the products (ionization and dissociative energies are discussed in \cite{Liu2009}). We note that the exact energies of the products of dissociation reactions depend on the vibrational level of the $\text{D}_2$ molecule or $\text{D}_2^+$ ion. Considering the excitation of a $\text{D}_2$ molecule in a given initial state, the set of vibrational levels accessible for the molecule in the final state are the ones lying within the region of the potential energy surface accessed by that particular vertical transition, known as the Franck-Condon region. The mean energy of the reaction products is thus the average over the Franck-Condon region, taking into account all accessible vibrational states.

In the present work, we model the products of dissociative reactions by considering that they are reemitted isotropically in the reference frame of the incoming massive particle ($\text{D}_2$ or $\text{D}_2^+$), thus approximating their velocity distribution as a Maxwellian centered at the velocity of the incoming $\text{D}_2$ or $\text{D}_2^+$. The temperature of the Maxwellian, together with the average electron energy loss for each process, are obtained from the values presented in \cite{Janev1987}. Since these energies depend on the intermediate excited state of the $\text{D}_2$ or $\text{D}_2^+$ particle, different values are found for different channels within the same dissociative process. This requires that an average is performed over all possible excited states, taking into account the respective cross section of each process. We present these calculations in detail for each process, following \cite{Janev1987}. 

The energy loss and the energy of the reaction products may depend on the electronic levels ($n$) and sub-levels ($l$) of the reaction products, on the molecular orbital (MO) of the intermediate state, if bonding or antibonding, and on the energy of the incident electron. The energy values are experimentally determined for all relevant dissociation channels. These quantities are then averaged over all vibrational states $v$ of the $\text{D}_2$ molecules or $\text{D}_2^+$ ion and over the Franck-Condon region, from \cite{Janev1987}.

We start by considering the dissociation of $\text{D}_2$ molecules, i.e.

\begin{equation}
\begin{aligned}
\text{e}^- + \text{D}_2 \rightarrow \text{e}^- + \text{D} + \text{D}.
\end{aligned}
\end{equation}

\noindent For this reaction, the values of the electron energy loss, $\left\langle \Delta E_{\text{e}} \right\rangle$, and reaction product energies, $\left\langle E_{\text{D}} \right\rangle$, depend significantly on the electronic state of the products. Hence, considering that there are $i=1,...,N$ electronic states of the reaction products and, associated, $N$ different sub-processes contributing to the dissociation of $\text{D}_2$, the average electron energy loss $\left\langle \Delta E_{\text{e}} \right\rangle$ is obtained by performing a weighed average of $\left\langle \Delta E_{\text{e}} \right\rangle_i$, the energy loss for the sub-process $i$, based on the $\left\langle \sigma v \right\rangle_i$ reaction rate, yielding

\begin{equation}\label{enloss}
\begin{aligned}
\left\langle \Delta E_{\text{e}} \right\rangle = \frac{\Sigma_{i=1}^N \left[\left\langle \sigma v \right\rangle_i \left\langle \Delta E_{\text{e}} \right\rangle_i\right]}{\Sigma_{i=1}^N \left[\left\langle \sigma v \right\rangle_i\right]},
\end{aligned}
\end{equation}

\noindent For simplicity, we evaluate all quantities at the reference temperature, $T_{\text{e}}=20\text{eV}$. Similarly, the average value for the energy of the reaction products is obtained as

\begin{equation}\label{enprod}
\begin{aligned}
\left\langle E_{\text{D}} \right\rangle = \frac{\Sigma_{i=1}^N \left[\left\langle \sigma v \right\rangle_i \left\langle E_{\text{D}} \right\rangle_i\right]}{\Sigma_{i=1}^N \left[\left\langle \sigma v \right\rangle_i\right]},
\end{aligned}
\end{equation}

\noindent with $\left\langle E_{\text{D}} \right\rangle_i$ the average energy of the products for the sub-process $i$. 

The values of $\left\langle \sigma v \right\rangle_i$, $\left\langle \Delta E_{\text{e}} \right\rangle_i$, $\left\langle E_{\text{D}} \right\rangle_i$ are presented in Table \ref{tab_diss_D2} for all sub-processes. The additional information between brackets refers to the minimum and maximum of the range of energies accessible to $\left\langle \Delta E_{\text{e}} \right\rangle_i$ and $\left\langle E_{\text{D}} \right\rangle_i$, following the values listed in \cite{Janev1987}. We highlight that $\text{D}(1\text{s})$ denotes a $\text{D}$ atom in the fundamental state (electron at the lowest orbital $1\text{s}$), while $\text{D}^*(2\text{s})$ and $\text{D}^*(2\text{p})$ denote an atom in the excited state $n = 2$ with the electron in an orbital of type $s$ or $p$, respectively, and $\text{D}^*(n=3)$ represents an atom in the excited state $n = 3$. Following \cite{Janev1987}, we assume that the energy is equally distributed over the reaction products, regardless of the fact that their electronic states are the same. Based on the values in Table \ref{tab_diss_D2}, from Eqs. (\ref{enloss}) and (\ref{enprod}), we obtain $\left\langle \Delta E_{\text{e}} \right\rangle \simeq 14.3 \text{eV}$ and $\left\langle E_{\text{D}} \right\rangle \simeq 1.95 \text{eV}$, respectively, at $T_{\text{e}} = 20 \text{eV}$. These are the values mentioned in Table \ref{energies}.

\begin{table}[!ht]
\begin{indented}
\item[]\caption{\label{tab_diss_D2}$\left\langle \sigma v_{\text{e}} \right\rangle$ product, average electron energy loss and average energy of reaction products for each sub-process of $\text{D}_2$ dissociation.}
\item[]\begin{tabular}{@{}llll} 
\br
\textbf{Reaction} & $\left\langle \sigma v_{\text{e}} \right\rangle_i$  & $\left\langle \Delta E_{\text{e}} \right\rangle_i$ & $\left\langle E_{\text{D}} \right\rangle_i$ \\ 
\mr
$\text{e}^- + \text{D}_2 \rightarrow \text{e}^- + \text{D}(1\text{s}) + \text{D}(1\text{s})$ & $3.8 \times 10^{-9} \text{cm}^3/\text{s}$ &  $10.5 \text{eV}$ & $3 \text{eV}$  \\ 
$\text{e}^- + \text{D}_2 \rightarrow \text{e}^- + \text{D}(1\text{s}) + \text{D}^*(2\text{s})$ &  $5.3 \times 10^{-9} \text{cm}^3/\text{s}$ & $15.3 \text{eV}$ & $0.3 \text{eV}$ \\ 
$\text{e}^- + \text{D}_2 \rightarrow \text{e}^- + \text{D}^*(2\text{p}) + \text{D}^*(2\text{s})$ &  $9.2 \times 10^{-10} \text{cm}^3/\text{s}$ & $34.6 \text{eV}$ & $4.85 \text{eV}$ \\ 
$\text{e}^- + \text{D}_2 \rightarrow \text{e}^- + \text{D}(1\text{s}) + \text{D}^*(\text{n}=3)$ & $5.7 \times 10^{-10} \text{cm}^3/\text{s}$ & $21.5 \text{eV}$ &  $5.7 \text{eV}$\\ 
\br
\end{tabular}
\end{indented}
\end{table}

Focusing now on the dissociative-ionization of $\text{D}_2$, 

\begin{equation}
\begin{aligned}
\text{e}^- + \text{D}_2 \rightarrow \text{D} + \text{D}^+ + 2\text{e}^-,
\end{aligned}
\end{equation}

\noindent we consider three cases. If the incoming electron has an energy $E_{\text{e}} < E_{\text{th}(\text{g})}$, with $E_{\text{th}(\text{g})} = 18 \text{eV}$, no dissociation takes place. If $E_{\text{th}(\text{g})} < E_{\text{e}} < E_{\text{th}(\text{u})}$, with $E_{\text{th}(\text{u})} = 26 \text{eV}$, the electron can ionize the molecule, resulting in an unstable $\text{D}_2^+$ ion, which then dissociates into a $\text{D}$ atom and a $\text{D}^+$ ion. The short-lived $\text{D}_2^+$ has the electron in a bonding molecular orbital (MO) with $\sigma$-symmetry, thus exhibiting $\it{gerade}$ (g) symmetry (German for even) state, denoted as $\text{D}_2^+(\Sigma_{\text{g}})$. If $E_{\text{e}} > E_{\text{th}(\text{u})}$, the intermediate $\text{D}_2^+$ ion has the electron in a higher-energy antibonding MO with $\sigma$-symmetry, which exhibits $\it{ungerade}$ (u) symmetry (German for odd), thus denoted as $\text{D}_2^+(\Sigma_{\text{u}})$. As a result of the different energy levels of the intermediate $\text{D}_2^+$ ion, the energy of the final products will also be different, as well as the average electron energy loss. According to the results presented in \cite{Janev1987}, these energies still depend on the energy of the incoming electron within each sub-process. To simplify the evaluation of the $\left\langle \Delta E_{\text{e}} \right\rangle$ and the energy of the products, we consider the energy to be evenly distributed by the reaction products ($\text{D}$ and $\text{D}^+$) and we consider the two cases separately. For $E_{\text{th}(\text{g})} < E_{\text{e}} < E_{\text{th}(\text{u})}$, all dissociative-ionization events originate an intermediate state $\text{D}_2^+(\Sigma_{\text{g}})$, while for $E_{\text{e}} > E_{\text{th}(\text{u})}$ all events generate an intermediate state $\text{D}_2^+(\Sigma_{\text{u}})$. The values for the electron energy loss and reaction product energies being considered for each case are evaluated for \cite{Janev1987} and listed in Table \ref{tab_diss_iz_D2}. We note that this is just an approximation, as even with $T_{\text{e}} < E_{\text{th}(\text{u})}$ there are electrons with energies superior to the threshold that will generate a $\text{D}_2^+$ ion in a $\text{D}_2^+(\Sigma_{\text{u}})$ state, and vice-versa. Nevertheless, this approximation avoids us to evaluate $\left\langle \Delta E_{\text{e}} \right\rangle$ and $\left\langle E_{\text{D}} \right\rangle$ at every single value of $T_{\text{e}}$. 

\begin{table}[!ht]
\caption{\label{tab_diss_iz_D2}Average electron energy loss and average energy of reaction products for the two cases of dissociative-ionization of $\text{D}_2$.}
\begin{indented}
\item[]\begin{tabular}{@{}lll} 
\br
\textbf{Reaction} & $\left\langle \Delta E_{\text{e}} \right\rangle $ & $\left\langle E_{\text{D}} \right\rangle = \left\langle E_{\text{D}^+} \right\rangle$ \\ 
\mr
$\text{e}^- + \text{D}_2 \rightarrow \text{e}^- + \left[\text{D}_2^+(\Sigma_{\text{g}}) + \text{e}^- \right] \rightarrow \text{D} + \text{D}^+ + 2\text{e}^-$ & $18.25 \text{eV}$ & $0.25 \text{eV}$ \\ 
$\text{e}^- + \text{D}_2 \rightarrow \text{e}^- + \left[\text{D}_2^+(\Sigma_{\text{u}}) + \text{e}^- \right] \rightarrow \text{D} + \text{D}^+ + 2\text{e}^-$ & $33.6 \text{eV}$ & $7.8 \text{eV}$ \\ 
\br
\end{tabular}
\end{indented}
\end{table}

For the dissociation of $\text{D}_2^+$, i.e. 

\begin{equation}
\begin{aligned}
\text{e}^- + \text{D}_2^+ \rightarrow \text{D}^+ + \text{D} + \text{e}^-,
\end{aligned}
\end{equation}

\noindent different sub-processes are taken into account, following an approach similar to the one adopted to treat the dissociation of $\text{D}_2$. We perform a weighed average of the electron energy loss and the reaction products energy by using Eqs. (\ref{enloss}) and (\ref{enprod}), respectively. The values of $\left\langle \sigma v_{\text{e}} \right\rangle_i$, $\left\langle \Delta E_{\text{e}} \right\rangle_i$ and $\left\langle E_{\text{D}} \right\rangle = \left\langle E_{\text{D}^+} \right\rangle_i$ for each sub-process are presented in Table \ref{tab_diss_D2p}. The weighed averaged values for the electron energy loss and reaction products energy at the reference temperature, $T_{\text{e}} = 20 \text{eV}$, yield $\left\langle \Delta E_{\text{e}} \right\rangle = 13.7 \text{eV}$ and $\left\langle E_{\text{D}} \right\rangle = \left\langle E_{\text{D}^+} \right\rangle = 3.0 \text{eV}$, as listed in Table \ref{tab_diss_D2p}.

\begin{table}[!ht]
\begin{indented}
\item[]\begin{tabular}{@{}llll} 
\br
\textbf{Reaction} & $\left\langle \sigma v_{\text{e}} \right\rangle_i$  & $\left\langle \Delta E_{\text{e}} \right\rangle_i$ & $\left\langle E_{\text{D}} \right\rangle = \left\langle E_{\text{D}^+} \right\rangle_i$ \\ 
\mr
$\text{e}^- + \text{D}_2^+ \rightarrow \text{D}^+ + \text{D}(1\text{s}) + \text{e}^-$ & $1.2 \times 10^{-7} \text{cm}^3/\text{s}$ & $10.5 \text{eV}$ & $4.3 \text{eV}$   \\ 
$\text{e}^- + \text{D}_2^+ \rightarrow \text{D}^+ + \text{D}^*(\text{n}=2) + \text{e}^-$ & $1.0 \times 10^{-7} \text{cm}^3/\text{s}$ & $17.5 \text{eV}$ & $1.5 \text{eV}$ \\ 
\br
\end{tabular}
\caption{$\left\langle \sigma v_{\text{e}} \right\rangle$ product, average electron energy loss and average energy of reaction products for each sub-process of $\text{D}_2^+$ dissociation.}
\label{tab_diss_D2p}
\end{indented}
\end{table}

Regarding the dissociative-ionization of $\text{D}_2^+$, i.e.

\begin{equation}
\begin{aligned}
\text{e}^- + \text{D}_2^+ \rightarrow \text{D}^+ + \text{D}^+ + 2\text{e}^-,
\end{aligned}
\end{equation}

\noindent we follow \cite{Janev1987}, where the average energy of the resulting $\text{D}^+$ ions is obtained from an average performed over all vibrational states ($v=0-9$) of the $\text{D}_2^+$ ion and over the Franck-Condon region. This yields $\left\langle E_{\text{D}^+}\right\rangle = 0.4 \text{eV}$, while the average electron energy loss is $\left\langle \Delta E_{\text{e}} \right\rangle = 15.5 \text{eV}$.

We finally focus on the dissociative-recombination of $\text{D}_2^+$, which generates a $\text{D}$ atom in the fundamental state (electron in orbital $1\text{s}$) and a $\text{D}$ atom in an excited state (electron with principal quantum number $\text{n}\ge 2$), i.e.

\begin{equation}
\begin{aligned}
\text{e}^- + \text{D}_2^+ \rightarrow \text{D}(1\text{s}) + \text{D}^*(\text{n}\ge 2).
\end{aligned}
\end{equation}

\noindent We assume that the energy of the products is evenly distributed among the two $\text{D}$ atoms and is given by 

\begin{equation}
\begin{aligned}
\left\langle E_{\text{D}(1\text{s})}\right\rangle \simeq \left\langle E_{\text{D}^*(\text{n}\ge 2)}\right\rangle \simeq \frac{1}{2} \left(E_\text{e} + \frac{\text{Ry}}{n^2}\right),
\end{aligned}
\end{equation}

\noindent with $\text{Ry} = 13.6\text{eV}$ the Rydberg unit of energy (corresponding to the electron binding energy in a hydrogen atom in the fundamental state). Since this expression depends on the energy of the incoming electron, $E_\text{e}$, and the electronic level $n$ of the excited atom, $\text{D}^*$, we assume an energy of the incident electron of $E_{\text{e}} \simeq 20 \text{eV}$, the typical value in the region around the LCFS at the HFS, and consider that these atoms are most likely in the accessible state of lowest energy $\text{n}=2$ (considering a higher excited state would not change the value of the energy of the products by a significant amount). Under these assumptions, we get $\left\langle E_{\text{D}(1\text{s})}\right\rangle \simeq \left\langle E_{\text{D}^*(\text{n}\ge 2)}\right\rangle \simeq 11.7 \text{eV}$.

\section*{Appendix B: Zhdanov collisional closure}

We focus on the derivation of the parallel friction forces and the parallel heat fluxes, denoted respectively by $R_{\| \alpha} = \mathbf{R_{\alpha}} \cdot \mathbf{b}$ and $q_{\| \alpha} = \mathbf{q_{\alpha}} \cdot \mathbf{b}$ for a given species $\alpha$, with $\mathbf{R_{\alpha}} = \int m_{\alpha} \mathbf{v'} C_{\alpha} d\mathbf{v}$ and $\mathbf{q_{\alpha}} = \int (m_{\alpha} v'^2)/2 \mathbf{v'} f_{\alpha} d\mathbf{v}$, where we introduce the thermal component of the velocity, $\mathbf{v}'=\mathbf{v}-\mathbf{v}_\alpha$, with $\mathbf{v}_\alpha = \int \mathbf{v} f_{\alpha} d\mathbf{v}$ the fluid velocity of the $\alpha$ species, and the collision operator $C_{\alpha} = \Sigma_{\beta} C_{\alpha \beta}(f_{\alpha},f_{\beta})$, with $C_{\alpha \beta}$ describing collisions of species $\alpha$ with species $\beta$. We consider the collisional closure derived by Zhdanov in \cite{Zhdanov2002}, relying on the approach proposed in \cite{Bufferand2019} and discussed in \cite{Raghunathan2021} for its numerical implementation. 

Following \cite{Zhdanov2002}, the parallel component of the friction forces and heat fluxes of the species $\alpha$ is related to the parallel gradients of the temperature and parallel velocity of all species through 

\begin{equation}\label{Zhdanov_base}
\begin{aligned}
    \begin{bmatrix}
    q_{\|\alpha} \\
    R_{\|\alpha}
    \end{bmatrix}
    =
    \sum_{\beta} Z_{\alpha\beta}
    \begin{bmatrix}
    \nabla_{\|} T_{\beta} \\
    w_{\|\beta}
    \end{bmatrix},
\end{aligned}
\end{equation}

\noindent where $T_{\beta}$ denotes the temperature of plasma species $\beta$ and $w_{\|\beta}$ is the parallel component of the fluid velocity of species $\beta$ with respect to the center of mass of the plasma, $\mathbf{w_{\beta}} = \mathbf{v_{\beta}}-\mathbf{v_{\text{CM}}}$, with $\mathbf{v_{\text{CM}}} = \left(\sum_{\beta} n_{\beta} m_{\beta} \mathbf{v_{\beta}}\right)/\left(\sum_{\beta} n_{\beta} m_{\beta} \right)$. The matrix $Z_{\alpha\beta}$ relates the parallel heat fluxes and friction forces with the parallel gradients of temperature and parallel velocity. We remark that Eq. (\ref{Zhdanov_base}) simplifies the general result obtained by Zhdanov \cite{Zhdanov2002} to the case of singly-ionized states, neglecting possible multiplicity of charge states for the chemical species present in the plasma.   

In order to compute the matrix $Z_{\alpha\beta}$, we consider the $21 N$-moment approximation of the distribution function \cite{Zhdanov2002}, thus including the moments up to the fifth order moment. We first express $\mathbf{R}_{\alpha}$ and $\mathbf{q}_{\alpha}$ in terms of these moments of the distribution function, namely the first order moment, $\mathbf{w}_{\alpha}$, the third order moment, $\mathbf{h}_{\alpha} = q_{\| \alpha}$, and the fifth order moment, $\mathbf{r}_{\alpha} = m_{\alpha}/4 \int (c^4 - 14 c^2/\gamma_{\alpha} + 35 \gamma_{\alpha}) \mathbf{c} f_{\alpha} d\mathbf{c}$, where we introduce the velocity with respect to the center of mass of the plasma, $\mathbf{c} = \mathbf{v} - \mathbf{v_{\text{CM}}}$, and the parameter $\gamma_{\alpha} = m_{\alpha}/(k T_{\alpha})$, with $T_{\alpha} = \int (m_{\alpha} v'^2/2) f_{\alpha} d\mathbf{v}$. Since only the expressions for the parallel component of the friction forces and heat fluxes are needed, we consider only the parallel component of these equations. The heat flux, $q_{\| \alpha}$, simply corresponds to the third order moment, $h_{\| \alpha}$, while the friction forces, $R_{\| \alpha}$, are obtained in terms of $w_{\| \alpha}$, $h_{\| \alpha}$ and $r_{\| \alpha}$ \cite{Zhdanov1980}, yielding

\begin{equation}\label{q_alpha}
\begin{aligned}
    q_{\| \alpha} = h_{\| \alpha},
\end{aligned}
\end{equation}

\begin{equation}\label{R_alpha}
\begin{aligned}
    R_{\| \alpha} = \sum_{\beta} \left[G_{\alpha \beta}^{(1)} \left(w_{\| \alpha}-w_{\| \beta}\right) + \frac{\mu_{\alpha \beta}}{kT} G_{\alpha \beta}^{(2)} \left(\frac{h_{\|\alpha}}{m_{\alpha} n_{\alpha}}-\frac{h_{\| \beta}}{m_{\beta} n_{\beta}}\right) + \left(\frac{\mu_{\alpha \beta}}{k T}\right)^2 G_{\alpha \beta}^{(8)}\left(\frac{r_{\alpha}}{m_{\alpha} n_{\alpha}}-\frac{r_{\beta}}{m_{\beta} n_{\beta}}\right) \right].
\end{aligned}
\end{equation}    

\noindent where $m_{\alpha}$ and $n_{\alpha}$ are respectively the mass and density of species $\alpha$, $\mu_{\alpha \beta} = (m_{\alpha} m_{\beta})/(m_{\alpha} + m_{\beta})$ is the reduced mass, and $G_{\alpha \beta}^{(\text{n})}$ are polynomial functions of the local plasma density and temperature, their exact expressions being presented in \cite{Zhdanov2002} (chapter 8.1, pp. 163-164). Eqs. (\ref{q_alpha}) and (\ref{R_alpha}) can then be written in matrix form as

\begin{equation}\label{qR_matrix}
\begin{aligned}
    \begin{bmatrix}
    q_{\|\alpha} \\
    R_{\|\alpha}
    \end{bmatrix}
    =
    \sum_{\beta} A_{\alpha\beta}
    \begin{bmatrix}
    h_{\|\beta} \\
    r_{\|\beta}
    \end{bmatrix}
    +
    \sum_{\beta} B_{\alpha\beta}
    \begin{bmatrix}
    \nabla_{\|} T_{\beta} \\
    w_{\|\beta}
    \end{bmatrix},
\end{aligned}
\end{equation}

\noindent where the matrices $A$ and $B$ are defined to satisfy Eqs. (\ref{q_alpha}) and (\ref{R_alpha}). We now aim at expressing the moments $\mathbf{h}_{\alpha}$ and $\nabla r_{\alpha}$ in terms of $\mathbf{w}_{\alpha}$ and $\nabla T_{\alpha}$. This can be achieved by solving a system of moment equations similar to the one presented in \cite{Zhdanov2002} (chapter 8.1, pp. 162-163), including the time evolution of the moments ($\mathbf{w}_{\alpha}$, $\mathbf{h}_{\alpha}$ and $\nabla r_{\alpha}$) and the time evolution of basic thermodynamic variables ($\rho$, $\mathbf{v_{\text{CM}}}$ and $T$). We neglect time derivatives and nonlinear terms. For simplicity, we also assume that, for two massive particle species $\text{D}^+$ and $\text{D}_2^+$, the condition $|T_{\text{D}_2^+} - T_{\text{D}^+}| \ll T_{\text{D}_2^+}$ is fulfilled, which allows us to write $T_{\text{D}_2^+} = T_{\text{D}^+} = T$. Moreover, as long as $T_{\text{e}}/T_{\text{D}^+} \gg m_{\text{e}}/m_{\text{D}^+}$ is verified, $T$ can also be replaced by $T_{\text{e}}$, following \cite{Zhdanov2002} (the simulation results shown in Fig. \ref{plasma} meet these conditions). We therefore impose $T_{\text{D}_2^+} = T_{\text{D}^+} = T_{\text{e}} = T$, while no assumption is made on the temperature and pressure gradients, i.e. temperature gradients can be different from species to species \cite{Zhdanov2002}. 

The parallel projection of the system of moment equations can then be written as (see \cite{Zhdanov1980}) 

\begin{equation}\label{eq_h_r_1}
\begin{aligned}
& \frac{5}{2}n_{\|\alpha} k \nabla T_{\alpha} = \sum_{\beta} \left[ \frac{5}{2} \frac{\mu_{\alpha \beta}}{m_{\alpha}}G_{\alpha \beta}^{(2)} \left(w_{\|\alpha}-w_{\|\beta}\right) + G_{\alpha \beta}^{(5)} \frac{h_{\|\alpha}}{p_{\alpha}}\right. \\
& \left.+ G_{\alpha \beta}^{(6)} \frac{h_{\|\beta}}{p_{\beta}} + \frac{\mu_{\alpha \beta}}{k T} \left(G_{\alpha \beta}^{(9)} \frac{r_{\|\alpha}}{p_{\alpha}} + G_{\alpha \beta}^{(10)} \frac{r_{\|\beta}}{p_{\beta}}\right) \right],
\end{aligned}
\end{equation}

\begin{equation}\label{eq_h_r_2}
\begin{aligned}
& 0 = \sum_{\beta} \left[ \frac{35}{2} \left(\frac{\mu_{\alpha \beta}}{m_{\alpha}}\right)^2 G_{\alpha \beta}^{(8)} \left(w_{\|\alpha}-w_{\|\beta}\right) + 7\frac{\mu_{\alpha \beta}}{m_{\alpha}}\left(G_{\alpha \beta}^{(9)} \frac{h_{\|\alpha}}{p_{\alpha}} + G_{\alpha \beta}^{(10)} \frac{h_{\beta}}{p_{\beta}}\right)\right. \\
& \left.+ \frac{m_{\alpha}}{k T} G_{\alpha \beta}^{(11)} \frac{r_{\|\alpha}}{p_{\alpha}} + \frac{m_{\beta}}{k T} G_{\alpha \beta}^{(12)} \frac{r_{\|\beta}}{p_{\beta}} \right],
\end{aligned}
\end{equation}

\noindent where $p_{\alpha}$ is the pressure of species $\alpha$. Rewriting Eqs. (\ref{eq_h_r_1}-\ref{eq_h_r_2}) in matrix form, one obtains

\begin{equation}\label{matrix_h_q_w_T}
\begin{aligned}
    \sum_{\gamma} P_{\alpha \gamma}
    \begin{bmatrix}
    \nabla_{\|} T_{\gamma} \\
    w_{\|\gamma}
    \end{bmatrix}
    =
    \sum_{\beta} M_{\alpha\beta}
    \begin{bmatrix}
    h_{\|\beta} \\
    r_{\|\beta}
    \end{bmatrix},
\end{aligned}
\end{equation}

\noindent which can be inverted to express the parallel third and fourth order fluid moments in terms of the parallel gradient of temperature and relative parallel velocity as

\begin{equation}\label{matrix_inv}
\begin{aligned}
    \begin{bmatrix}
    h_{\|\beta} \\
    r_{\|\beta}
    \end{bmatrix}
    =
    \sum_{\alpha} \sum_{\gamma} 
    M_{\alpha\beta}^{-1}
    P_{\alpha \gamma}
    \begin{bmatrix}
    \nabla_{\|} T_{\gamma} \\
    w_{\|\gamma}
    \end{bmatrix}.
\end{aligned}
\end{equation}

\noindent Finally, making use of Eq. (\ref{matrix_inv}) to express $h_{\| \alpha}$ and $r_{\| \alpha}$ in Eq. (\ref{qR_matrix}) in terms of the parallel temperature gradients and relative velocities, one obtains the expressions for the parallel heat flux and friction forces in the matrix form presented in Eq. (\ref{Zhdanov_base}), that is

\begin{equation}\label{qR_final}
\begin{aligned}
    \begin{bmatrix}
    q_{\|\alpha} \\
    R_{\|\alpha}
    \end{bmatrix}
    =
    \left(A_{\alpha \lambda} M_{\gamma \lambda}^{-1} P_{\gamma \beta} + B_{\alpha\beta}\right)
    \begin{bmatrix}
    \nabla_{\|} T_{\beta} \\
    w_{\|\beta}
    \end{bmatrix}.
\end{aligned}
\end{equation}

Since the matrices $A$, $B$, $P$ and $M$ are fully determined by Eqs. (\ref{q_alpha}), (\ref{R_alpha}), (\ref{eq_h_r_1}) and (\ref{eq_h_r_2}), the expressions of the parallel heat flux and friction forces can be found. Following Zhdanov \cite{Zhdanov2002}, these matrices can be expressed in terms of the local values of plasma quantities, namely densities $n_{\text{e}}$, $n_{\text{D}^+}$ and $n_{\text{D}_2^+}$ and temperatures $T_{\text{e}}$ and $T_{\text{D}^+}$ (we again assume $T_{\text{D}_2^+} = T_{\text{D}^+}$, mass ratios and characteristic time scales $\tau_{\text{eD}}$ and $\tau_{\text{DD}}$, with $\tau_{\text{eD}}$ defined as the inverse of the collision frequency for momentum transfer between electrons and $\text{D}^+$ ions, and $\tau_{\text{DD}}$ the ion timescale defined as the inverse of the collision frequency for momentum transfer between $\text{D}^+$ ions. We retain only terms of leading order in $\sqrt{m_{\text{e}}/m_{\text{D}}}$, while terms proportional to the fast electron timescale $\tau_{\text{eD}}$ are neglected when compared to terms proportional to $\tau_{\text{DD}}$, which considerably simplifies the final expressions. We also highlight that, besides imposing the quasi-neutrality relation $n_{\text{e}} = n_{\text{D}^+} + n_{\text{D}_2^+}$, we take into account the fact that the density of the molecular ion species is much smaller than the density of the main ion species $\text{D}^+$ for typical tokamak boundary conditions, i.e. $n_{\text{D}_2^+}/n_{\text{D}^+} \ll 1$, keeping therefore only leading order terms in $n_{\text{D}_2^+}/n_{\text{D}^+}$. As a result, the friction forces between molecular ions and other species are neglected, as well as molecular ion temperature gradient terms, while friction and thermal force contributions involving $\text{D}^+$ and $\text{e}^-$ species are kept in the expressions of the parallel components of the heat fluxes and friction forces. The expressions obtained for the friction forces and heat fluxes finally yield

\begin{equation}\label{qR_brute}
\begin{aligned}
    & q_{\|e} = -\frac{3.16 n_{\text{e}} T_{\text{e}} \tau_{\text{eD}}}{m_{\text{e}}} \nabla_{\|} T_{\text{e}} + 0.71 n_{\text{e}} T_{\text{e}} (v_{\| \text{e}}-v_{\|\text{D}^+}),\\ 
    & q_{\|\text{D}^+} = -\frac{4.52 n_{\text{e}} T_{\text{D}^+} \tau_{\text{DD}}}{m_{\text{D}}} \nabla_{\|} T_{\text{D}^+},\\ 
    & q_{\|\text{D}_2^+} = -\frac{1.80 n_{\text{e}} T_{\text{D}^+} \tau_{\text{DD}}}{m_{\text{D}}} \nabla_{\|} T_{\text{D}^+},\\
    & R_{\|e} = - 0.71 n_{\text{e}} \nabla_{\|} T_{\text{e}} - \frac{0.51 m_{\text{e}} n_{\text{e}}}{\tau_{\text{eD}}} (v_{\| \text{e}}-v_{\|\text{D}^+}),\\
    & R_{\|\text{D}^+} = 0.71 n_{\text{e}} \nabla_{\|} T_{\text{e}} - \frac{0.51 m_{\text{e}} n_{\text{e}}}{\tau_{\text{eD}}} (v_{\|\text{D}^+}-v_{\| \text{e}}),\\ 
    & R_{\|\text{D}_2^+} = 0,\\
\end{aligned}
\end{equation}

The expressions in Eqs. (\ref{qR_brute}) can be simplified by applying the relation between the electron and ion characteristic times,

\begin{equation}\label{tau_eD_DD}
\begin{aligned}
    \frac{\tau_{\text{DD}}}{\tau_{\text{eD}}} = \frac{1}{\sqrt{2}}\sqrt{\frac{m_{\text{D}}}{m_{\text{e}}}}\left(\frac{T_{\text{e}}}{T_{\text{D}^+}}\right) \sim \frac{1}{\sqrt{2}}\sqrt{\frac{m_{\text{D}}}{m_{\text{e}}}},
\end{aligned}
\end{equation}

\noindent having again assumed $T_{\text{D}^+} \sim T_{\text{e}}$. This enables one to write $\tau_{\text{DD}}$ appearing in Eq. (\ref{qR_brute}) in terms of $\tau_{\text{eD}}$. Following Braginskii's approach \cite{Braginskii1965} and considering that the the electron characteristic time is $\tau_{\text{e}} = \tau_{\text{eD}}$, we then write Eqs. (\ref{qR_brute}) in terms of the resistivity, defined as \cite{Ricci2012,Halpern2016}

\begin{equation}\label{nu_tau_eD}
\begin{aligned}
    \nu = 0.51 \frac{m_{\text{e}}}{m_{\text{D}}}\frac{R_0}{c_{\text{s}0}}\frac{1}{n_{\text{e}} \tau_{\text{eD}}},
\end{aligned}
\end{equation}

\noindent The parallel friction forces and heat fluxes, as they appear in Eqs. (\ref{vpare}-\ref{vparD2}) and Eqs. (\ref{Te}-\ref{TD2}), respectively, are therefore written in normalized units as

\begin{equation}\label{qR_expressions}
\begin{aligned}
    & R_{\|e} = - 0.71 n_{\text{e}} \nabla_{\|} T_{\text{e}} - \nu n_{\text{e}} (v_{\| \text{e}}-v_{\|\text{D}^+}),\\ 
    & R_{\|\text{D}^+} = 0.71 n_{\text{e}} \nabla_{\|} T_{\text{e}} - \nu n_{\text{e}} (v_{\|\text{D}^+}-v_{\| \text{e}}),\\
    & R_{\|\text{D}_2^+} = 0,\\ 
    & q_{\|e} = -\frac{1.62}{\nu} n_{\text{e}} T_{\text{e}} \nabla_{\|} T_{\text{e}} + 0.71 n_{\text{e}} T_{\text{e}} (v_{\| \text{e}}-v_{\|\text{D}^+}),\\ 
    & q_{\|\text{D}^+} = -\frac{2.32}{\sqrt{2} \nu} \sqrt{\frac{m_{\text{e}}}{m_{\text{D}}}} n_{\text{e}} T_{\text{D}^+} \nabla_{\|} T_{\text{D}^+},\\ 
    & q_{\|\text{D}_2^+} = -\frac{0.92}{\sqrt{2} \nu} \sqrt{\frac{m_{\text{e}}}{m_{\text{D}}}} n_{\text{e}} T_{\text{D}^+} \nabla_{\|} T_{\text{D}^+}.
\end{aligned}
\end{equation}

\noindent We note that, similarly to the single-ion species model implemented in GBS \cite{Ricci2012}, the ohmic heating terms are neglected. 

\section*{Appendix C: List of kernel functions}

The kernels used in Eqs. (\ref{nD2}-\ref{GammaD}) for $n_{\text{D}_2}$, $\Gamma_{\text{out,D}_2}$, $n_{\text{D}}$ and $\Gamma_{\text{D}}$ are defined as

      \begin{flalign}
      &\begin{aligned}
      &K_{p\rightarrow p}^{\text{D}_2,\text{D}_2^+}(\mathbf{x}_\perp,\mathbf{x}'_\perp) = 
      K_{p\rightarrow p,\text{dir}}^{\text{D}_2,\text{D}_2^+}(\mathbf{x}_\perp,\mathbf{x}'_\perp)
      + \alpha_{\text{refl}} K_{p\rightarrow p,\text{refl}}^{\text{D}_2,\text{D}_2^+}(\mathbf{x}_\perp,\mathbf{x}'_\perp),
      \end{aligned}&
      \end{flalign}

      \begin{flalign}
      &\begin{aligned}
      &K_{b\rightarrow p}^{\text{D}_2\text{,reem}}(\mathbf{x}_\perp,\mathbf{x'_{\perp \text{b}}}) = K_{b\rightarrow p,\text{dir}}^{\text{D}_2\text{,reem}}(\mathbf{x}_\perp,\mathbf{x'_{\perp \text{b}}})
      + \alpha_{\text{refl}} K_{b\rightarrow p,\text{refl}}^{\text{D}_2\text{,reem}}(\mathbf{x}_\perp,\mathbf{x'_{\perp \text{b}}}),
      \end{aligned}&
      \end{flalign}

      \begin{flalign}
      &\begin{aligned}
      &K_{b\rightarrow p}^{\text{D}_2\text{,refl}}(\mathbf{x}_\perp,\mathbf{x'_{\perp \text{b}}}) = K_{b\rightarrow p,\text{dir}}^{\text{D}_2\text{,refl}}(\mathbf{x}_\perp,\mathbf{x'_{\perp \text{b}}})
      + \alpha_{\text{refl}} K_{b\rightarrow p,\text{refl}}^{\text{D}_2\text{,refl}}(\mathbf{x}_\perp,\mathbf{x'_{\perp \text{b}}}),
      \end{aligned}&
      \end{flalign}
      
      \begin{flalign}
      &\begin{aligned}
      &K_{p\rightarrow b}^{\text{D}_2,\text{D}_2^+}(\mathbf{x_{\perp \text{b}}},\mathbf{x}'_\perp) = K_{p\rightarrow b,\text{dir}}^{\text{D}_2,\text{D}_2^+}(\mathbf{x_{\perp \text{b}}},\mathbf{x}'_\perp)
      + \alpha_{\text{refl}} K_{p\rightarrow b,\text{refl}}^{\text{D}_2,\text{D}_2^+}(\mathbf{x_{\perp \text{b}}},\mathbf{x}'_\perp),
      \end{aligned}&
      \end{flalign}
      
      \begin{flalign}
      &\begin{aligned}
      &K_{b\rightarrow b}^{\text{D}_2\text{,reem}}(\mathbf{x_{\perp \text{b}}},\mathbf{x'_{\perp \text{b}}}) = K_{b\rightarrow b,\text{dir}}^{\text{D}_2\text{,reem}}(\mathbf{x_{\perp \text{b}}},\mathbf{x'_{\perp \text{b}}}) + \alpha_{\text{refl}} K_{b\rightarrow b,\text{refl}}^{\text{D}_2\text{,reem}}(\mathbf{x_{\perp \text{b}}},\mathbf{x'_{\perp \text{b}}}),
      \end{aligned}&
      \end{flalign}
      
      \begin{flalign}
      &\begin{aligned}
      &K_{b\rightarrow b}^{\text{D}_2\text{,refl}}(\mathbf{x_{\perp \text{b}}},\mathbf{x'_{\perp \text{b}}}) = K_{b\rightarrow b,\text{dir}}^{\text{D}_2\text{,refl}}(\mathbf{x_{\perp \text{b}}},\mathbf{x'_{\perp \text{b}}}) + \alpha_{\text{refl}} K_{b\rightarrow b,\text{refl}}^{\text{D}_2\text{,refl}}(\mathbf{x_{\perp \text{b}}},\mathbf{x'_{\perp \text{b}}}),
      \end{aligned}&
      \end{flalign}      
      
      \begin{flalign}
      &\begin{aligned}
      &K_{p\rightarrow p}^{\text{D,D}^+}(\mathbf{x}_\perp,\mathbf{x}'_\perp) = K_{p\rightarrow p,\text{dir}}^{\text{D,D}^+}(\mathbf{x}_\perp,\mathbf{x}'_\perp) + \alpha_{\text{refl}} K_{p\rightarrow p,\text{refl}}^{\text{D,D}^+}(\mathbf{x}_\perp,\mathbf{x}'_\perp),
      \end{aligned}&
      \end{flalign}

      \begin{flalign}
      &\begin{aligned}
      & K_{p\rightarrow p}^{\text{D,D}_2^+}(\mathbf{x}_\perp,\mathbf{x}'_\perp) = K_{p\rightarrow p,\text{dir}}^{\text{D,D}_2^+}(\mathbf{x}_\perp,\mathbf{x}'_\perp) + \alpha_{\text{refl}} K_{p\rightarrow p,\text{refl}}^{\text{D,D}_2^+}(\mathbf{x}_\perp,\mathbf{x}'_\perp),
      \end{aligned}&
      \end{flalign}
      
      \begin{flalign}
      &\begin{aligned}
      & K_{p\rightarrow p}^{\text{D,diss}\left(\text{D}_2^+\right)}(\mathbf{x}_{\perp},\mathbf{x}'_\perp) = K_{b\rightarrow b,\text{dir}}^{\text{D,reem}}(\mathbf{x_{\perp \text{b}}},\mathbf{x'_{\perp \text{b}}}) + \alpha_{\text{refl}} K_{b\rightarrow b,\text{refl}}^{\text{D,reem}}(\mathbf{x_{\perp \text{b}}},\mathbf{x'_{\perp \text{b}}}),
      \end{aligned}&
      \end{flalign}
      
      \begin{flalign}
      &\begin{aligned}
      & K_{p\rightarrow p}^{\text{D,diss-rec}\left(\text{D}_2^+\right)}(\mathbf{x}_{\perp},\mathbf{x}'_\perp) = K_{b\rightarrow b,\text{dir}}^{\text{D,reem}}(\mathbf{x_{\perp \text{b}}},\mathbf{x'_{\perp \text{b}}}) + \alpha_{\text{refl}} K_{b\rightarrow b,\text{refl}}^{\text{D,reem}}(\mathbf{x_{\perp \text{b}}},\mathbf{x'_{\perp \text{b}}}),
      \end{aligned}&
      \end{flalign}
      
      \begin{flalign}
      &\begin{aligned}
      & K_{p\rightarrow p}^{\text{D,diss}\left(\text{D}_2\right)}(\mathbf{x}_{\perp},\mathbf{x}'_\perp) = K_{b\rightarrow b,\text{dir}}^{\text{D,reem}}(\mathbf{x_{\perp \text{b}}},\mathbf{x'_{\perp \text{b}}}) + \alpha_{\text{refl}} K_{b\rightarrow b,\text{refl}}^{\text{D,reem}}(\mathbf{x_{\perp \text{b}}},\mathbf{x'_{\perp \text{b}}}),
      \end{aligned}&
      \end{flalign}
      
      \begin{flalign}
      &\begin{aligned}
      & K_{p\rightarrow p}^{\text{D,diss-iz}\left(\text{D}_2\right)}(\mathbf{x}_{\perp},\mathbf{x}'_\perp) = K_{b\rightarrow b,\text{dir}}^{\text{D,reem}}(\mathbf{x_{\perp \text{b}}},\mathbf{x'_{\perp \text{b}}}) + \alpha_{\text{refl}} K_{b\rightarrow b,\text{refl}}^{\text{D,reem}}(\mathbf{x_{\perp \text{b}}},\mathbf{x'_{\perp \text{b}}}),
      \end{aligned}&
      \end{flalign}
      
      \begin{flalign}
      &\begin{aligned}
      & K_{b\rightarrow p}^{\text{D,reem}}(\mathbf{x}_\perp,\mathbf{x'_{\perp \text{b}}}) = K_{b\rightarrow b,\text{dir}}^{\text{D,reem}}(\mathbf{x_{\perp \text{b}}},\mathbf{x'_{\perp \text{b}}}) + \alpha_{\text{refl}} K_{b\rightarrow b,\text{refl}}^{\text{D,reem}}(\mathbf{x_{\perp \text{b}}},\mathbf{x'_{\perp \text{b}}}),
      \end{aligned}&
      \end{flalign}
      
      \begin{flalign}
      &\begin{aligned}
      &K_{b\rightarrow p}^{\text{D,refl}}(\mathbf{x}_\perp,\mathbf{x'_{\perp \text{b}}}) = K_{b\rightarrow b,\text{dir}}^{\text{D,reem}}(\mathbf{x_{\perp \text{b}}},\mathbf{x'_{\perp \text{b}}}) + \alpha_{\text{refl}} K_{b\rightarrow b,\text{refl}}^{\text{D,reem}}(\mathbf{x_{\perp \text{b}}},\mathbf{x'_{\perp \text{b}}}),
      \end{aligned}&
      \end{flalign}

      \begin{flalign}\label{p_to_b}
      &\begin{aligned}
      & K_{p\rightarrow b}^{\text{D,D}^+}(\mathbf{x_{\perp \text{b}}},\mathbf{x}'_\perp) = K_{b\rightarrow b,\text{dir}}^{\text{D,reem}}(\mathbf{x_{\perp \text{b}}},\mathbf{x'_{\perp \text{b}}}) + \alpha_{\text{refl}} K_{b\rightarrow b,\text{refl}}^{\text{D,reem}}(\mathbf{x_{\perp \text{b}}},\mathbf{x'_{\perp \text{b}}}),
      \end{aligned}&
      \end{flalign}

      \begin{flalign}
      &\begin{aligned}
      & K_{p\rightarrow b}^{\text{D,D}_2^+}(\mathbf{x_{\perp \text{b}}},\mathbf{x}'_\perp) = K_{b\rightarrow b,\text{dir}}^{\text{D,reem}}(\mathbf{x_{\perp \text{b}}},\mathbf{x'_{\perp \text{b}}}) + \alpha_{\text{refl}} K_{b\rightarrow b,\text{refl}}^{\text{D,reem}}(\mathbf{x_{\perp \text{b}}},\mathbf{x'_{\perp \text{b}}}),
      \end{aligned}&
      \end{flalign}
      
      \begin{flalign}
      &\begin{aligned}
      & K_{p\rightarrow b}^{\text{D,diss}\left(\text{D}_2^+\right)}(\mathbf{x_{\perp \text{b}}},\mathbf{x}'_\perp) = K_{b\rightarrow b,\text{dir}}^{\text{D,reem}}(\mathbf{x_{\perp \text{b}}},\mathbf{x'_{\perp \text{b}}}) + \alpha_{\text{refl}} K_{b\rightarrow b,\text{refl}}^{\text{D,reem}}(\mathbf{x_{\perp \text{b}}},\mathbf{x'_{\perp \text{b}}}),
      \end{aligned}&
      \end{flalign}
      
      \begin{flalign}
      &\begin{aligned}
      & K_{p\rightarrow b}^{\text{D,diss-rec}\left(\text{D}_2^+\right)}(\mathbf{x_{\perp \text{b}}},\mathbf{x}'_\perp) = K_{b\rightarrow b,\text{dir}}^{\text{D,reem}}(\mathbf{x_{\perp \text{b}}},\mathbf{x'_{\perp \text{b}}}) + \alpha_{\text{refl}} K_{b\rightarrow b,\text{refl}}^{\text{D,reem}}(\mathbf{x_{\perp \text{b}}},\mathbf{x'_{\perp \text{b}}}),
      \end{aligned}&
      \end{flalign}
      
      \begin{flalign}
      &\begin{aligned}
      & K_{p\rightarrow b}^{\text{D,diss}\left(\text{D}_2\right)}(\mathbf{x_{\perp \text{b}}},\mathbf{x}'_\perp) = K_{b\rightarrow b,\text{dir}}^{\text{D,reem}}(\mathbf{x_{\perp \text{b}}},\mathbf{x'_{\perp \text{b}}}) + \alpha_{\text{refl}} K_{b\rightarrow b,\text{refl}}^{\text{D,reem}}(\mathbf{x_{\perp \text{b}}},\mathbf{x'_{\perp \text{b}}}),
      \end{aligned}&
      \end{flalign}
      
      \begin{flalign}
      &\begin{aligned}
      & K_{p\rightarrow b}^{\text{D,diss-iz}\left(\text{D}_2\right)}(\mathbf{x_{\perp \text{b}}},\mathbf{x}'_\perp) = K_{b\rightarrow b,\text{dir}}^{\text{D,reem}}(\mathbf{x_{\perp \text{b}}},\mathbf{x'_{\perp \text{b}}}) + \alpha_{\text{refl}} K_{b\rightarrow b,\text{refl}}^{\text{D,reem}}(\mathbf{x_{\perp \text{b}}},\mathbf{x'_{\perp \text{b}}}),
      \end{aligned}&
      \end{flalign}
      
      \begin{flalign}
      &\begin{aligned}
      & K_{b\rightarrow b}^{\text{D,reem}}(\mathbf{x_{\perp \text{b}}},\mathbf{x'_{\perp \text{b}}}) = K_{b\rightarrow b,\text{dir}}^{\text{D,reem}}(\mathbf{x_{\perp \text{b}}},\mathbf{x'_{\perp \text{b}}}) + \alpha_{\text{refl}} K_{b\rightarrow b,\text{refl}}^{\text{D,reem}}(\mathbf{x_{\perp \text{b}}},\mathbf{x'_{\perp \text{b}}}),
      \end{aligned}&
      \end{flalign}
      
      \begin{flalign}
      &\begin{aligned}
      &K_{b\rightarrow b}^{\text{D,refl}}(\mathbf{x_{\perp \text{b}}},\mathbf{x'_{\perp \text{b}}}) = K_{b\rightarrow b,\text{dir}}^{\text{D,refl}}(\mathbf{x_{\perp \text{b}}},\mathbf{x'_{\perp \text{b}}}) + \alpha_{\text{refl}} K_{b\rightarrow b,\text{refl}}^{\text{D,reem}}(\mathbf{x_{\perp \text{b}}},\mathbf{x'_{\perp \text{b}}}),
      \end{aligned}&
      \end{flalign}
      
\noindent where the kernel functions for a given $\text{path} = \{\text{dir},\text{refl}\}$ are defined as 
      
      \begin{flalign}
      &\begin{aligned}
      &K_{p\rightarrow p,\text{path}}^{\text{D}_2,\text{D}_2^+}(\mathbf{x}_\perp,\mathbf{x}'_\perp) = \int_0^\infty \frac{1}{r'_{\perp}} \Phi_{\perp\left[\mathbf{v_{\perp,\text{D}_2^+}},T_{\text{D}_2^+}\right]}(\mathbf{x'_{\perp}},\mathbf{v_{\perp}}) 
      \text{exp}\left[-\frac{1}{v_\perp}\int_0^{r'_{\perp}}\nu_{\text{eff,D}_2}(\mathbf{x}''_\perp)dr''_\perp\right]dv_\perp,
      \end{aligned}&
      \end{flalign}

      \begin{flalign}
      &\begin{aligned}
      &K_{b\rightarrow p,\text{path}}^{\text{D}_2\text{,reem}}(\mathbf{x}_\perp,\mathbf{x'_{\perp \text{b}}}) = \int_0^\infty \frac{v_\perp}{r'_{\perp}}\text{cos}\theta' \chi_{\perp,\text{in,D}_2}(\mathbf{x'_{\perp \text{b}}},\mathbf{v}_\perp)
      \text{exp}\left[-\frac{1}{v_\perp}\int_0^{r'_{\perp}}\nu_{\text{eff,D}_2}(\mathbf{x}''_\perp)dr''_\perp\right]dv_\perp,
      \end{aligned}&
      \end{flalign}
      
      \begin{flalign}
      &\begin{aligned}
      &K_{b\rightarrow p,\text{path}}^{\text{D}_2\text{,refl}}(\mathbf{x}_\perp,\mathbf{x'_{\perp \text{b}}}) = \int_0^\infty \frac{1}{r'_{\perp}} \Phi_{\perp\left[\mathbf{v}_{\text{refl}\left(\text{D}_2^+\right)},T_{\text{D}_2^+}\right]}(\mathbf{x}',\mathbf{v})
      \text{exp}\left[-\frac{1}{v_\perp}\int_0^{r'_{\perp}}\nu_{\text{eff,D}_2}(\mathbf{x}''_\perp)dr''_\perp\right]dv_\perp,
      \end{aligned}&
      \end{flalign}

      \begin{flalign}
      &\begin{aligned}
      &K_{p\rightarrow b,\text{path}}^{\text{D}_2,\text{D}_2^+}(\mathbf{x_{\perp \text{b}}},\mathbf{x}'_\perp) = \int_0^\infty \frac{v_\perp}{r'_{\perp}} \text{cos}\theta \Phi_{\perp\left[\mathbf{v_{\perp,\text{D}_2^+}},T_{\text{D}_2^+}\right]}(\mathbf{x'_{\perp}},\mathbf{v_{\perp}})
      \text{exp}\left[-\frac{1}{v_\perp}\int_0^{r'_{\perp}}\nu_{\text{eff,D}_2}(\mathbf{x}''_\perp)dr''_\perp\right]dv_\perp,
      \end{aligned}&
      \end{flalign}

      \begin{flalign}
      &\begin{aligned}
      &K_{b\rightarrow b,\text{path}}^{\text{D}_2\text{,reem}}(\mathbf{x_{\perp \text{b}}},\mathbf{x'_{\perp \text{b}}}) = \int_0^\infty \frac{v_\perp^2}{r'_{\perp}}\text{cos}\theta\text{cos}\theta' \chi_{\perp,\text{in,D}_2}(\mathbf{x'_{\perp \text{b}}},\mathbf{v}_\perp)
      \text{exp}\left[-\frac{1}{v_\perp}\int_0^{r'_{\perp}}\nu_{\text{eff,D}_2}(\mathbf{x}''_\perp)dr''_\perp\right]dv_\perp,
      \end{aligned}&
      \end{flalign}
      
      \begin{flalign}
      &\begin{aligned}
      &K_{b\rightarrow b,\text{path}}^{\text{D}_2\text{,refl}}(\mathbf{x_{\perp \text{b}}},\mathbf{x'_{\perp \text{b}}}) = \int_0^\infty \frac{v_\perp}{r'_{\perp}}\text{cos}\theta \Phi_{\perp\left[\mathbf{v}_{\text{refl}\left(\text{D}_2^+\right)},T_{\text{D}_2^+}\right]}(\mathbf{x}',\mathbf{v})
      \text{exp}\left[-\frac{1}{v_\perp}\int_0^{r'_{\perp}}\nu_{\text{eff,D}_2}(\mathbf{x}''_\perp)dr''_\perp\right]dv_\perp,
      \end{aligned}&
      \end{flalign}
      
      \begin{flalign}
      &\begin{aligned}
      &K_{p\rightarrow p,\text{path}}^{\text{D,D}^+}(\mathbf{x}_\perp,\mathbf{x}'_\perp) = \int_0^\infty \frac{1}{r'_{\perp}} \Phi_{\perp\left[\mathbf{v_{\perp,\text{D}^+}},T_{\text{D}^+}\right]}(\mathbf{x'_{\perp}},\mathbf{v_{\perp}})
      \text{exp}\left[-\frac{1}{v_\perp}\int_0^{r'_{\perp}}\nu_{\text{eff,D}}(\mathbf{x}''_\perp)dr''_\perp\right]dv_\perp,
      \end{aligned}&
      \end{flalign}
      
      \begin{flalign}
      &\begin{aligned}
      & K_{p\rightarrow p,\text{path}}^{\text{D,D}_2^+}(\mathbf{x}_\perp,\mathbf{x}'_\perp) = \int_0^\infty \frac{1}{r'_{\perp}} \Phi_{\perp\left[\mathbf{v_{\perp,\text{D}_2^+}},T_{\text{D}_2^+}\right]}(\mathbf{x'_{\perp}},\mathbf{v_{\perp}})
      \text{exp}\left[-\frac{1}{v_\perp}\int_0^{r'_{\perp}}\nu_{\text{eff,D}}(\mathbf{x}''_\perp)dr''_\perp\right]dv_\perp,
      \end{aligned}&
      \end{flalign}
      
      \begin{flalign}
      &\begin{aligned}
      & K_{p\rightarrow p,\text{path}}^{\text{D,diss}\left(\text{D}_2^+\right)}(\mathbf{x}_{\perp},\mathbf{x}'_\perp) = \int_0^\infty \frac{1}{r'_{\perp,}} \Phi_{\perp\left[\mathbf{v}_{\perp,\text{D}^+_2},T_{\text{D,diss}\left(D_2^+\right)}\right]}(\mathbf{x'_{\perp}},\mathbf{v_{\perp}})
      \text{exp}\left[-\frac{1}{v_\perp}\int_0^{r'_{\perp}}\nu_{\text{eff,D}}(\mathbf{x}''_\perp)dr''_\perp\right]dv_\perp,
      \end{aligned}&
      \end{flalign}
      
      \begin{flalign}
      &\begin{aligned}
      & K_{p\rightarrow p,\text{path}}^{\text{D,diss-rec}\left(\text{D}_2^+\right)}(\mathbf{x}_{\perp},\mathbf{x}'_\perp) = \int_0^\infty \frac{1}{r'_{\perp}} \Phi_{\perp\left[\mathbf{v}_{\perp,\text{D}^+_2},T_{\text{D,diss-rec}\left(D_2^+\right)}\right]}(\mathbf{x'_{\perp}},\mathbf{v_{\perp}})
      \text{exp}\left[-\frac{1}{v_\perp}\int_0^{r'_{\perp}}\nu_{\text{eff,D}}(\mathbf{x}''_\perp)dr''_\perp\right]dv_\perp,
      \end{aligned}&
      \end{flalign}
      
      \begin{flalign}
      &\begin{aligned}
      & K_{p\rightarrow p,\text{path}}^{\text{D,diss}\left(\text{D}_2\right)}(\mathbf{x}_{\perp},\mathbf{x}'_\perp) = \int_0^\infty \frac{1}{r'_{\perp}} \Phi_{\perp\left[\mathbf{v}_{\perp,\text{D}_2},T_{\text{D,diss}\left(D_2\right)}\right]}(\mathbf{x'_{\perp}},\mathbf{v_{\perp}})
      \text{exp}\left[-\frac{1}{v_\perp}\int_0^{r'_{\perp}}\nu_{\text{eff,D}}(\mathbf{x}''_\perp)dr''_\perp\right]dv_\perp,
      \end{aligned}&
      \end{flalign}
      
      \begin{flalign}
      &\begin{aligned}
      & K_{p\rightarrow p,\text{path}}^{\text{D,diss-iz}\left(\text{D}_2\right)}(\mathbf{x}_{\perp},\mathbf{x}'_\perp) = \int_0^\infty \frac{1}{r'_{\perp}} \Phi_{\perp\left[\mathbf{v}_{\perp,\text{D}_2},T_{\text{D,diss-iz}\left(D_2\right)}\right]}(\mathbf{x'_{\perp}},\mathbf{v_{\perp}})
      \text{exp}\left[-\frac{1}{v_\perp}\int_0^{r'_{\perp}}\nu_{\text{eff,D}}(\mathbf{x}''_\perp)dr''_\perp\right]dv_\perp,
      \end{aligned}&
      \end{flalign}
      
      \begin{flalign}
      &\begin{aligned}
      & K_{b\rightarrow p,\text{path}}^{\text{D,reem}}(\mathbf{x}_\perp,\mathbf{x'_{\perp \text{b}}}) = \int_0^\infty \frac{v_\perp}{r'_{\perp}}\text{cos}\theta' \chi_{\perp,\text{in,D}}(\mathbf{x'_{\perp \text{b}}},\mathbf{v}_\perp)
      \text{exp}\left[-\frac{1}{v_\perp}\int_0^{r'_{\perp}}\nu_{\text{eff,D}}(\mathbf{x}''_\perp)dr''_\perp\right]dv_\perp,
      \end{aligned}&
      \end{flalign}
      
      \begin{flalign}
      &\begin{aligned}
      &K_{b\rightarrow p,\text{path}}^{\text{D,refl}}(\mathbf{x}_\perp,\mathbf{x'_{\perp \text{b}}}) = \int_0^\infty \frac{1}{r'_{\perp}} \Phi_{\perp\left[\mathbf{v}_{\text{refl}\left(\text{D}^+\right)},T_{\text{D}^+}\right]}(\mathbf{x}',\mathbf{v})
      \text{exp}\left[-\frac{1}{v_\perp}\int_0^{r'_{\perp}}\nu_{\text{eff,D}}(\mathbf{x}''_\perp)dr''_\perp\right]dv_\perp,
      \end{aligned}&
      \end{flalign}

      \begin{flalign}\label{p_to_b_path}
      &\begin{aligned}
      & K_{p\rightarrow b,\text{path}}^{\text{D,D}^+}(\mathbf{x_{\perp \text{b}}},\mathbf{x}'_\perp) = \int_0^\infty \frac{v_\perp}{r'_{\perp}}\text{cos}\theta \Phi_{\perp\left[\mathbf{v_{\perp,\text{D}^+}},T_{\text{D}^+}\right]}(\mathbf{x'_{\perp}},\mathbf{v_{\perp}})
      \text{exp}\left[-\frac{1}{v_\perp}\int_0^{r'_{\perp}}\nu_{\text{eff,D}}(\mathbf{x}''_\perp)dr''_\perp\right]dv_\perp,
      \end{aligned}&
      \end{flalign}

      \begin{flalign}
      &\begin{aligned}
      & K_{p\rightarrow b,\text{path}}^{\text{D,D}_2^+}(\mathbf{x_{\perp \text{b}}},\mathbf{x}'_\perp) = \int_0^\infty \frac{v_\perp}{r'_{\perp}}\text{cos}\theta \Phi_{\perp\left[\mathbf{v_{\perp,\text{D}_2^+}},T_{\text{D}_2^+}\right]}(\mathbf{x'_{\perp}},\mathbf{v_{\perp}})
      \text{exp}\left[-\frac{1}{v_\perp}\int_0^{r'_{\perp}}\nu_{\text{eff,D}}(\mathbf{x}''_\perp)dr''_\perp\right]dv_\perp,
      \end{aligned}&
      \end{flalign}
      
      \begin{flalign}
      &\begin{aligned}
      & K_{p\rightarrow b,\text{path}}^{\text{D,diss}\left(\text{D}_2^+\right)}(\mathbf{x_{\perp \text{b}}},\mathbf{x}'_\perp) = \int_0^\infty \frac{v_\perp}{r'_{\perp}}\text{cos}\theta \Phi_{\perp\left[\mathbf{v}_{\perp,\text{D}^+_2},T_{\text{D,diss}\left(D_2^+\right)}\right]}(\mathbf{x'_{\perp}},\mathbf{v_{\perp}})
      \text{exp}\left[-\frac{1}{v_\perp}\int_0^{r'_{\perp}}\nu_{\text{eff,D}}(\mathbf{x}''_\perp)dr''_\perp\right]dv_\perp,
      \end{aligned}&
      \end{flalign}
      
      \begin{flalign}
      &\begin{aligned}
      & K_{p\rightarrow b,\text{path}}^{\text{D,diss-rec}\left(\text{D}_2^+\right)}(\mathbf{x_{\perp \text{b}}},\mathbf{x}'_\perp) = \int_0^\infty \frac{v_\perp}{r'_{\perp}}\text{cos}\theta \Phi_{\perp\left[\mathbf{v}_{\perp,\text{D}^+_2},T_{\text{D,diss-rec}\left(D_2^+\right)}\right]}(\mathbf{x'_{\perp}},\mathbf{v_{\perp}})
      \text{exp}\left[-\frac{1}{v_\perp}\int_0^{r'_{\perp}}\nu_{\text{eff,D}}(\mathbf{x}''_\perp)dr''_\perp\right]dv_\perp,
      \end{aligned}&
      \end{flalign}
      
      \begin{flalign}
      &\begin{aligned}
      & K_{p\rightarrow b,\text{path}}^{\text{D,diss}\left(\text{D}_2\right)}(\mathbf{x_{\perp \text{b}}},\mathbf{x}'_\perp) = \int_0^\infty \frac{v_\perp}{r'_{\perp}}\text{cos}\theta \Phi_{\perp\left[\mathbf{v}_{\perp,\text{D}_2},T_{\text{D,diss}\left(D_2\right)}\right]}(\mathbf{x'_{\perp}},\mathbf{v_{\perp}})
      \text{exp}\left[-\frac{1}{v_\perp}\int_0^{r'_{\perp}}\nu_{\text{eff,D}}(\mathbf{x}''_\perp)dr''_\perp\right]dv_\perp,
      \end{aligned}&
      \end{flalign}
      
      \begin{flalign}
      &\begin{aligned}
      & K_{p\rightarrow b,\text{path}}^{\text{D,diss-iz}\left(\text{D}_2\right)}(\mathbf{x_{\perp \text{b}}},\mathbf{x}'_\perp) = \int_0^\infty \frac{v_\perp}{r'_{\perp}}\text{cos}\theta \Phi_{\perp\left[\mathbf{v}_{\perp,\text{D}_2},T_{\text{D,diss-iz}\left(D_2\right)}\right]}(\mathbf{x'_{\perp}},\mathbf{v_{\perp}})
      \text{exp}\left[-\frac{1}{v_\perp}\int_0^{r'_{\perp}}\nu_{\text{eff,D}}(\mathbf{x}''_\perp)dr''_\perp\right]dv_\perp,
      \end{aligned}&
      \end{flalign}
      
      \begin{flalign}
      &\begin{aligned}
      & K_{b\rightarrow b,\text{path}}^{\text{D,reem}}(\mathbf{x_{\perp \text{b}}},\mathbf{x'_{\perp \text{b}}}) = \int_0^\infty \frac{v_\perp^2}{r'_{\perp}}\text{cos}\theta\text{cos}\theta' \chi_{\perp, \text{in,D}}(\mathbf{x'_{\perp \text{b}}},\mathbf{v}_\perp)
      \text{exp}\left[-\frac{1}{v_\perp}\int_0^{r'_{\perp}}\nu_{\text{eff,D}}(\mathbf{x}''_\perp)dr''_\perp\right]dv_\perp,
      \end{aligned}&
      \end{flalign}
      
      \begin{flalign}
      &\begin{aligned}
      &K_{b\rightarrow b,\text{path}}^{\text{D,refl}}(\mathbf{x_{\perp \text{b}}},\mathbf{x'_{\perp \text{b}}}) = \int_0^\infty \frac{v_\perp}{r'_{\perp}}\text{cos}\theta \Phi_{\perp\left[\mathbf{v}_{\text{refl}\left(\text{D}^+\right)},T_{\text{D}^+}\right]}(\mathbf{x}',\mathbf{v})
      \text{exp}\left[-\frac{1}{v_\perp}\int_0^{r'_{\perp}}\nu_{\text{eff,D}}(\mathbf{x}''_\perp)dr''_\perp\right]dv_\perp.
      \end{aligned}&
      \end{flalign}
      
\noindent We remark that all velocity distributions given by a Maxwellian or a Knudsen cosine law are integrated along the parallel velocity, that is

\begin{flalign}
&\begin{aligned}
\Phi_{\perp\left[\mathbf{v_{\perp,\text{D}_2^+}},T_{\text{D}_2^+}\right]}(\mathbf{x'_{\perp}},\mathbf{v_{\perp}}) = \int_0^\infty \Phi_{\left[\mathbf{v_{\perp,\text{D}_2^+}},T_{\text{D}_2^+}\right]}(\mathbf{x'_{\perp}},\mathbf{v_{\perp}}) dv_{\|},
\end{aligned}&
\end{flalign}

\begin{flalign}
&\begin{aligned}
\Phi_{\perp\left[\mathbf{v_{\perp,\text{D}^+}},T_{\text{D}^+}\right]}(\mathbf{x'_{\perp}},\mathbf{v_{\perp}}) = \int_0^\infty \Phi_{\left[\mathbf{v_{\perp,\text{D}^+}},T_{\text{D}^+}\right]}(\mathbf{x'_{\perp}},\mathbf{v_{\perp}}) dv_{\|},
\end{aligned}&
\end{flalign}

\begin{flalign}
&\begin{aligned}
\Phi_{\perp\left[\mathbf{v}_{\perp,\text{D}_2},T_{\text{D,diss}\left(\text{D}_2\right)}\right]}(\mathbf{x'_{\perp}},\mathbf{v_{\perp}}) = \int_0^\infty \Phi_{\mathbf{v_{\perp,\text{D}_2}},T_{\text{D,diss}\left(\text{D}_2\right)}}(\mathbf{x'_{\perp}},\mathbf{v_{\perp}}) dv_{\|}.,
\end{aligned}&
\end{flalign}

\begin{flalign}
&\begin{aligned}
\Phi_{\perp\left[\mathbf{v}_{\perp,\text{D}_2},T_{\text{D,diss-iz}\left(\text{D}_2\right)}\right]}(\mathbf{x'_{\perp}},\mathbf{v_{\perp}}) = \int_0^\infty \Phi_{\left[\mathbf{v_{\perp,\text{D}_2}},T_{\text{D,diss-iz}\left(\text{D}_2\right)}\right]}(\mathbf{x'_{\perp}},\mathbf{v_{\perp}}) dv_{\|},
\end{aligned}&
\end{flalign}

\begin{flalign}
&\begin{aligned}
\Phi_{\perp\left[\mathbf{v}_{\perp,\text{D}^+_2},T_{\text{D,diss}\left(D_2^+\right)}\right]}(\mathbf{x'_{\perp}},\mathbf{v_{\perp}}) = \int_0^\infty \Phi_{\left[\mathbf{v_{\perp,\text{D}^+_2}},T_{\text{D,diss}\left(D_2^+\right)}\right]}(\mathbf{x'_{\perp}},\mathbf{v_{\perp}}) dv_{\|},
\end{aligned}&
\end{flalign}

\begin{flalign}
&\begin{aligned}
\Phi_{\perp\left[\mathbf{v}_{\perp,\text{D}^+_2},T_{\text{D,diss-rec}\left(\text{D}_2^+\right)}\right]}(\mathbf{x'_{\perp}},\mathbf{v_{\perp}}) = \int_0^\infty \Phi_{\left[\mathbf{v_{\perp,\text{D}^+_2}},T_{\text{D,diss-rec}\left(\text{D}_2^+\right)}\right]}(\mathbf{x'_{\perp}},\mathbf{v_{\perp}}) dv_{\|},
\end{aligned}&
\end{flalign}

\begin{flalign}
&\begin{aligned}
\Phi_{\perp\left[\mathbf{v}_{\text{refl}\left(\text{D}^+\right)},T_{\text{D}^+}\right]}(\mathbf{x}',\mathbf{v}) = \int_0^{\infty} \Phi_{\left[\mathbf{v}_{\text{refl}\left(\text{D}^+\right)},T_{\text{D}^+}\right]}(\mathbf{x}',\mathbf{v}) dv_{\|},
\end{aligned}&
\end{flalign}

\begin{flalign}
&\begin{aligned}
\Phi_{\perp\left[\mathbf{v}_{\text{refl}\left(\text{D}_2^+\right)},T_{\text{D}_2^+}\right]}(\mathbf{x}',\mathbf{v}) = \int_0^{\infty} \Phi_{\left[\mathbf{v}_{\text{refl}\left(\text{D}_2^+\right)},T_{\text{D}_2^+}\right]}(\mathbf{x}',\mathbf{v}) dv_{\|},
\end{aligned}&
\end{flalign}

\begin{flalign}
&\begin{aligned}
\chi_{\perp,\text{in,D}_2}(\mathbf{x'_{\perp,\text{b}}},\mathbf{v_{\perp}}) = \int_0^\infty \chi_{\text{in,D}_2}(\mathbf{x'_{\perp,\text{b}}},\mathbf{v_{\perp}}) dv_{\|},
\end{aligned}&
\end{flalign}

\begin{flalign}
&\begin{aligned}
\chi_{\perp,\text{in,D}}(\mathbf{x'_{\perp,\text{b}}},\mathbf{v_{\perp}}) = \int_0^\infty \chi_{\text{in,D}}(\mathbf{x'_{\perp,\text{b}}},\mathbf{v_{\perp}}) dv_{\|}.
\end{aligned}&
\end{flalign}

\section*{Appendix D: Numerical solution of the neutral equations}

The coupled neutral equations for $\text{D}_2$ and $\text{D}$, Eqs. (\ref{nD2}-\ref{GammaD}), may be discretized as a linear matrix system, $\mathbf{x} = A \mathbf{x} + \mathbf{b}$, with the unknown $\mathbf{x}$ representing the density and boundary flux of the $\text{D}_2$ and $\text{D}$ species. Indicating with $N_P$ the number of points that discretize the poloidal plane and $N_B$ the number of points discretizing the boundary, $\mathbf{x}$ is a vector of size $2(N_P+N_B)$, $A$ is a $2(N_P+N_B)\times 2(N_P+N_B)$ matrix and $\mathbf{b}$ is a $2(N_P+N_B)$ vector that includes all contributions not proportional to the neutral density or flux, namely the effect of recombination of $\text{D}^+$ and $\text{D}_2^+$ with electrons, the effect of dissociative processes to which $\text{D}_2^+$ ions are subject and the contributions from the flux of $\text{D}^+$ and $\text{D}_2^+$ ions to the boundary.

The matrix $M$, and the vectors $\mathbf{x}$ and $\mathbf{b}$ can then be written as

\begin{equation}\label{mat_sys}
\mathbf{x} 
=
\begin{bmatrix}
n_{\text{D}} \\
\Gamma_{\text{out,D}} \\
n_{\text{D}_2} \\
\Gamma_{\text{out,D}_2}
\end{bmatrix},
M 
=
\begin{bmatrix}
M_{11} \ \ \ M_{12} \ \ \ M_{13} \ \ \ M_{14} \\
M_{21} \ \ \ M_{22} \ \ \ M_{23} \ \ \ M_{24} \\
M_{31} \ \ \ M_{32} \ \ \ M_{33} \ \ \ M_{34} \\
M_{41} \ \ \ M_{42} \ \ \ M_{43} \ \ \ M_{44} 
\end{bmatrix},
\mathbf{b}
=
\begin{bmatrix}
b_1 \\
b_2 \\
b_3 \\
b_4
\end{bmatrix},
\end{equation}

where $M_{11}$ is a matrix of size $N_P\times N_P$, 

      \begin{flalign}
      &\begin{aligned}
      & M_{11} = \nu_{\text{cx,D}} K_{p\rightarrow p}^{\text{D,D}^+},
      \end{aligned}&
      \end{flalign}
      
\noindent that discretizes the kernel $K_{p\rightarrow p}^{\text{D,D}^+}$ defined in Eq. (\ref{kernels_dr}) at the spatial points where $n_{\text{D}}$ is evaluated. The matrix

      \begin{flalign}
      &\begin{aligned}
      & M_{21} = \nu_{\text{cx,D}} K_{p\rightarrow b}^{\text{D,D}^+},
      \end{aligned}&
      \end{flalign}
      
\noindent has size $N_B\times N_P$ and discretizes the kernel $K_{p\rightarrow b}^{\text{D,D}^+}$ defined in Eq. (\ref{p_to_b}) at the points where $\Gamma_{\text{D}}$ is evaluated. The other matrices appearing in the definition of $M$ are defined similarly,
      
      \begin{flalign}
      &\begin{aligned}
      & M_{31} = \left[\frac{n_{\text{D}_2^+}}{n_{\text{D}}}\nu_{\text{cx,D}_2^+-\text{D}}\right] K_{p\rightarrow p}^{\text{D}_2,\text{D}_2^+},
      \end{aligned}&
      \end{flalign}

      \begin{flalign}
      &\begin{aligned}
      & M_{41} = \left[\frac{n_{\text{D}_2^+}}{n_{\text{D}}}\nu_{\text{cx,D}_2^+-\text{D}}\right] K_{p\rightarrow b}^{\text{D}_2,\text{D}_2^+},
      \end{aligned}&
      \end{flalign}
      
      \begin{flalign}
      &\begin{aligned}
      & M_{12} = (1-\alpha_{\text{refl}}) (1-\beta_{\text{assoc}}) K_{b\rightarrow p}^{\text{D,reem}},
      \end{aligned}&
      \end{flalign}

      \begin{flalign}
      &\begin{aligned}
      & M_{22} = (1-\alpha_{\text{refl}}) (1-\beta_{\text{assoc}}) K_{b\rightarrow b}^{\text{D,reem}},
      \end{aligned}&
      \end{flalign}
      
      \begin{flalign}
      &\begin{aligned}
      & M_{32} = (1-\alpha_{\text{refl}}) \frac{\beta_{\text{assoc}}}{2} K_{b\rightarrow p}^{\text{D}_2\text{,reem}},
      \end{aligned}&
      \end{flalign}

      \begin{flalign}
      &\begin{aligned}
      & M_{42} = (1-\alpha_{\text{refl}}) \frac{\beta_{\text{assoc}}}{2} K_{b\rightarrow b}^{\text{D}_2\text{,reem}},
      \end{aligned}&
      \end{flalign}

      \begin{flalign}
      &\begin{aligned}
      & M_{13} = \nu_{\text{cx,D}_2-\text{D}^+} K_{p\rightarrow p}^{\text{D,D}^+} + \nu_{\text{diss,D}_2} K_{p\rightarrow p}^{\text{D,diss}\left(\text{D}_2\right)} + \nu_{\text{diss-iz,D}_2} K_{p\rightarrow p}^{\text{D,diss-iz}\left(\text{D}_2\right)},
      \end{aligned}&
      \end{flalign}

      \begin{flalign}
      &\begin{aligned}
      & M_{23} = \nu_{\text{cx,D}_2-\text{D}^+} K_{p\rightarrow b}^{\text{D,D}^+} + \nu_{\text{diss,D}_2} K_{p\rightarrow b}^{\text{D,diss}\left(\text{D}_2\right)} + \nu_{\text{diss-iz,D}_2} K_{p\rightarrow b}^{\text{D,diss-iz}\left(\text{D}_2\right)},
      \end{aligned}&
      \end{flalign}
      
      \begin{flalign}
      &\begin{aligned}
      & M_{33} = \nu_{\text{cx,D}_2} K_{p\rightarrow p}^{\text{D}_2,\text{D}_2^+},
      \end{aligned}&
      \end{flalign}

      \begin{flalign}
      &\begin{aligned}
      & M_{43} = \nu_{\text{cx,D}_2} K_{p\rightarrow b}^{\text{D}_2,\text{D}_2^+},
      \end{aligned}&
      \end{flalign}
      
      \begin{flalign}
      &\begin{aligned}
      & M_{14} = 0,
      \end{aligned}&
      \end{flalign}

      \begin{flalign}
      &\begin{aligned}
      & M_{24} = 0,
      \end{aligned}&
      \end{flalign}
      
      \begin{flalign}
      &\begin{aligned}
      & M_{34} = (1-\alpha_{\text{refl}}) K_{b\rightarrow p}^{\text{D}_2},
      \end{aligned}&
      \end{flalign}

      \begin{flalign}
      &\begin{aligned}
      & M_{44} = (1-\alpha_{\text{refl}}) K_{b\rightarrow b}^{\text{D}_2},
      \end{aligned}&
      \end{flalign}

The vector $\mathbf{b}$ is defined through the vectors $b_1$ and $b_3$ of size $N_P$,

      \begin{flalign}
      &\begin{aligned}
      & b_1 = n_{\text{D}[\text{rec}(\text{D}^+)]}(\mathbf{x}_\perp) + n_{\text{D}[\text{diss}(\text{D}_2^+)]}(\mathbf{x}_\perp)  + n_{\text{D}[\text{out}(\text{D}^+)]}(\mathbf{x}_\perp),
      \end{aligned}&
      \end{flalign}

      \begin{flalign}
      &\begin{aligned}
      & b_3 = n_{\text{D}_2[\text{rec}(\text{D}_2^+)]}(\mathbf{x}_\perp) + n_{\text{D}_2[\text{out}(\text{D}_2^+)]}(\mathbf{x}_\perp) + n_{\text{D}_2[\text{out}(\text{D}^+)]}(\mathbf{x}_\perp),
      \end{aligned}&
      \end{flalign}
      
\noindent and the vector $b_2$ and $b_4$ of size $N_B$,
      
      \begin{flalign}
      &\begin{aligned}
      & b_2 = \Gamma_{\text{out},\text{D}[\text{rec}(\text{D}^+)]}(\mathbf{x}_\perp) + \Gamma_{\text{out},\text{D}[\text{diss}(\text{D}_2^+)]}(\mathbf{x}_\perp) + \Gamma_{\text{out,D}[\text{out}(\text{D}^+)]}(\mathbf{x}_\perp),
      \end{aligned}&
      \end{flalign}
      
      \begin{flalign}
      &\begin{aligned}
      & b_4 = \Gamma_{\text{out,D}_2[\text{rec}(\text{D}_2^+)]}(\mathbf{x}_\perp) + \Gamma_{\text{out,D}_2[\text{out}(\text{D}_2^+)]}(\mathbf{x}_\perp) + \Gamma_{\text{out,D}_2[\text{out}(\text{D}^+)]}(\mathbf{x}_\perp).
      \end{aligned}&
      \end{flalign}

It is remarked that the vector $\mathbf{b}$ can also be written as $\mathbf{b} = N \mathbf{x}_{\text{i}}$, where $\mathbf{x}_{\text{i}}$ refers to the densities and boundary fluxes of the $\text{D}^+$ and $\text{D}_2^+$ ion species, 

\begin{equation}\label{vecxi}
\mathbf{x}_{\text{i}}=
\begin{bmatrix}
n_{\text{D}^+} \\
\Gamma_{\text{out,D}^+} \\
n_{\text{D}_2^+} \\
\Gamma_{\text{out,D}_2^+}
\end{bmatrix},
\end{equation}

\noindent and the matrix $N$ can be expressed as

\begin{equation}\label{matN}
N
=
\begin{bmatrix}
N_{11} \ \ \ N_{12} \ \ \ N_{13} \ \ \ N_{14} \\
N_{21} \ \ \ N_{22} \ \ \ N_{23} \ \ \ N_{24} \\
N_{31} \ \ \ N_{32} \ \ \ N_{33} \ \ \ N_{34} \\
N_{41} \ \ \ N_{42} \ \ \ N_{43} \ \ \ N_{44} 
\end{bmatrix},
\end{equation}

\noindent with entries

      \begin{flalign}
      &\begin{aligned}
      & N_{11} = \nu_{\text{rec,D}^+} K_{p\rightarrow p}^{\text{D,D}^+},
      \end{aligned}&
      \end{flalign}

      \begin{flalign}
      &\begin{aligned}
      & N_{21} = \nu_{\text{rec,D}^+} K_{p\rightarrow b}^{\text{D,D}^+},
      \end{aligned}&
      \end{flalign}
      
      \begin{flalign}
      &\begin{aligned}
      & N_{31} = \nu_{\text{rec,D}_2^+} K_{p\rightarrow p}^{\text{D}_2,\text{D}_2^+},
      \end{aligned}&
      \end{flalign}

      \begin{flalign}
      &\begin{aligned}
      & N_{41} = \nu_{\text{rec,D}_2^+} K_{p\rightarrow b}^{\text{D}_2,\text{D}_2^+},
      \end{aligned}&
      \end{flalign}
      
      \begin{flalign}
      &\begin{aligned}
      & N_{12} = (1-\alpha_{\text{refl}}) (1-\beta_{\text{assoc}}) K_{b\rightarrow p}^{\text{D,reem}} + \alpha_{\text{refl}} K_{b\rightarrow p}^{\text{D,refl}},
      \end{aligned}&
      \end{flalign}

      \begin{flalign}
      &\begin{aligned}
      & N_{22} = (1-\alpha_{\text{refl}}) (1-\beta_{\text{assoc}}) K_{b\rightarrow b}^{\text{D,reem}} + \alpha_{\text{refl}} K_{b\rightarrow b}^{\text{D,refl}},
      \end{aligned}&
      \end{flalign}
      
      \begin{flalign}
      &\begin{aligned}
      & N_{32} = (1-\alpha_{\text{refl}}) \frac{\beta_{\text{assoc}}}{2} K_{b\rightarrow p}^{\text{D}_2\text{,reem}},
      \end{aligned}&
      \end{flalign}

      \begin{flalign}
      &\begin{aligned}
      & N_{42} = (1-\alpha_{\text{refl}}) \frac{\beta_{\text{assoc}}}{2} K_{b\rightarrow b}^{\text{D}_2\text{,reem}},
      \end{aligned}&
      \end{flalign}

      \begin{flalign}
      &\begin{aligned}
      & N_{13} = \nu_{\text{diss,D}_2^+} K_{p\rightarrow p}^{\text{D,diss}\left(\text{D}_2^+\right)} + 2 \nu_{\text{diss-rec,D}_2^+} K_{p\rightarrow p}^{\text{D,diss-rec}\left(\text{D}_2^+\right)},
      \end{aligned}&
      \end{flalign}

      \begin{flalign}
      &\begin{aligned}
      & N_{23} = \nu_{\text{diss,D}_2^+} K_{p\rightarrow b}^{\text{D,diss}\left(\text{D}_2^+\right)} + 2 \nu_{\text{diss-rec,D}_2^+} K_{p\rightarrow b}^{\text{D,diss-rec}\left(\text{D}_2^+\right)},
      \end{aligned}&
      \end{flalign}
      
      \begin{flalign}
      &\begin{aligned}
      & N_{33} = \nu_{\text{rec,D}_2^+} K_{p\rightarrow p}^{\text{D}_2,\text{D}_2^+},
      \end{aligned}&
      \end{flalign}

      \begin{flalign}
      &\begin{aligned}
      & N_{43} = \nu_{\text{rec,D}_2^+} K_{p\rightarrow b}^{\text{D}_2,\text{D}_2^+},
      \end{aligned}&
      \end{flalign}
      
      \begin{flalign}
      &\begin{aligned}
      & N_{14} = 0,
      \end{aligned}&
      \end{flalign}

      \begin{flalign}
      &\begin{aligned}
      & N_{24} = 0,
      \end{aligned}&
      \end{flalign}
      
      \begin{flalign}
      &\begin{aligned}
      & N_{34} = (1-\alpha_{\text{refl}}) K_{b\rightarrow p}^{\text{D}_2\text{,reem}},
      \end{aligned}&
      \end{flalign}

      \begin{flalign}
      &\begin{aligned}
      & N_{44} = (1-\alpha_{\text{refl}}) K_{b\rightarrow b}^{\text{D}_2\text{,reem}} + \alpha_{\text{refl}} K_{b\rightarrow b}^{\text{D}_2\text{,refl}}.
      \end{aligned}&
      \end{flalign}
      
We remark that a convergence study to estimate the error introduced by the discretization of the neutral equation was carried out for a single neutral species model and it is reported in \cite{Giacomin2021}.

\section*{ACKNOWLEDGEMENTS}

We would like to thank C. Theiler, C. Wersal, D. Mancini, K. Verhaegh and M. Wensing for useful discussions. The  simulations presented herein were carried out in part at CSCS (Swiss National Supercomputing Center) under the project ID s882 and in part on the CINECA Marconi supercomputer under the GBSedge project. This work has been carried out within the framework of the EUROfusion Consortium and has received funding from the Fond National Suisse de la Recherche Scientifique and from the Euratom research and training programme 2014-2018 and 2019-2020 under grant agreement No 633053. The views and opinions expressed herein do not necessarily reflect those of the European Commission.

\section*{DATA AVAILABILITY}
The data that support the findings of this study are available from the corresponding authors upon reasonable request.

\section*{REFERENCES}

\bibliographystyle{unsrt}
\bibliography{article}


\end{document}